\documentclass[useAMS, twocolumn, usenatbib]{mn2e}
\usepackage{graphicx}
\usepackage{amsmath, amssymb}
\usepackage{natbib}
\usepackage{pdflscape}
\usepackage{color}

\def\vec#1{\mbox{\boldmath $#1$}}

\newcommand{\be}{\begin{equation}}
\newcommand{\ee}{\end{equation}}
\newcommand       \bea          {\begin{eqnarray}}
\newcommand       \eea          {\end{eqnarray}}

\begin{document}

\title[Empirical Rates of Stellar Tidal Disruption]{Rates of Stellar Tidal Disruption as Probes of the Supermassive Black Hole Mass Function}
\author[Stone \& Metzger]{Nicholas C.~Stone$^{1}$, Brian D.~Metzger$^{1}$
\\$^{1}$Columbia Astrophysics Laboratory, Columbia University, New York, NY, 10027}
\maketitle

\begin{abstract}
Rates of stellar tidal disruption events (TDEs) by supermassive black holes (SMBHs) due to two-body relaxation are calculated using a large galaxy sample ($N \approx 200$) in order to explore the sensitivity of the TDE rates to observational uncertainties, such as the parametrization of galaxy light profiles and the stellar mass function.  The largest uncertainty arises due to the poorly constrained occupation fraction of SMBHs in low-mass galaxies, which otherwise dominate the total TDE rate.  The detection rate of TDE flares by optical surveys is calculated as a function of SMBH mass and other observables for several physically-motivated models of TDE emission.  We also quantify the fraction of galaxies that produce deeply penetrating disruption events.  If the majority of the detected events are characterized by super-Eddington luminosities (such as disk winds, or synchrotron radiation from an off-axis relativistic jet), then the measured SMBH mass distribution will tightly constrain the low-end SMBH occupation fraction.  If Eddington-limited emission channels dominate, however, then the occupation fraction sensitivity is much less pronounced in a flux-limited survey (although still present in a volume-complete event sample).  The SMBH mass distribution of the current sample of TDEs, though highly inhomogeneous and encumbered by selection effects, already suggests that Eddington-limited emission channels dominate.  Even our most conservative rate estimates appear to be in tension with much lower observationally inferred TDE rates, and we discuss several possible resolutions to this discrepancy.

\end{abstract}

\section{Introduction}

Stars are tidally disrupted in galactic nuclei when orbital perturbations reduce their angular momentum and place them on nearly radial orbits.  Once the stellar pericenter is reduced below a critical value, a strong tidal encounter with the central supermassive black hole (SMBH) destroys it on an orbital time \citep{Hills75}.  Roughly half of the stellar mass falls back onto the SMBH, circularizing into an accretion disk and powering a luminous flare \citep{Rees88}.  

Approximately two dozen of these tidal disruption events (TDEs) have been observed, found with a diverse mixture of optical \citep{vanVelzen+11, Cenko+12b, Gezari+12, Arcavi+14, Chorno+14}, UV \citep{Gezari+06, Gezari+08, Gezari+09}, X-ray \citep{Bade+96, KomGre99, Maksym+13}, and gamma ray \citep{Bloom+11, Levan+11, Zauderer+11, Cenko+12} detections (see \citealt{Gezari13} for a review).  The mass fallback curves of these events encode information on the mass, radius, and structure of the disrupted star \citep{Lodato+09, GuiRam13}, the SMBH mass \citep{Rees88, EvaKoc89}, and, more subtly, the pericenter of disruption \citep{GuiRam13, Stone+13}.  TDE light curves and spectra can in principle be used to measure these dynamical quantities, although it is currently unclear how mass fallback rates translate into luminosities.  The spin of the SMBH may also be imprinted into these observables \citep{StoLoe12, Kesden12a, Lei+13, SheMat14}.  Individual TDEs are valuable tools for probing SMBHs in distant galactic nuclei, but in this paper we focus on the information that can be obtained from large statistical samples of TDEs, and in particular from the rates of these events and their distributions of parameters, such as SMBH mass.   

The rates of stellar TDEs are currently uncertain; typically a value $\sim 10^{-5}~{\rm yr}^{-1}$ per galaxy is inferred from X-ray (e.g.~\citealt{Donley+02}), UV (\citealt{Gezari+08}) and optically (e.g., ~\citealt{vanVelzen&Farrar14}) selected events.  These observed rates are generally an order of magnitude lower than previous theoretical predictions of $\gtrsim 10^{-4}$ yr$^{-1}$ gal$^{-1}$ (e.g., the two-body relaxation calculations of \citealt{MagTre99,Wang&Merritt04}), although selection effects, small number statistics, and the possible influence of dust or photoelectric extinction could all contribute to this disagreement.  The rates of TDEs accompanied by relativistic jets appear to be smaller still (e.g.~\citealt{Bower+13}, \citealt{VanVelzen+13}).  Many candidate TDE flares furthermore possess much higher optical luminosities than predicted by previous models for TDE emission (e.g.,~\citealt{Gezari+12,Arcavi+14}); some of the tension in rates could thus be alleviated if the detected events represent only the brightest tail of the luminosity function.  Conversely, two body relaxation is generally thought to set a conservative floor on the true TDE rate, which can be enhanced by more exotic dynamical processes.  This would worsen the existing tension between observed and theoretical rates.

Fortunately, the observed sample of candidate TDEs is expanding rapidly, especially at optical frequencies, due to the advent of wide-field sensitive surveys such as Pan-STARRs (\citealt{Kaiser+02}) and the Palomar Transient Factory (soon to be Zwicky Transient Facility; \citealt{Kulkarni12}).  The study of TDEs will be further revolutionized in the next decade by the Large Synoptic Survey Telescope (LSST) and by the wide-field X-ray satellite eROSITA (\citealt{Merloni+12}), which could detect hundreds or thousands of TDEs per year \citep{Gezari+08, StrQua09, vanVelzen+11, Khabibulli+14}.  One potential drawback of the future LSST/eROSITA era is that the large rate of detections could overwhelm what resources are available for photometric or spectroscopic follow-up observations.  This situation enhances the value of statistical studies of TDEs, such as how the TDE rate varies with galaxy type or SMBH mass.  

Past theoretical work has estimated TDE rates by deprojecting high resolution surface brightness profiles for small samples of galaxies, and making simplifying assumptions to obtain stellar densities and distribution functions (\citealt{MagTre99, Wang&Merritt04}).  A key conclusion of these works is that the rate of TDEs is dominated by the lowest mass galaxies that host black holes.  This is due to a combination of three factors: the larger numbers of small galaxies, the negative correlation between SMBH mass and central stellar density, and the nontrivial dynamical result that the TDE rate is higher for lower SMBH masses in steeply sloped (``cuspy'') galaxies\footnote{\citet{Wang&Merritt04} find that in cuspy stellar profiles, the TDE rate scales roughly inversely with SMBH mass.}.  Although the intrinsic TDE rate is thus likely to be highest among the smallest extant SMBHs, this is not necessarily true of the {\it detected} TDE rate, because the latter also depends on how the TDE luminosity in a given waveband scales with the SMBH mass.  

In this paper, we update past theoretical estimates of stellar tidal disruption rates, using the newest calibration of the $M_\bullet-\sigma$ relation and a much larger sample of galaxy surface brightness profiles than was employed in the past ($\S\ref{sec:TDErates}$).  We examine the robustness of the TDE rate to a number of uncertainties, including the stellar mass function, the SMBH mass function, choices of galaxy scaling relations, and the (somewhat arbitrary) choice of surface brightness parametrization.  Whenever possible, we make a conservative choice of assumptions, to test the robustness of the disagreement between theory and observation.  Our main conclusions are that the poorly constrained occupation fraction of SMBHs in low mass galaxies represents the largest current uncertainty in the intrinsic TDE rate, and that the tension between theory and observation is persistent.

We briefly overview the observable quantities of interest for TDEs (\S \ref{sec:observables}), and then translate our volumetric rates into detection rates by future surveys ($\S\ref{sec:results}$), considering several different models for the optical light curves of TDEs.   An analytic parameterization of the SMBH occupation fraction is used to clarify how TDE samples can be used to constrain the ubiquity of low mass SMBHs.  We also estimate for the first time the relative abundances of deeply penetrating and grazing tidal disruptions.  We next discuss our results ($\S\ref{sec:discussion}$) in the context of the current TDE sample and describe possible resolutions to the ``rates dilemma" discussed above.  Finally ($\S\ref{sec:conclusions}$), we provide a bulleted summary of our conclusions.    

Appendix \ref{sec:analytic} provides a derivation of closed-form analytic expressions for several theoretical quantities of interest (e.g. orbit-averaged diffusion coefficients and per-energy flux of stars into the loss cone) in limiting regimes.  In Appendix \ref{sec:optical} we review four different optical emission mechanisms, and in several cases update or improve existing theoretical models for these.  Our full results are tabulated in Appendix \ref{sec:fullResults}.

\section{TDE Rates}
\label{sec:TDErates}

Although many different dynamical processes may contribute to observed rates of tidal disruption, the most robust and ubiquitous is the collisional two-body relaxation of stars into the phase space ``loss cone,'' the region of $\{\vec{x}, \vec{v}\}$ space where orbital pericenters $r_{\rm p}$ are less than the tidal radius $r_{\rm t}$ and stars are destroyed on a dynamical time (e.g.~\citealt{Alexander12} for a review).

Other processes may contribute to, and in some cases, dominate, observed TDE rates.  Secular resonances in the vicinity of an SMBH may lead to ``resonant relaxation'' of angular momentum \citep{RauTre96}, although this is likely a subdominant contributor to the total TDE rate \citep{RauIng98, HopAle06, Madiga+11}.  Massive perturbers, such as intermediate mass black holes (IMBHs), giant molecular clouds, or infalling globular clusters, can strongly perturb stellar orbits and lead to rapid refilling of an empty loss cone \citep{Perets+07}.  Analogous dynamical processes in the vicinity of an SMBH binary can temporarily enhance the TDE rate \citep{Ivanov+05, Chen+09, Chen+11, StoLoe11}, potentially by several orders of magnitude, but SMBH mergers are sufficiently rare that this channel nonetheless contributes subdominantly to the total TDE rate \citep{WegBod14}.  Finally, in nonspherical stellar systems, non-conservation of angular momentum allows stars to ergodically explore a large portion of phase space, and some will wander into the loss cone even absent collisional relaxation.  This modestly enhances TDE rates in axisymmetric systems \citep{MagTre99, VasMer13} and can increase them dramatically in triaxial ones \citep{MerPoo04}.  In general, all of these processes involve much greater observational and theoretical uncertainty than does simple two-body relaxation, and we therefore neglect them in the remainder of this paper.  Neglecting these additional processes is in part justified by the fact that the observed TDE rate is already significantly less than the minimum rate estimated from two-body relaxation alone.    

TDE rates in spherical star clusters containing massive black holes were first estimated in an analytic way \citep{FraRee76}, which was quickly supplemented by semi-analytic calculations \citep{LigSha77, Cohn&Kulsrud78} treating angular momentum diffusion as a Fokker-Planck process.  The results of these semi-analytic calculations were used to calibrate analytic approximations, which enable accurate calculation of the TDE rate using integrals over moments of the stellar distribution function \citep{MagTre99, Wang&Merritt04}.  Increasingly, direct N-body simulations are used to estimate tidal disruption rates \citep{Brocka+11, VasMer13, Zhong+14}.  A detailed comparison of diffusion in N-body simulations finds good agreement with the Fokker-Planck approximation \citep[Fig. 7]{VasMer13} except in regions inside the SMBH influence radius, where resonant relaxation may enhance diffusion coefficients above their two-body values.  In the following subsections, we follow the analytic prescriptions of \citet[hereafter WM04]{Wang&Merritt04} to compute TDE rates in a large sample of observed galaxies.

\subsection{Galaxy Sample and Parametrization}

Unfortunately, it is not feasible to measure the distribution function of stars in distant galactic nuclei.  Instead, 2D surface brightness profiles $I(R)$ are measured as a function of the projected radial distance from the center of light $R$.   Under various assumptions, these can be used to determine the 3D stellar density profile and the implied phase space distribution ($\S\ref{sec:losscone}$).  

Observed isophotes can be fit by many different parametrizations.  These include the Sersic model \citep{Sersic68}, 
\begin{equation}
I_{\rm S}(R) = I_{\rm S}(0) \exp(-b_{\rm n}(R/R_{\rm e})^{1/n}),
\end{equation}
where $I_{\rm S}(0)$ is the central intensity, $R_{\rm e}$ is the galaxy half-light radius, $n$ is a parameter encoding the curvature of the profile, and $b_{\rm n}$ is a constant given by the solution to the equation $2\Gamma (2n, b_{\rm n}) = \Gamma (2n),$ where $\Gamma(x)$ and $\Gamma(x, y)$ are complete and incomplete\footnote{Specifically, we define $\Gamma (x, y)=\int_0^y t^{x-1} \exp(-t)dt$.} Gamma functions, respectively.

A more complex variant of this is the core-Sersic model \citep{Graham+03}, given by
\begin{equation}
I_{\rm CS}(R) = \tilde{I}_{\rm CS}(1+(R_{\rm b}/R)^\alpha)^{\Gamma/\alpha} \exp(-b((R^\alpha + R_{\rm b}^\alpha)/R_{\rm e}^\alpha)^{1/(n\alpha)}).
\end{equation}
This profile behaves like a standards Sersic profile at radii $R\gg R_{\rm b}$, but interior to this break radius it obeys a power law with slope $\Gamma$.  The sharpness of the transition at $R_{\rm b}$ is mediated by the parameter $\alpha$, while the normalization is given by $\tilde{I}_{\rm CS} = 2^{-\Gamma/\alpha} I_{\rm b} \exp(2^{1/(n \alpha)} b (R_{\rm b}/R_{\rm e})^{1/n})$, where $I_{\rm b}$ is the surface brightness at the break radius.  We will only consider the $\alpha=\infty$ limit; if we make the further (reasonable) assumption that $R_{\rm b} \ll R_{\rm e}$, then $b$ is a constant given by the solution of $
\Gamma(2n)+\Gamma(2n, b(R_{\rm b}/R_{\rm e})^{1/n}) = 2\Gamma(2n, b)$.

The final parametrization that we consider is the ``Nuker'' profile \citep{Lauer+95}, which is essentially a broken power law with inner slope $\gamma$ and outer slope $\beta$:
\begin{equation}
I_{\rm N}(R) = I_{\rm b} 2^{(\beta - \gamma)/\alpha} (R/R_{\rm b})^{-\gamma} \left( 1 + (R/R_{\rm b})^\alpha \right)^{-(\beta-\gamma)/\alpha}.
\label{eq:nuker}
\end{equation}

Our primary galaxy sample is from \citet{Lauer+07a}, who provide distances, luminosities $L_{\rm gal}$, and a complete set of Nuker parameters for 219 galaxies, of which 137 also have tabulated half-light radii $R_{\rm e}$.  Tabulated velocity dispersions $\sigma$ for every galaxy in this sample are taken from \citet{Lauer+07b}.  Moreover, 21 galaxies in this sample overlap with the tabulated Sersic and core-Sersic parameters given in \citet{Trujil+04}, allowing us to perform a detailed comparison between the TDE rate calculated from these three $I(R)$ parametrizations.  When $R_{\rm e}$ is available, we calculate the mass-to-light ratio as
\begin{equation}
\Upsilon = \frac{2\sigma^2R_{\rm e}}{GL_{\rm gal}},
\label{eq:upsilon}
\end{equation}
but when $R_{\rm e}$ is not available we instead use the empirically derived galaxy scaling relationship \citep{Magorr+98}
\begin{equation}
\frac{\Upsilon_{V}}{\Upsilon_\odot} = 4.9 \left( \frac{L_{\rm V}}{10^{10}L_\odot} \right)^{0.18}.
\end{equation}
Here $L_{\rm V}$ is the {\it V}-band luminosity; this can be calculated from the {\it V}-band absolute magnitudes tabulated in \citet{Lauer+07a}.

Whenever possible, SMBH masses are computed for all galaxies in our sample using the $M_\bullet-\sigma$ relation, 
\be M_{\bullet} = M_{200}(\sigma/200\,{\rm km\,s^{-1}})^{p}, 
\label{eq:msigma}
\ee
 where in most cases we adopt the recent calibration of \citet{McConnell&Ma13} ($M_{200} = 10^{8.32}M_{\odot}; p = 5.64$).  However, for the sake of comparison to WM04, we also consider the older relation of \citet{MerFer01}, where $M_{200} = 10^{8.17}M_{\odot}$ and $ p = 4.65$ (technically, this is the updated version of the \citealt{MerFer01} result used in WM04).  However, $\approx 10\%$ of our galaxies lack tabulated $\sigma$ values; for these we estimate SMBH masses using the $M_\bullet - L_{\rm V}$ relation of \citet{McConnell&Ma13}.

We note that many low-mass galaxies possess nuclear star clusters, which are specifically excluded from the Nuker power-law fits in \citet{Lauer+07a}.  Neglecting these central overdensities is a conservative choice in our two-body rate estimates, which would in general be enhanced by the presence of additional central density.

\subsection{Loss Cone Dynamics}
\label{sec:losscone}

A star of mass $M_\star$ and radius $R_\star$ is tidally disrupted if it passes within a tidal radius,
\begin{equation}
r_{\rm t} = R_\star \left( \frac{M_\bullet}{M_\star} \right)^{1/3},
\label{eq:rt}
\end{equation}
of an SMBH with mass $M_\bullet$.  Geometrically this criterion can be thought of as creating a loss cone in the six dimensional phase space of stellar orbits.  The loss cone is defined as the set of orbits with pericenters $r_{\rm p} < r_{\rm t}$, a condition which translates into a specific angular momentum $J$ less than the loss-cone value $J_{\rm LC} = \sqrt{2GM_{\bullet} r_{\rm t}}$ if one assumes a spherical potential and considers only highly eccentric orbits.  In calculating the TDE rate we also demand that $r_{\rm p} > 2r_{\rm g}$ so as not to count stars swallowed whole, where $r_{\rm g} = GM_{\bullet}/c^{2}$ is the black hole gravitational radius.   This criterion is appropriate for non-spinning SMBHs; corrections due to SMBH spin are modest \citep{Kesden12b} for black holes significantly beneath the Hills mass, $M_{\rm H}^{2/3}=R_{\star}c^2/(2GM_{\star}^{1/3})$.  As we shall see, these heavier SMBHs contribute only marginally to the total TDE rate, so we neglect spin corrections to the Hills mass and tidal radius for the remainder of this paper

Giant stars can be disrupted by much larger SMBHs, but their greatly increased tidal radii also greatly reduce the rate of mass fallback onto (and thus luminosity of) the black hole.  Furthermore, the long timescales associated with these events can make them difficult to capture in a time-limited transient survey.  For these reasons we neglect giant disruptions in this paper, but we note that the rates of these events can be comparable to main sequence TDE rates \citep{SyeUlm99, MacLeo+12}, and of course dominate the total TDE rate in galaxies with $M_\bullet > M_{\rm H}$.

Assuming spherical symmetry, the observed surface brightness profile can be deprojected into a 3D stellar density profile $\rho_{\star}(r)$ using an Abel inversion,
\begin{equation}
\rho_\star(r) = -\frac{\Upsilon}{\pi}\int_r^\infty \frac{{\rm d}I}{{\rm d}R}\frac{{\rm d}R}{\sqrt{R^2-r^2}},
\end{equation}
where $\Upsilon$ is the mass-to-light ratio from Eq.~\eqref{eq:upsilon}.  Under the same assumptions, the total gravitational potential is calculated according to\footnote{We adopt the stellar dynamics definition of the potential, which is the negative of the more commonly used definition.}
\begin{equation}
\psi(r) = \frac{GM_\bullet}{r} + \frac{GM_\star(r)}{r} + 4\pi G \int_r^\infty \rho_\star(r') r' {\rm d}r',
\end{equation}
where $M_\star(r)$ is the stellar mass enclosed within a radius $r$.  

From the stellar density profile and $\psi(r)$, the stellar distribution function (DF) $f(\epsilon)$ is calculated using Eddington's formula,
\begin{equation}
f(\epsilon) = \frac{1}{8^{1/2}\pi^2 M_\star} \frac{{\rm d}}{{\rm d}\epsilon} \int_0^\epsilon \frac{{\rm d}\rho_\star}{{\rm d}\psi} \frac{{\rm d}\psi}{\sqrt{\epsilon - \psi}},
\label{eq:Eddington}
\end{equation}
where we have made the additional assumption of isotropic velocities and here $\epsilon$ is the negative of the specific orbital energy.  The use of Eddington's formula is only justified if it produces a positive-definite value of $f(\epsilon)$.  When not the case, this indicates that the assumption of velocity isotropy is incompatible with $\rho_\star (r)$.  We discard such galaxies from our sample, as occurs in practice for a relatively small number with very flat interior slopes ($\gamma \approx 0$ in the Nuker parametrization).

The DF can be used to calculate the orbit-averaged angular momentum diffusion coefficient for highly eccentric orbits,
\begin{equation}
\bar{\mu}(\epsilon) = \frac{2}{P(\epsilon)} \int_{r_{\rm p}}^{r_{\rm a}} \frac{{\rm d}r}{v_{\rm r}(r)} \lim_{R\rightarrow0} \frac{\langle ( \Delta R)^2 \rangle}{2R},
\end{equation}
where $R \equiv J^{2}/J_{\rm c}^{2}$ and $J_{\rm c}(\epsilon)$ is the angular momentum of a circular orbit with energy $\epsilon.$  Here, $r_{\rm p}$, $r_{\rm a}$, $P(\epsilon) = 2\int_0^{r_{\rm a}(\epsilon)} $d$r/\sqrt{2(\psi - \epsilon)}$, and $v_{\rm r}$ are the pericenter radius, apocenter radius, orbital period, and radial velocity of the orbit.  The local diffusion coefficient 
\be
 \lim_{R\rightarrow0} \frac{\langle ( \Delta R)^2 \rangle}{2R} = \frac{32 \pi^2 r^2 G^2 \langle M_\star^2 \rangle \ln \Lambda}{3J_{\rm c}^2(\epsilon)} \left( 3I_{1/2}(\epsilon) - I_{3/2} (\epsilon) + 2 I_0 (\epsilon) \right), 
\label{eq:diffusion}
\ee
can be expressed in terms of the DF moments
\bea
I_0(\epsilon) \equiv & \int_0^\epsilon f(\epsilon ') {\rm d}\epsilon \\
I_{n/2}(\epsilon) \equiv & \left[2(\psi(r) - \epsilon) \right]^{-n/2} \\
& \times  \int_\epsilon^{\psi(r)} \left[2(\psi(r) - \epsilon ') \right]^{n/2} f(\epsilon ') {\rm d}\epsilon ', \notag
\eea
where ln$\Lambda \approx \ln (0.4 M_\bullet / M_{\star})$ is the Coulomb logarithm \citep{SpiHar71} and
\be
\langle M_{\star}^{2} \rangle \equiv \int \frac{dN_{\star}}{dM_{\star}}M_{\star}^{2}dM_{\star}
\label{eq:Mstardiff}
\ee
averages the contributions of different stars to orbital diffusion over the stellar mass distribution ${\rm d}N_\star/{\rm d}M_\star$.  For most mass functions, the largest stars generally dominate the diffusion coefficients.

The flux of stars that scatter into the loss cone per unit time and energy is given by:
\begin{equation}
\mathcal{F}(\epsilon){\rm d}\epsilon = 4\pi^2 J_{\rm c}^2(\epsilon) \bar{\mu}(\epsilon) \frac{f(\epsilon)}{\ln R_0^{-1}}{\rm d}\epsilon,
\label{eq:flux}
\end{equation}
where $R_{0}(\epsilon)< R_{\rm LC}(\epsilon)$ defines the angular momentum below which no stars remain.  $R_{0}(\epsilon)$ generally resides inside the nominal loss cone because stars can scatter into and out of the $R < R_{\rm LC}$ parts of phase space during a single orbit.  \citet{Cohn&Kulsrud78} show that
\begin{equation}
R_0(\epsilon) = R_{\rm LC}(\epsilon)
\begin{cases}
\exp (-q), &q>1 \\ 
\exp (-0.186q-0.824q^{1/2}), &q<1,
\end{cases}
\end{equation}
where 
\begin{equation}
q(\epsilon) = \bar{\mu}(\epsilon) \frac{P(\epsilon)}{R_{\rm LC}(\epsilon)},
\label{eq:q}
\end{equation}
is the dimensionless ratio of the per-orbit change in $R$ to its loss cone value.

Physically, $q(\epsilon)$ can be thought of as demarcating two different regimes of loss cone refilling.  When $q \gg 1$ ($R_0 \ll R_{\rm LC}$), as applies to orbits far from the SMBH, stars wander in and out of the loss cone many times during the course of a single orbit (the so-called ``pinhole'' limit).  Very near the SMBH, $q \ll 1$ ($R_0 \approx R_{\rm LC}$) and stars instead diffuse into the loss cone over many orbits (the so-called ``diffusion" limit).  The observational consequences of this are potentially of interest; pinhole-dominated galaxies are capable of producing TDEs with large penetration parameter $\beta \equiv r_{\rm t}/r_{\rm p}$.  In particular, $\dot{N}_{\rm TDE}(\beta) \propto \beta^{-1}$ in the pinhole limit.  For diffusion-dominated galaxies, however, virtually all TDEs will have $\beta \approx 1$.   

In practice, $\mathcal{F}(\epsilon)$ is a very sharply peaked function, and the vast majority of flux into the loss cone comes from energies near $\epsilon_{\rm crit}$, where $q(\epsilon_{\rm crit})\equiv 1$.  In coordinate terms this corresponds to a critical radius from the SMBH, $\psi(r_{\rm crit}) \equiv \epsilon_{\rm crit}$, that sources most TDEs in a given galaxy.  For the lower-mass galaxies that produce observable TDEs, the SMBH influence\footnote{Defined here as the radius that contains a mass in stars equal to $M_\bullet$.} radius $r_{\rm infl} \sim r_{\rm crit}$.  However, this is largely a coincidence: as we show in Appendix \ref{sec:fullResults}, when $M_\bullet \gtrsim 10^8 M_\odot$, $r_{\rm crit} \gtrsim 10 r_{\rm infl}$, typically.  

We apply the above procedure to all Nuker galaxies in our sample, discarding only those which cannot be spherically deprojected ($\gamma<0$) and those whose DFs are incompatible with the assumption of isotropic velocities ($\gamma \lesssim 0.05$, although this criteria varies from galaxy to galaxy).  The TDE rate per galaxy is calculated by integrating the total flux into the loss cone $\dot{N}_{\rm TDE} = \int \mathcal{F}(\epsilon) {\rm d}\epsilon$ for stars of a given mass, and then by integrating over the stellar mass function ($\S\ref{sec:PDMF}$).  Although the functions of interest to this calculation (e.g. $f(\epsilon), q(\epsilon)$, $\mathcal{F}(\epsilon)$) are in general too complex to be written in closed form, in Appendix \ref{sec:analytic} we derive analytic expressions for them in limiting regimes.  These expressions are useful for checking the results of numerical integration.

\subsection{Stellar Mass Function}
\label{sec:PDMF}

In addition to the properties of the galaxy, the TDE rate depends on the present day mass function $dN_{\star}/dM_{\star}$ (PDMF) of stars.  \citet{MagTre99} considered a PDMF resulting from a Salpeter initial mass function (IMF)
\begin{equation}
\chi_{\rm Sal} = \left.\frac{dN_{\star}}{dM_{\star}}\right|_{\rm Sal} =
\begin{cases}
0.046(M_\star/M_{\odot})^{-2.35}&, M_\star^{\rm min}<M_\star<M_\star^{\rm max} \\
0&, {\rm otherwise},
\end{cases}
\end{equation}
truncated at a maximum mass $M_\star^{\rm max} = M_{\odot}$, where $M_\star^{\rm min}=0.08M_{\odot}$.  The upper truncation was chosen to approximate an old stellar population, and we keep it for the same reason.  A further motivation for this upper truncation is that it is a conservative choice with regards to the total TDE rate, for reasons described below.  We also consider a second PDMF, derived by applying the same $1M_\odot$ cutoff to the Kroupa IMF, viz.~ 
\begin{equation}
\chi_{\rm Kro} = \left.\frac{dN_{\star}}{dM_{\star}}\right|_{\rm Kro} =
\begin{cases}
0.98(M_\star/M_{\odot})^{-1.3}&, M_\star^{\rm min}<M_\star<0.5 M_\odot \\
2.4(M_\star/M_{\odot})^{-2.3}&, 0.5M_\odot<M_\star<M_\star^{\rm max}, \\
0&, {\rm otherwise}.
\label{eq:Kroupa}
\end{cases}
\end{equation}
where $M_\star^{\rm min}$ and $M_\star^{\rm max}$ have the same values as the Salpeter PDMF.

As compared to a monochromatic distribution of $M_\star = M_\odot$ stars, including a realistic PDMF increases the TDE rate due to the greater number of sub-solar mass stars, but decreases it because of the reduced angular momentum diffusion coefficients $\bar{\mu}\propto \langle M_{\star}^{2}\rangle$ (Eq.~\ref{eq:Mstardiff}).  For high mass black holes, the TDE rate is also reduced because of the smaller tidal radii $r_{\rm t} \propto R_{\star}M_{\star}^{-1/3} \propto M_{\star}^{0.47}$  (Eq.~\ref{eq:rt}) of low mass stars, which reduce the Hills mass $M_{\rm H}$.  Here we have used the fact that $R_\star \propto M_\star^{0.8}$ on the lower main-sequence.

Both the Salpeter and Kroupa PDMFs give comparable rate enhancements (relative to the monochromatic stellar population) of 1.63 and 1.53, respectively.  Interestingly, TDE rates depend on the age of the nuclear stellar population, as the diffusion coefficients are largely set by the most massive extant stars.  If we extend the Kroupa IMF to values of $M_\star^{\rm max}/M_\odot=\{2, 5, 10 \}$ we find enhancements (again relative to a monochromatic $M_\star=M_\odot$ stellar population) of $\{1.91, 2.75, 3.80\}$, corresponding to stellar populations of age $t_{\rm age}/{\rm yr} = \{1.77 \times10^9, 1.79\times 10^8, 3.16\times 10^7 \}$.

In addition to main sequence stars, scattering by stellar remnants (white dwarfs, neutron stars, and stellar mass black holes) may contribute to the TDE rate; white dwarfs can also themselves be disrupted by smaller massive black holes.\footnote{A white dwarf can be disrupted by a SMBH with a mass as high as $\sim 10^{6} M_{\odot}$ if the SMBH is nearly maximally spinning.}  White dwarfs and neutron stars, being both less common and comparable in mass to solar type stars, make little difference for the total rate of angular momentum diffusion.  Stellar mass black holes, on the other hand, possess considerably greater masses than the Sun and hence could contribute if their number densities are sufficiently high.  

To explore the possible influence of stellar mass black holes, we calculate the enhancement to the diffusion coefficients from inclusion of the black hole mass functions in \citet[Fig. 1]{Belczy+10}.  We consider three different mass functions for stellar remnants, corresponding to metallicities $Z=\{Z_\odot, 0.3Z_\odot, 0.01Z_\odot\}$.  All three tabulated mass functions (Belczynski, private communication) have slightly different minimum masses, $M_{\rm SN} \approx 7M_\odot$, for the onset of supernovae.  Stars drawn from the IMF with zero-age-main-sequence masses $1M_\odot < M_{\rm ZAMS} < M_{\rm SN}$ are assigned final masses of $0.5M_\odot$; these white dwarf stars are of minimal importance for the diffusion coefficients.  We find that in all three models, the stellar mass black holes dominate the total relaxation rate, with the high, medium, and low metallicity cases increasing diffusion coefficients by factors of 1.3, 2.8, and 4.9, respectively (for a Kroupa IMF).  The total loss cone flux increases by roughly the same amount, because the normalization of the PDMF is relatively unchanged (i.e. the black holes dominate $\int M_\star^2 {\rm d}N_\star$ but not $\int M_\star {\rm d}N_\star$).  

This enhancement to the TDE rate can be prevented if mass segregation moves most black holes inward from the critical radius that dominates flux into the loss cone.  The mass segregation timescale for a stellar mass black hole of mass $M_{\rm SBH}$ in a stellar population with average mass $\langle M_\star \rangle$ is $t_{\rm seg}(r) = (M_{\rm SBH}/\langle M_\star \rangle)t_{\rm rel}(r)$, where the energy relaxation timescale
\begin{equation}
t_{\rm rel}(r) = 0.34 \frac{\sigma^3(r)}{G^2 \sqrt{\langle M_{\star}^2\rangle}\rho_\star(r) \ln \Lambda}.
\end{equation}

\begin{figure}
\includegraphics[width=85mm]{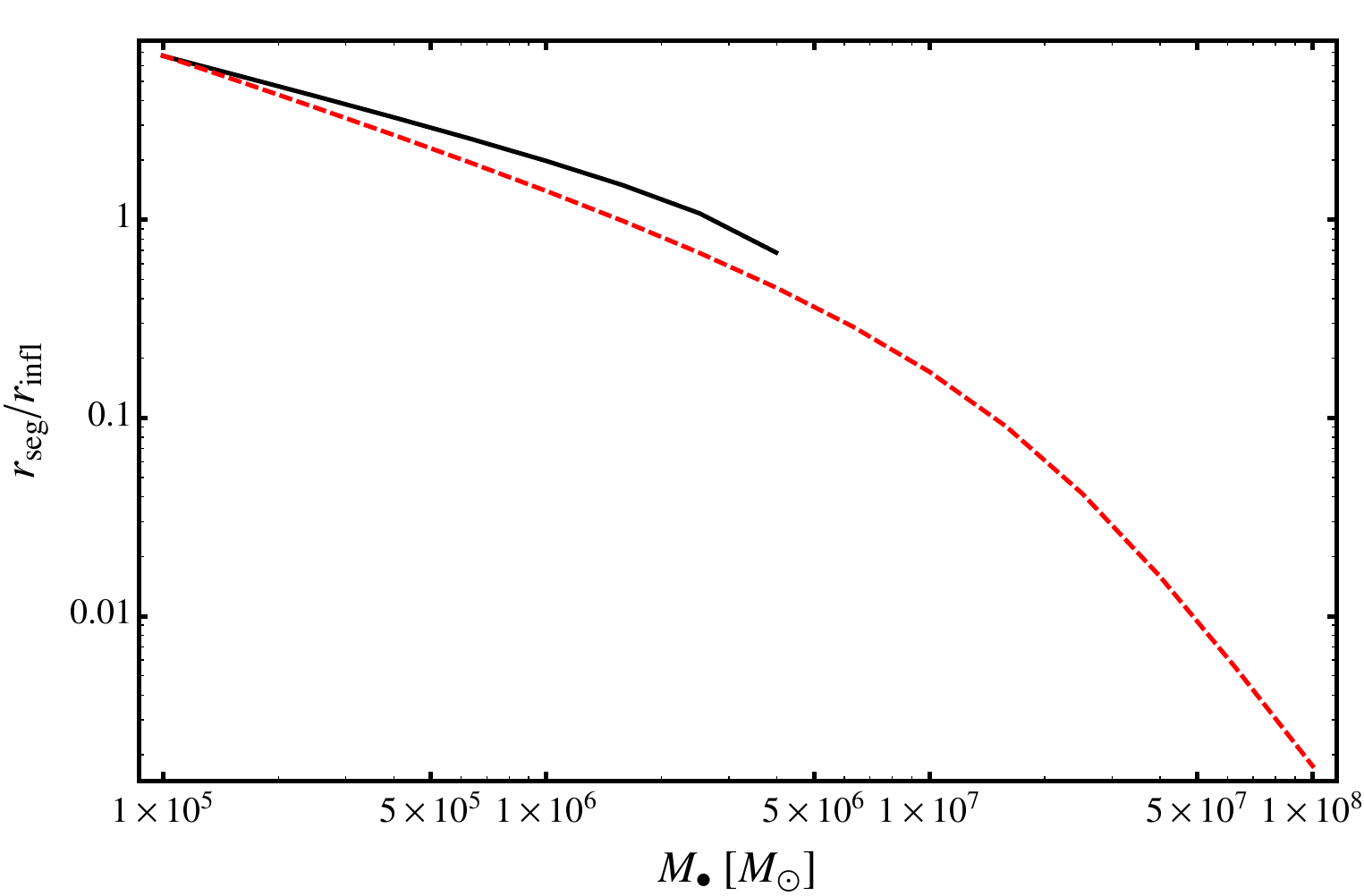}
\caption{The radius $r_{\rm seg}$ out to which stellar mass black holes of mass $M_{\rm SBH}=10M_\odot$ can reach energy equipartition at radii $\sim r_{\rm infl}$, in units of the SMBH influence radius $r_{\rm infl}$, for stellar density profiles $\rho_\star(r) \propto r^{-g}$.  The solid black curve is for core galaxies with $g=1$, and the dashed red curve is for cusp galaxies with $g=2$.  Because the tidal disruption critical radius $r_{\rm crit} \approx r_{\rm infl}$ in the small galaxies that dominate the TDE rate, this plot indicates that stellar remnants will increase TDE rates by a factor of a few for cusp galaxies with $M_\bullet \gtrsim 10^6 M_\odot$.  Below this mass, stellar remnants will mass segregate into a smaller subcluster in cusp galaxies.  The rate enhancement from stellar remnants likely occurs in all core galaxies; when $g \le 1.5$, relaxation times do not monotonically decrease with decreasing radius.  The $g=1$ curve cuts off at a finite value of $M_\bullet$ because of this effect (i.e. larger SMBHs in core profiles will see {\it no} mass segregation in a Hubble time).}
\label{fig:massSegregation}
\end{figure}

Figure \ref{fig:massSegregation} shows the segregation radius $r_{\rm seg}$, interior to which black holes will have mass segregated from the sphere of influence radius $r_{\rm infl}$ into a more compact subcluster within the Hubble time, as a function of SMBH mass.  Cusp galaxies with low mass SMBHs ($M_\bullet \lesssim 10^6 M_\odot$) have $r_{\rm seg} > r_{\rm infl}$ and hence will have stellar mass BHs removed from radii $\sim r_{\rm infl}$ that dominate the TDE rate.  For cusp galaxies with larger SMBHs or for core galaxies, however, black hole segregation is generally unimportant.  This implies that stellar remnants will indeed enhance the TDE rate by factors of a few in most galaxies.  Because of astrophysical uncertainties in the metallicity distribution of stars in distant galactic nuclei, as well as our approximate treatment of mass segregation, we neglect this enhancement for the remainder of the paper, but we emphasize that all but the smallest cusp galaxies could see a rate enhancement $\gtrsim 1.5$ due to stellar remnants.  This enhancement could be even larger if top-heavy IMFs are common in galactic nuclei, as has been suggested for some stars in the center of the Milky Way \citep{Bartko+10}.  Although this simplified discussion of mass segregation largely agrees with simulations of two-component stellar systems \citep{Watters00}, star clusters with a realistic mass spectrum see a more complicated evolution of their stellar mass black holes towards mass segregation, e.g. \citet{Baumga+04}.

More speculatively, a large population of freely floating gas giant planets could also contribute to the total TDE rate, especially at low SMBH masses for which the tidal radius $r_{\rm t} \approx 50r_{\rm g}(M_{\bullet}/10^{6}M_{\odot})^{-2/3}(M_{\rm p}/M_{\rm J})^{-1/3}$ (Eq.~\ref{eq:rt}) of a planet of mass $M_{\rm p}$ exceeds the gravitational radius $r_{\rm g} = GM_{\bullet}/c^{2}$, where $M_{\rm J}$ is the mass of Jupiter and we have assumed a constant planetary radius of $R_{\rm p} = R_{\rm J} = 7\times 10^{9}$ cm.  Although freely floating planets may be common relative to main sequence stars  (e.g.~\citealt{Sumi+11}), the much lower accretion rates produced by the disruption of a planet $\propto M_p^{2}$ (Eq.~\ref{eq:Mdotpeak}) implies much dimmer events which are unlikely to contribute appreciably to the {\it detected} TDE rate.    

\section{TDE Observables}
\label{sec:observables}

Following tidal disruption, the gaseous stellar debris travels on approximately geodesic trajectories, with a ``frozen-in'' specific energy spread \citep{Rees88} given by
\begin{equation}
\Delta \epsilon = \frac{GM_\bullet R_\star}{r_{\rm t}^2}.
\end{equation}
If one assumes a simple top-hat distribution of debris mass with respect to specific energy (width $\Delta \epsilon$), then the most tightly bound debris returns to pericenter after a time (e.g.~\citealt{Stone+13})
\begin{equation}
t_{\rm fall} = 3.5\times 10^6~{\rm s}~ M_6^{1/2}m_\star^{-1}r_\star^{3/2},
\label{eq:tfb}
\end{equation}
where $M_6=M_\bullet/10^6M_\odot$, $m_\star=M_\star/M_\odot$, and $r_\star=R_\star/R_\odot$.  The peak mass fallback rate occurs at this time, and has an Eddington-normalized value of 
\begin{equation}
\frac{\dot{M}_{\rm peak}}{\dot{M}_{\rm edd}} = 133 \eta_{-1} M_6^{-3/2}m_\star^2r_\star^{-3/2},
\label{eq:Mdotpeak}
\end{equation}
where $\dot{M}_{\rm edd} \equiv L_{\rm edd}/\eta c^{2}$ is the Eddington accretion rate, $L_{\rm edd} \simeq 1.5\times 10^{46}M_{8}$ erg s$^{-1}$ is the Eddington luminosity and $\eta = 0.1\eta_{-1}$ is the constant accretion efficiency.  Assuming the initial mass fallback rate is super-Eddington, it will remain so for a timescale
\begin{equation}
t_{\rm edd} = 6.6\times 10^7~{\rm s}~ \eta_{-1}^{3/5}M_6^{-2/5}m_\star^{1/5}r_\star^{3/5}.
\label{eq:tedd}
\end{equation}
The above equations have the correct parameter scalings but can have errors $\approx 2$ in the prefactors due to the debris mass distribution not actually being a top-hat, with weak dependences on stellar structure and $\beta$ \citep{GuiRam13} that we neglect here.  In our rate calculations we assume $r_{\star} = m_{\star}^{0.8}$, as is appropriate for low mass main sequence stars.   

\begin{figure}
\includegraphics[width=85mm]{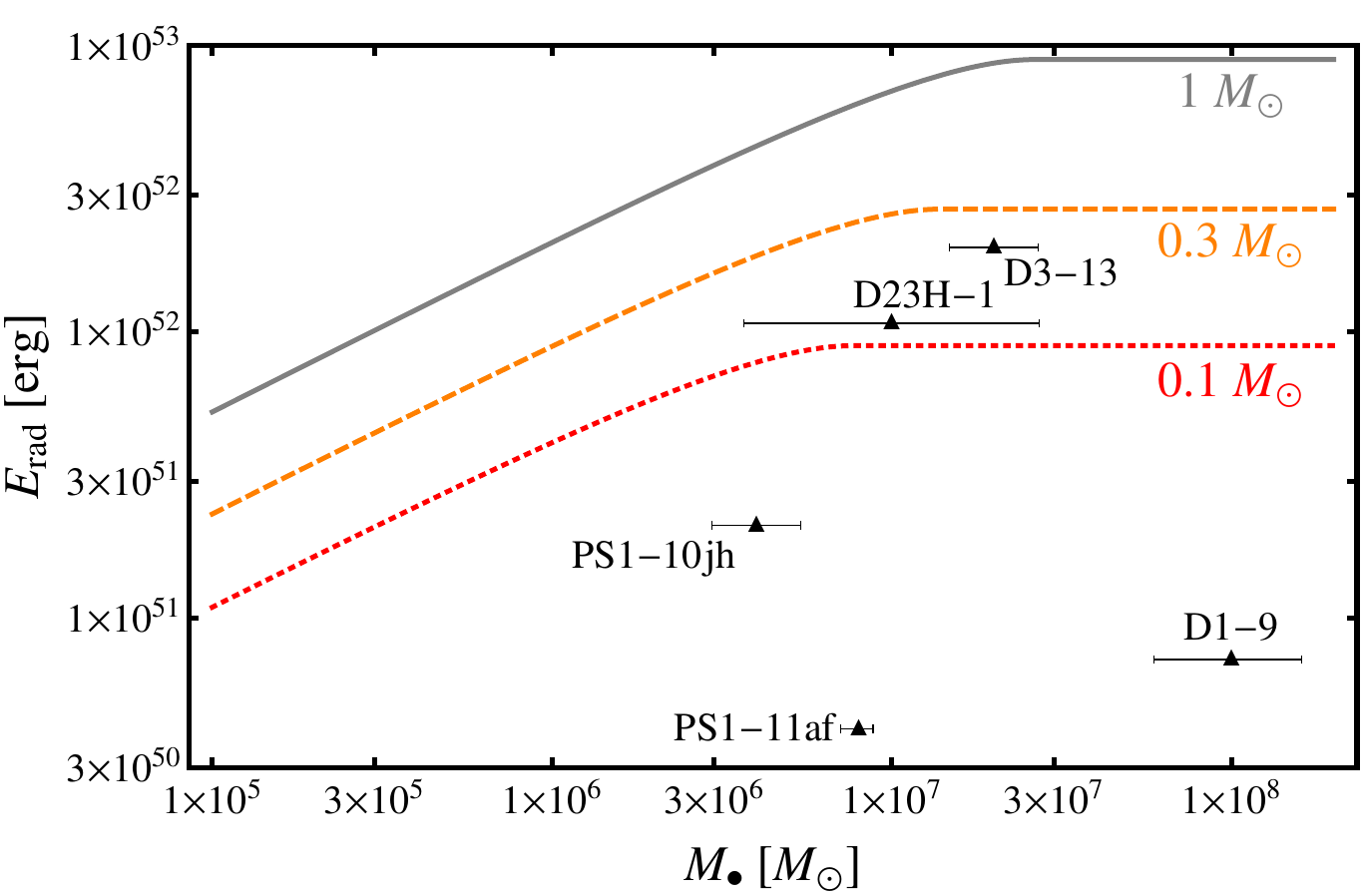}
\caption{Total energy radiated in a TDE, $E_{\rm rad}$, as a function of SMBH mass $M_{\bullet}$, calculated assuming the radiated luminosity is limited to the Eddington luminosity of the SMBH, and that radiative efficiency $\eta=0.1$.  Gray solid, orange dashed, and red dotted curves are for stellar masses of $1M_\odot$, $0.3M_\odot$, and $0.1M_\odot$, respectively.  Low mass black holes with highly super-Eddington peak accretion rates radiate an order of magnitude less total energy than will TDEs from high mass SMBHs (Eq.~\ref{eq:Mdotpeak}).  We also plot lower limits on the radiated energy for all five TDE candidates selected in the optical or UV with published $E_{\rm rad}$ estimates: D1-9, D3-13 \citep{Gezari+08}, D23H-1 \citep{Gezari+09}, PS1-10jh \citep{Gezari+12}, and PS1-11af \citep{Chorno+14}.  In all five events we take SMBH mass estimates and error bars from the discovery papers.}
\label{fig:ERad}
\end{figure}

Another observable in TDE flares is the total radiated energy, $E_{\rm rad}$.  The exact value of $E_{\rm rad}$ will depend on the particular emission model, but we can gain intuition by considering a very simple model for the luminosity $L(t)$ of initially super-Eddington TDEs, where $L=L_{\rm edd}$ if $t<t_{\rm edd}$, and $L=\eta \dot{M}c^{2}$ if $t>t_{\rm edd}$.  Assuming $\eta$ is constant in time, we then have
\bea
E_{\rm rad} &=& L_{\rm edd}(\tilde{t}_{\rm edd}-t_{\rm fall}) + \frac{M_\star\eta c^2}{2} \left(\frac{t_{\rm fall}}{\tilde{t}_{\rm edd}} \right)^{2/3} \notag \\
 &\approx& \begin{cases} 8.1\times 10^{51}~{\rm erg}~\eta_{-1}^{3/5}M_6^{3/5}m_\star^{1/5}r_\star^{3/5}, &t_{\rm edd} \gg t_{\rm fall} \\ 8.9\times 10^{52}~{\rm erg}~\eta_{-1}m_\star, &t_{\rm edd} \lesssim t_{\rm fall},  \end{cases} 
\label{eq:ERad}
\eea
where $\tilde{t}_{\rm edd}=\max( t_{\rm edd}, t_{\rm fall} )$.  Figure \ref{fig:ERad} shows the dependence of $E_{\rm rad}$ on $M_{\bullet}$, from which one observes that stars disrupted by low-mass SMBHs ($\sim 10^5 M_\odot$) radiate an order of magnitude less energy than their high mass, initially sub-Eddington counterparts.  In practice, $E_{\rm rad}$ is a difficult quantity to measure: X-ray selected TDEs (or TDEs found in SDSS) suffer from poor cadence and generally miss the peak of the light curve, where most of the energy is emitted.  Optically- or UV-selected TDEs avoid this problem, but unless followup observations reveal the peak of the SED, the bolometric correction to the observed light is quite uncertain.  We take five optically- or UV-selected TDE candidates and show their {\it lower} limits for $E_{\rm rad}$.  While two events (D3-13 and D23H-1) fall in the expected portion of parameter space, the other three (D1-9, PS1-10jh, and PS1-11af) possess $E_{\rm rad} \ll 0.1M_\odot \eta c^2$, for $\eta \approx 0.1$.  We list here four possible explanations for the low $E_{\rm rad}$ seen in these events:
\begin{enumerate}
\item All data points in Fig. \ref{fig:ERad} are lower limits, because of uncertainties in the bolometric correction.  If the brightness temperature is larger than the minimal value assumed, the true $E_{\rm rad}$ values will be higher.  This inherent underestimate may be worsened by dust extinction, particularly in events with UV data.
\item Partial tidal disruption of a stellar envelope (i.e. $\beta \lesssim 1$) will reduce the amount of mass fed to the SMBH, often by orders of magnitude \citep{GuiRam13}.
\item An accretion efficiency $\eta \ll 0.1$.  This could arise from the hydrodynamics of super-Eddington accretion flows, or (because $r_{\rm crit} \gg R_{\rm g}=GM_\bullet/c^2$) from the isotropic distribution of pre-disruption orbits with respect to the SMBH spin axis.  The subset of TDEs approaching rapidly spinning SMBHs on retrograde, roughly equatorial orbits will have accretion efficiencies $\eta \lesssim 0.01$.  
\item As mentioned above, the number of free-floating planets in the Milky Way may be comparable to or greater than the number of stars \citep{Sumi+11}.  Tidal disruption of a planet or brown dwarf would reduce the available mass budget for the TDE flare.
\end{enumerate}
Possibility (i) almost certainly is contributing at some level, but absent better followup observations of future TDEs (or a much better theoretical understanding of their optical emission mechanisms), its importance is difficult to quantify.  Possibility (ii) is promising, as partial tidal disruptions should be a nontrivial fraction of all TDEs.  In the diffusive regime of relaxation, partial disruptions will make up a large majority of TDEs; in the pinhole regime of relaxation (i.e. $q(\epsilon)> 1$; see \S \ref{sec:losscone}), they will make up a non-negligible fraction of events.  If we take $0.6<\beta < 1.0$ ($0.9<\beta<1.8$) as the parameter space for observable partial disruptions of polytropic $\gamma=5/3$ ($\gamma=4/3$) stars \citep[Fig. 3]{GuiRam13}, then $\approx 40 \%$ ($\approx 51\%$) of pinhole-regime TDEs from $M_\bullet \sim 10^6 M_\odot$ SMBHs will be partial disruptions.

The viability of the final two hypotheses is more ambiguous.  The retrograde orbits explanation for possibility (iii) is sufficiently fine-tuned that we can discount it on the basis of the five events in Fig. \ref{fig:ERad}, and recent general relativistic radiation hydrodynamic simulations of super-Eddington accretion suggest $\eta \sim 0.1$ is typical \citep{Sadows+14, Jiang+14}.  Finally, possibility (iv) is hard to disprove but fairly speculative.  A full explanation of the unexpectedly low $E_{\rm rad}$ seen in many TDEs is beyond the scope of this paper, but is an important topic for future work.

\subsection{Optical Emission Models}
\label{sec:opticalmodels}

Several physical processes may contribute to the optical emission from TDEs, which depend in different ways on the properties of the disrupted star and the SMBH.  These processes, as described in Appendix \ref{sec:optical}, include thermal emission from the (viscously spreading) accretion disk \citep{SheMat14}; super-Eddington outflows from the accretion disk \citep{StrQua09}; reprocessing of the accretion luminosity \citep{Guillochon+14} by an extended outer dense screen (e.g.,  nonvirialized tidal debris); and synchrotron emission from an off-axis relativistic jet that has decelerated to transrelativistic speeds through its interaction with the circumnuclear medium.  In $\S\ref{sec:detection}$ we calculate the observed rates of TDEs using several of these models.  We focus on the emission at optical wavebands because sensitive wide-field surveys, such as the Zwicky Transient Factory (ZTF) and LSST, are expected to greatly expand the current TDE sample in the near future.    

\begin{figure}
\includegraphics[width=85mm]{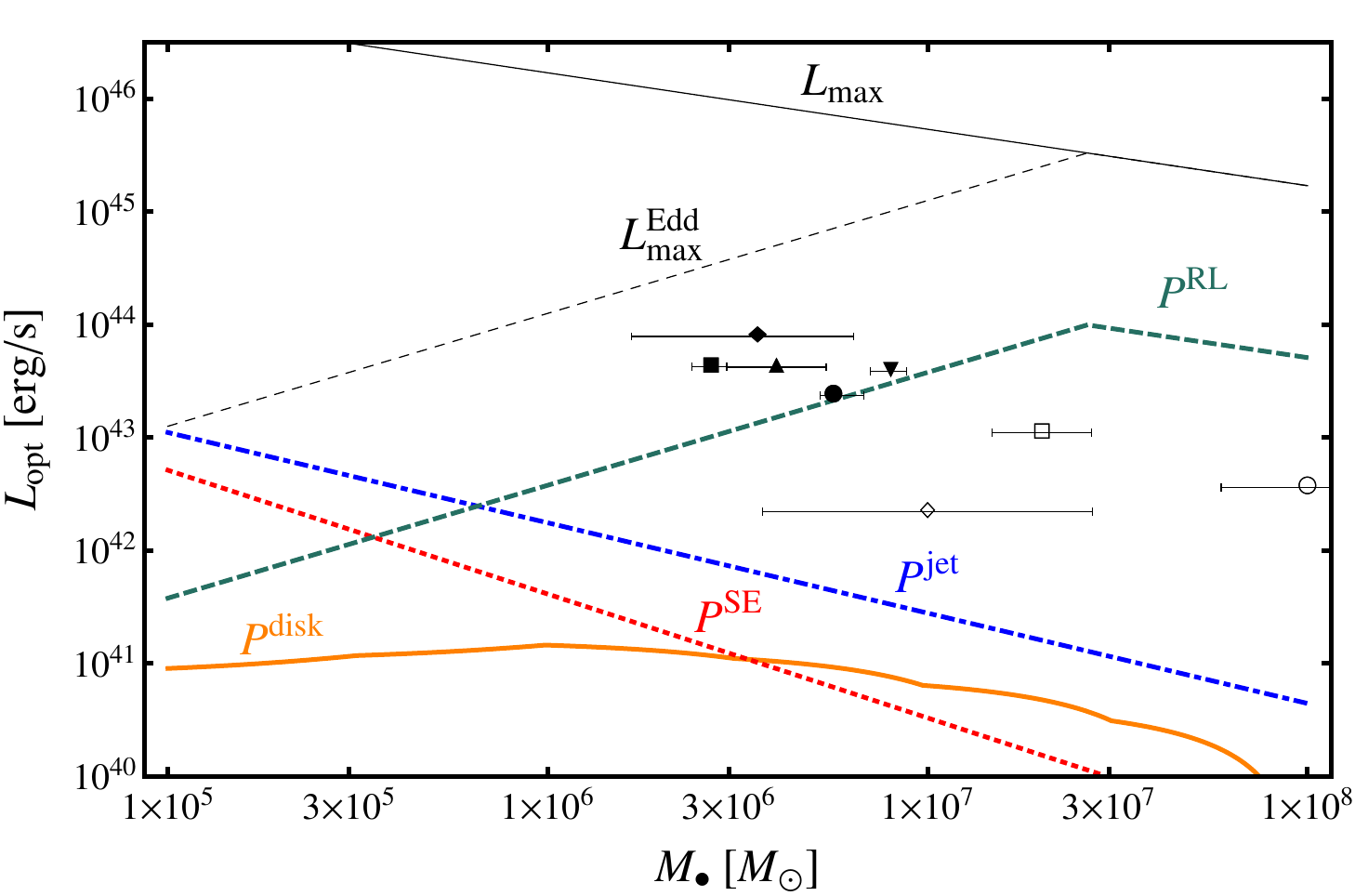}
\caption{Peak {\it g}-band optical luminosities $P$ as a function of SMBH mass $M_{\bullet}$ for different models of optical TDE emission (Appendix \ref{sec:optical}), assuming a $\beta = 1$ tidal disruption of a solar type star.  Models shown include thermal emission from the viscously spreading accretion disk ({\it orange, solid}; $\S\ref{sec:thermal}$); reprocessing by an extended outer layer of nonvirialized debris ({\it green, dashed}; $\S\ref{sec:reprocessing}$); super-Eddington outflows ({\it red, dotted}, $\S\ref{sec:SE}$); and off-axis emission from decelerating relativistic jet ({\it blue, dot-dashed}; $\S\ref{sec:jet}$).  Thin black lines represent upper limits: the solid, thin black line corresponds to the total bolometric $\dot{M}c^2$ power available (assuming a radiative efficiency of $\eta=0.1$), while the dashed, thin black line is the same, but Eddington-limited.  We also plot the peak {\it g}-band luminosities for all eight claimed TDE candidates that capture the peak of the optical light curve.  In particular, we plot PTF09ge ({\it circle}), PTF09axc ({\it square}), PTF09djl ({\it diamond}), PS1-10jh ({\it upward triangle}), PS1-11af ({\it downward triangle}), D1-9 ({\it open circle}), D3-13 ({\it open square}), and D1-9 ({\it open diamond}).}
\label{fig:LPeak}
\end{figure}

Figure \ref{fig:LPeak} shows the peak {\it g}-band optical luminosities predicted by each emission model as a function of SMBH mass for our fiducial choice of model parameters (see Appendix \ref{sec:optical}).  Both the spreading disk and the reprocessing models represent ``Eddington-limited" emission mechanisms, and hence their peak luminosities decrease for lower mass SMBHs (Eq.~\ref{eq:ERad}).  By contrast, emission from super-Eddington outflows or an off-axis jet are not limited to the Eddington luminosity and instead predicted higher peak luminosities for lower SMBH masses.  This difference has important implications for the sensitivity of the detected TDE fraction on the SMBH mass distribution ($\S\ref{sec:detection}$).  

Figure \ref{fig:LPeak} also shows observed peak {\it g}-band luminosities\footnote{PTF09axc and PTF09djl only have peak {\it r}-band magnitudes (Arcavi, private communication); we present the {\it g}-band extrapolation of a blackbody spectrum assuming both the {\it g}- and {\it r}-band are on the Rayleigh-Jeans tail of these events.  The peak {\it g}-band magnitudes of D1-9 and D3-13 are likewise calculated by correcting (observed) peak {\it r}-band magnitudes in \citet{Gezari+09}.} for five optically-selected TDE candidates, represented with filled points: PTF09ge, PTF09axc, PTF09djl \citep{Arcavi+14}, PS1-10jh \citep{Gezari+12}, and PS1-10af \citep{Chorno+14}.  Three UV-selected TDE candidates are plotted as open points: D1-9, D3-13 \citep{Gezari+08}, and D23H-1 \citep{Gezari+09}.  All five of the optically-selected events are tightly clustered in the diagram, at least relative to the huge uncertainties in the predicted optical emission of our four different mechanisms.  This is likely due to the flux limitations of PTF and Pan-STARRS.  Notably, all eight of these events appear compatible with only a single optical emission mechanism: an extended reprocessing layer that converts a fraction $\sim 10\%$ of the accretion power into optical luminosity.  The spreading disk and super-Eddington outflow scenarios are unable to produce optical luminosities comparable to those observed.\footnote{Technically, super-Eddington outflows can reach the observed peak luminosities $\sim 10^{43}~{\rm erg~s}^{-1}$, but only if all galaxies hosting these TDEs have SMBHs that are undermassive by a factor $\sim 30$.}  Our model for optical synchrotron emission from a decelerating jet possesses enough free parameters that it can be brought up into the observed luminosity range (and indeed, our choice of parameters was quite conservative), but none of these five events were seen to possess the nonthermal spectra characteristic of synchrotron radiation.

\section{Results}
\label{sec:results}

Following the prescriptions of the preceding section, we compute TDE rates $\dot{N}_{\rm TDE}$ for the 146 galaxies in our sample which can be deprojected and Eddington-inverted assuming spherical symmetry and isotropic velocities.  In this section, we present the raw results, provide power law fits for $\dot{N}_{\rm TDE}$ as a function of other galaxy parameters, and then use these fits to calculate distributions of TDE observables and rates of detectability by optical surveys.

\subsection{Rates}
\label{sec:rates}

\begin{figure}
\includegraphics[width=85mm]{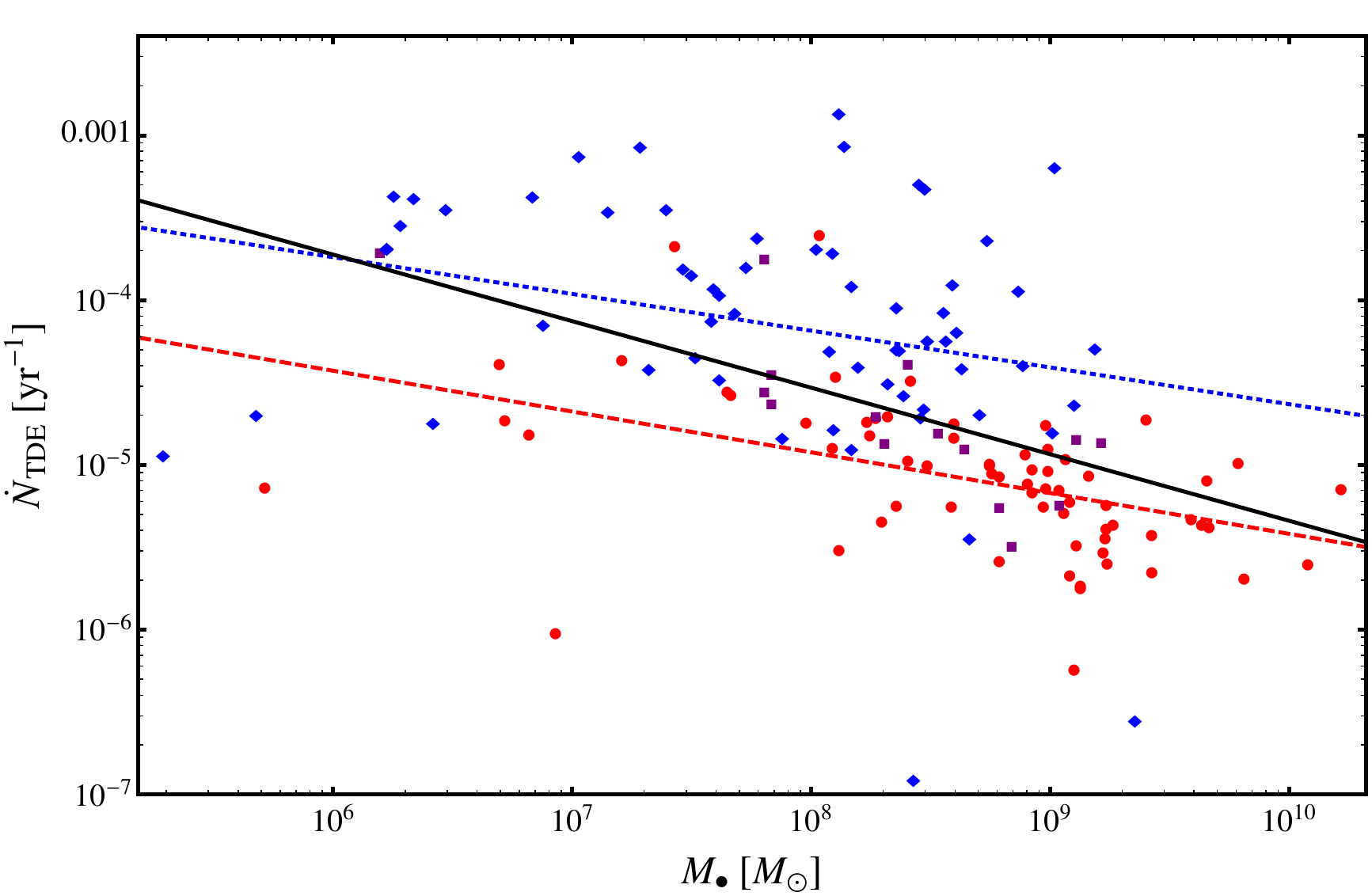}
\caption{Tidal disruption rates $\dot{N}_{\rm TDE}$ for every galaxy in our sample, measured in units of stars per year.  Results are plotted against SMBH mass $M_\bullet$.  Cusp galaxies are shown as blue diamonds, core galaxies are shown as red circles, and rare intermediate galaxies ($0.3\le \gamma \le 0.5$) are shown as purple squares.  The solid black line, dotted blue line, and dashed red line are best fit power laws to the full sample, the cusp subsample, and the core subsample, respectively.  The power law fit to the full sample is significantly steeper than the fits to the subsamples because of the transition from cusp to core galaxies with increasing $M_\bullet$.}
\label{fig:SampleNDotMBH}
\end{figure}

Table \ref{tab:rates} presents our results for TDE rates in every galaxy in our extended sample of Nuker galaxies, calculated under the assumption of a Kroupa PDMF and the latest calibration of the \citet[hereafter MM13]{McConnell&Ma13} $M_{\bullet}-\sigma$ relationship (this relation is broadly consistent with other recent analyses, e.g. \citealt{GraSco13}).  Because of its length, we relegate this table to a separate appendix, but analyze its data here.  Our most important findings are in Figure \ref{fig:SampleNDotMBH}, which shows how the TDE rate varies as a function of SMBH mass.  

Although we start with a sample of 219 galaxies, we are forced to discard 51 which cannot be deprojected assuming spherical symmetry ($\gamma<0$), as well as a further 22 whose distribution functions are incompatible with the assumption of velocity isotropy ($f(\epsilon)\ge 0$ is not everywhere satisfied)\footnote{Distribution functions unable to satisfy positivity generally occur for $\gamma < 0.05$.  Both of these rejection criteria generally occur for very massive galaxies with SMBHs incapable of disrupting main sequence stars.}.  We discard 2 more outlier galaxies with extremely small $\sigma$ values, leaving a final 144 galaxies with calculated TDE rates.  The best-fit regression of these results to an arbitrary power law gives
\begin{equation}
\dot{N}_{\rm TDE} = \dot{N}_0 \left( \frac{M_\bullet}{10^8M_\odot} \right)^{B},
\label{eq:bestfit}
\end{equation}
with $\dot{N}_0 = 2.9 \times 10^{-5}$ yr$^{-1}$ gal$^{-1}$ and $B = -0.404$, when we consider our entire galaxy sample.  If we limit ourselves to core (cusp) galaxies alone, we find $\dot{N}_0 = 1.2 \times 10^{-5}$ yr$^{-1}$ gal$^{-1}$ ($\dot{N}_0 = 6.5 \times 10^{-5}$ yr$^{-1}$ gal$^{-1}$) and $B=-0.247$ ($B=-0.223$).  The significantly steeper slope of our fit to the full sample occurs because of the transition from a core-dominated to a cusp-dominated galaxy population as one moves from high- to low-mass galaxies.  These fits are shown in Fig. \ref{fig:SampleNDotMBH}.  Our fit to the overall galaxy sample is comparable to that found using the results of WM04, which yield $\dot{N}_0 = 2.3\times 10^{-5}~{\rm yr}^{-1}$ and $B=-0.519$.

All of these fits are somewhat sensitive to the sample restrictions used in their calculation; beyond the choice of core, cusp, or both, we can also limit ourselves to galaxies under some Hills limit (say, $M_\bullet < 10^8 M_\odot$).  If we fit a power law to this subsample, we find almost no dependence of $\dot{N}_{\rm TDE}$ on SMBH mass, with $B=0.061$.  However, as we show in Appendix C, our most trustworthy results are for those galaxies with $M_\bullet \gtrsim 10^7 M_\odot$: for these galaxies, HST photometry resolves $r_{\rm crit}$, the radius from which the vast majority of loss cone flux originates.  Generally speaking, the critical radius is marginally resolved for $M_\bullet \approx 10^7 M_\odot$, and better resolved at larger SMBH masses.  For this reason, our fiducial model uses a power law fit to the entire galaxy sample ($B=-0.404$), but we will briefly comment on this choice at later points in the paper.

\begin{figure}
\includegraphics[width=85mm]{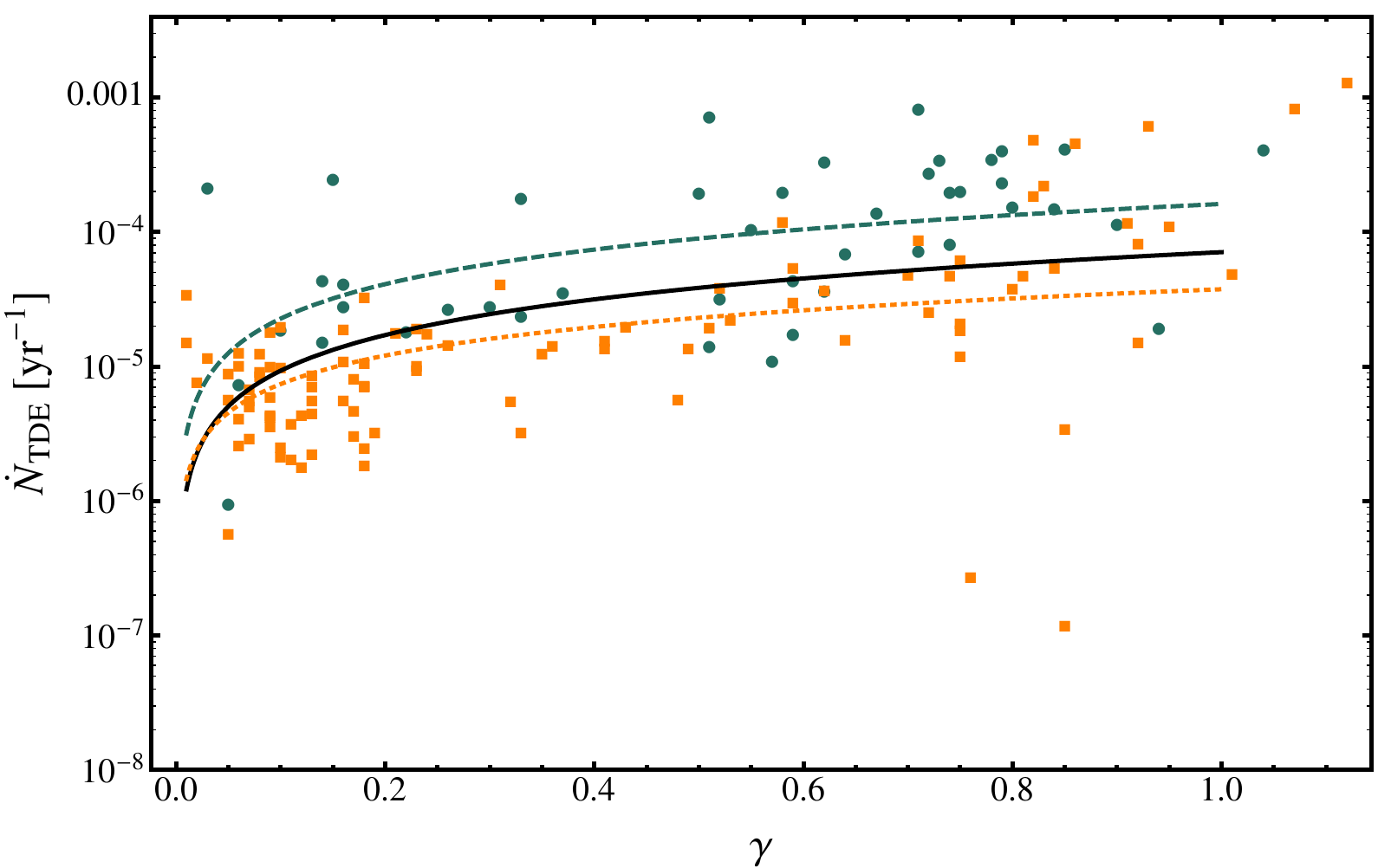}
\caption{Tidal disruption rates $\dot{N}_{\rm TDE}$ for every galaxy in our sample, measured in units of stars per year.  Results are plotted against the inner Nuker profile power law slope $\gamma$.  Green circles indicate galaxies with SMBH masses below the Hills mass for a solar type star; orange squares are larger galaxies with $M_\bullet$ above that Hills mass.  The solid black, dashed green, and dotted orange lines show best fit power laws for the full sample, the $M_\bullet < M_{\rm Hills}$ subsample, and the $M_\bullet > M_{\rm Hills}$ subsample, respectively.}
\label{fig:SampleNDotGamma}
\end{figure}

Figure \ref{fig:SampleNDotGamma} shows the TDE rate as a function of the inner stellar slope $\gamma$ of the Nuker profile.  Galaxies with larger $\gamma$ possess higher rates, as is expected because their denser central stellar populations naturally feature shorter relaxation times and faster rates of energy and angular momentum diffusion.  This correlation is a relatively strong one: employing our entire sample, we find $\dot{N}_{\rm TDE} \propto \gamma^{0.705}$.  No strong correlations exist between the per-galaxy TDE rate and the other Nuker structural parameters of $\alpha$, $\beta$, $R_{\rm b}$, and $I_{\rm b}$.

The TDE rate is relatively insensitive to assumptions about the parametrization of the galaxy profile.  Table \ref{tab:compare} compares our fiducial (Nuker, MM13) rates to rates calculated using instead the Sersic and Core-Sersic galaxy parameterizations, for a subsample of 21 galaxies from \citet{Trujil+04}.\footnote{Two galaxies in our sample, NGC1700 and NGC4478, do not have computable event rates for the Nuker parametrization because their best-fit $\gamma<0$.}  In general the Nuker parameterizations result in slightly higher TDE rates than the Sersic parameterization, but comparable to the more realistic Core-Sersic models.  This is not too surprising of a result, as the plots of fit residuals in \citet{Trujil+04}, Figs. 4-5, show very little difference between Nuker and Core-Sersic fits (pure Sersic fits sometimes do differ, and generally have greater residuals).

Our results are very insensitive to changes in the $M_\bullet-\sigma$ relation; we have rerun the WM04 sample with both the \citet{MerFer01} and the MM13 calibrations of this relation, and find that the mean change in individual rates $\dot{N}_{\rm TDE}$ is $15\%$.  However, some rates go down and others go up, with no systematic shift: if we take sample-averaged rates for both sets of $M_\bullet-\sigma$, the difference in these average rates is only $2\%$.

\begin{figure}
\includegraphics[width=85mm]{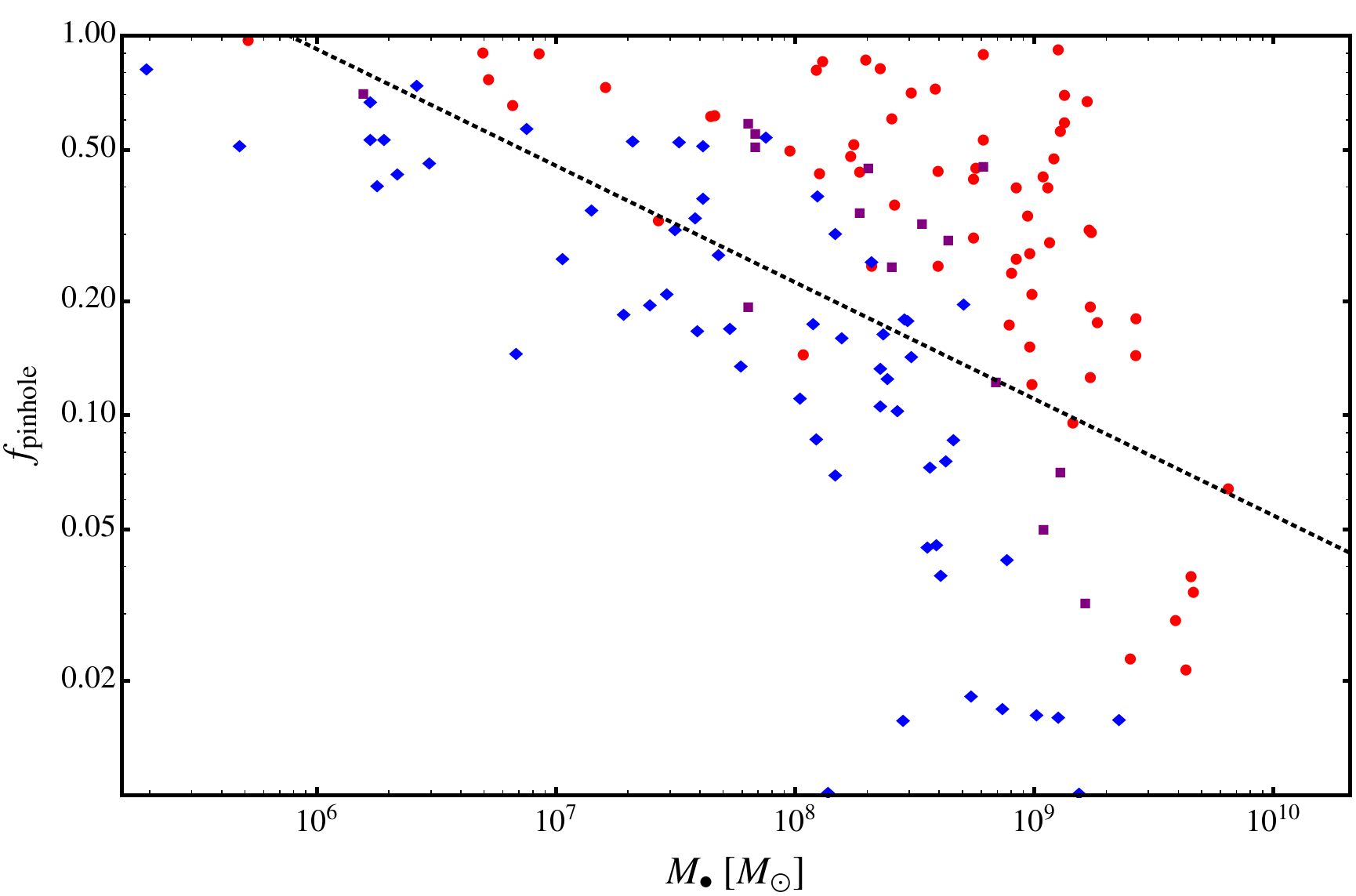}
\caption{The fraction, $f_{\rm pinhole}$, of all TDEs in a galaxy which fall into the pinhole regime of tidal disruption.  TDEs in this regime can access any $\beta$ value up to the maximum permitted by the size of the horizon, with a distribution $\dot{N}_{\rm TDE}(\beta) \propto \beta^{-1}$.  TDEs in the opposite, diffusive, regime, see $\beta \approx 1$ almost always, with higher $\beta$ values exponentially suppressed.  Data point styles are the same as in Fig. \ref{fig:SampleNDotMBH}.  Although $f_{\rm pinhole}$ increases with decreasing $M_\bullet$, and is highest among small cusp galaxies, at any given $M_\bullet$ value it is higher in core galaxies.}
\label{fig:SamplePinhole}
\end{figure}

Figure \ref{fig:SamplePinhole} shows $f_{\rm pinhole}$, the fraction of TDEs occuring in the pinhole regime for each galaxy, as a function of SMBH mass (see also the entries in Table \ref{tab:rates}).  As described in $\S\ref{sec:losscone}$, pinhole events possess a distribution $d\dot{N}_{\rm TDE}/d\beta \propto \beta^{-2}$ in penetration factor $\beta$, while non-pinhole (diffusive) TDEs instead possess $\beta \approx 1$, with the rate of higher $\beta$ strongly suppressed.  In other words, we expect the total $\beta$-distribution distribution to be approximately
\bea
\frac{d\dot{N}_{\rm TDE}}{d\beta}  &\approx& \begin{cases} f_{\rm pinhole}\beta^{-2} &\beta \gtrsim 1, \\ (1-f_{\rm pinhole}) & \beta \approx 1,  \end{cases} 
\label{BetaDist}
\eea
where $f_{\rm pinhole}$ is the TDE rate-weighted pinhole fraction, which we estimate to be $\approx 0.3$ based on our galaxy sample.  The maximum attainable $\beta$ for a given Schwarzschild SMBH is $\beta_{\rm max}\approx R_{\rm t}/(2R_{\rm g})$.  We find a strong correlation of the pinhole fraction with SMBH mass, and plot this in Fig.~\ref{fig:SamplePinhole}.  Specifically, $f_{\rm pinhole}$ rises with decreasing $M_\bullet$.  As we shall see, small SMBHs dominate the volumetric TDE rate, so this result indicates that high-$\beta$ events are relatively common among TDEs.  Our best fit power law is
\begin{equation}
f_{\rm pinhole} = 0.22 \left( \frac{M_\bullet}{10^8M_\odot} \right)^{-0.307}.
\end{equation}
Interestingly, at fixed $M_\bullet$, $f_{\rm pinhole}$ is largest for core galaxies.

In this section we have crudely approximated pinhole TDEs as occurring when $q > 1$, and diffusive TDEs as occurring when $q<1$; in practice, there is a large intermediate zone with $0.1 \lesssim q \lesssim 3$ where high-$\beta$ TDEs can occur, but at a rate that is somewhat suppressed relative to the logarithmically flat rate of Eq. \eqref{BetaDist}.  A formalism for more precise calculation of $\dot{N}(\beta)$ is presented in \citet{Strubb11}. 

\begin{table*}
 \begin{minipage}{180mm}
   \caption{Intrinsic tidal disruption event rates for the 21 galaxies parametrized in \citet{Trujil+04}.  We show event rates for the Nuker ($\dot{N}_{\rm TDE}^{N}$), Sersic ($\dot{N}_{\rm TDE}^{S}$), and core-Sersic ($\dot{N}_{\rm TDE}^{CS}$) surface brightness parametrizations, as well as the most relevant structural parameters.}
   \begin{tabular}{r|r|r|r|r|r|r|r|r|r|r|r} \\
     \hline
     Galaxy & Type$^{a}$ & $R_{\rm e}$$^{b}$ & $n$$^{c}$ & $R_{\rm b}$$^{d}$ & $\gamma$$^{e}$ & $M_\bullet$$^{f}$ & $\dot{N}_{\rm TDE}^{N}$$^{g}$ & $\dot{N}_{\rm TDE}^{S}$$^{h}$ & $\dot{N}_{\rm TDE}^{CS}$$^{i}$  \\
     \hline
     NGC2986 & $\cap$ & 3.66 & 3.29 & 226 & 0.18 & 8.96 & -5.15 & -4.92 & -4.91 \\  
     NGC3348 & $\cap$ & 3.95 & 3.09 & 199 & 0.09 & 8.76 & -5.00 & -5.03 & -5.00 \\  
     NGC4168 & $\cap$ & 3.88 & 2.68 & 365 & 0.17 & 8.14 & -5.52 & -5.27 & -5.36 \\  
     NGC4291 & $\cap$ & 1.99 & 3.75 & 72.7 & 0.01 & 9.13 & -4.76 & -4.31 & -4.33 \\  
     NGC5557 & $\cap$ & 5.16 & 3.74 & 204 & 0.02 & 8.95 & -5.11 & -4.74 & -4.73 \\  
     NGC5903 & $\cap$ & 5.13 & 2.96 & 257 & 0.13 & 8.52 & -5.35 & -5.22 & -5.26 \\  
     NGC5982 & $\cap$ & 4.19 &3.24 & 68.7 & 0.05 & 8.92 & -4.99 & -4.87 & -4.82 \\  
     NGC3613 & $\cap$ & 4.82 & 3.63 & 50.8 & 0.04 & 8.38 & -4.83 & -4.62 & -4.65 \\  
     NGC5077 & $\cap$ & 3.86 & 3.56 & 351 & 0.23 & 9.08 & -5.03 & -4.68 & -4.74 \\  
     NGC1426 & $\backslash$ & 4.15 & 4.95 & 5.10 & 0.26 & 7.69 & -4.58 & -4.22 & -3.68 \\  
     NGC1700 & $\cap$ & 7.39 & 5.99 & 11.8 & -0.10 & 8.80 & - & -4.27 & -4.27 \\  
     NGC2634 & $\backslash$ & 2.93 & 4.54 & 8.6 & 0.81 & 7.95 & -4.38 & -4.33 & -3.76 \\  
     NGC2872 & $\backslash$ & 4.29 & 4.56 & 11.5 & 1.01 & 9.18 & -4.31 & -4.37 & -3.93 \\  
     NGC3078 & $\backslash$ & 3.91 & 4.37 & 9.0 & 0.95 & 8.74 & -3.96 & -4.46 & -4.51 \\  
     NGC4458 & $\cap$ & 4.09 & 10.1 & 7.8 & 0.16 & 6.76 & -4.39 & -2.71 & -3.96 \\  
     NGC4478 & $\cap$ & 1.13 & 3.11 & 19.1 & -0.10 & 7.50 & - & -4.74 & -4.46 \\  
     NGC5017 & $\backslash$ & 1.91 & 5.11 & 9.8 & 1.12 & 7.98 & -2.89 & -3.99 & -3.42 \\  
     NGC5576 & $\cap$ & 3.96 & 4.74 & 550 & 2.73 & 0.4 & -4.47 & -4.09 & -4.11 \\  
     NGC5796 & $\wedge$  & 5.03 & 4.79 & 232 & 0.41 & 9.23 & -4.87 & -4.32 & -4.33 \\  
     NGC5831 & $\backslash$ & 3.36 & 4.72 & 7.0 & 0.33 & 7.89 & -4.63 & -4.11 & -4.08 \\  
     NGC5845 & $\backslash$ & 0.57 & 2.74 & 13.9 & 0.51 & 8.81 & -4.71 & -4.74 & -4.76
     \label{tab:compare}
   \end{tabular}
\\ $^{a}$ Galaxy type, with $\cap$, $\backslash$, and $\wedge$ denoting core, cusp, and intermediate galaxies, respectively; $^{b}$ half-light radius $R_{\rm e}$; $^{c}$ Sersic index $n$; $^{d}$ Nuker break radius $R_{\rm b}$; $^{e}$ inner power law slope $\gamma$; $^{f}$ SMBH mass $M_\bullet$ as computed from the $M_\bullet - \sigma$ relation; $^{g}$TDE rate, calculated using Nuker parameterization; $^{h}$TDE rate, calculated using Sersic parameterization; $^{i}$ TDE rate, calculated using core-Sersic parametrization.
 \end{minipage}
\end{table*}

\subsection{SMBH Occupation Fraction}

Volumetric TDE rates are obtained by combining well known galaxy scaling relations with our best-fit power law for $\dot{N}_{\rm TDE}(M_\bullet)$ (Eq.~\ref{eq:bestfit}).  The {\it R}-band luminosity $L_{\rm R}$ function of galaxies is assumed to follow the Schechter function \citep{Schechter76}
\begin{equation}
\phi(L_{\rm R}){\rm d}L_{\rm R} = \phi_\star \left( \frac{L_{\rm R}}{L_\star} \right)^{-1.1}\exp (-L_{\rm R}/L_\star){\rm d}L_{\rm R},
\end{equation}
where $\phi_\star = 4.9 \times 10^{-3} h_7^3~{\rm Mpc}^{-3}$, $L_\star = 2.9 \times 10^{10} h_7^{-2} L_{\odot}$, and we take the normalized Hubble constant $h_7 = 1$ \citep{Brown+01}.  Combining the Faber-Jackson law, $\sigma \approx 150~{\rm km~s}^{-1} ( L_{\rm R}/10^{10}L_{\odot})^{1/4}$, with the MM13 calibration of the $M_\bullet - \sigma$ relationship (Eq.~\ref{eq:msigma})
allows us to rewrite the Schechter function in terms of the SMBH mass,
\begin{align}
\phi (M_\bullet) {\rm d}M_\bullet = & 2.56 \phi_\star f_{\rm occ} \left( \frac{M_\bullet}{10^8M_\odot} \right)^{-1.07} \\
& \times \exp \left(-0.647 \left( \frac{M_\bullet}{10^8 M_\odot} \right)^{0.709} \right) {\rm d}M_\bullet, \notag
\end{align}
where $f_{\rm occ}(M_{\bullet})$ is the occupation fraction of SMBHs.  

\begin{figure}
\includegraphics[width=85mm]{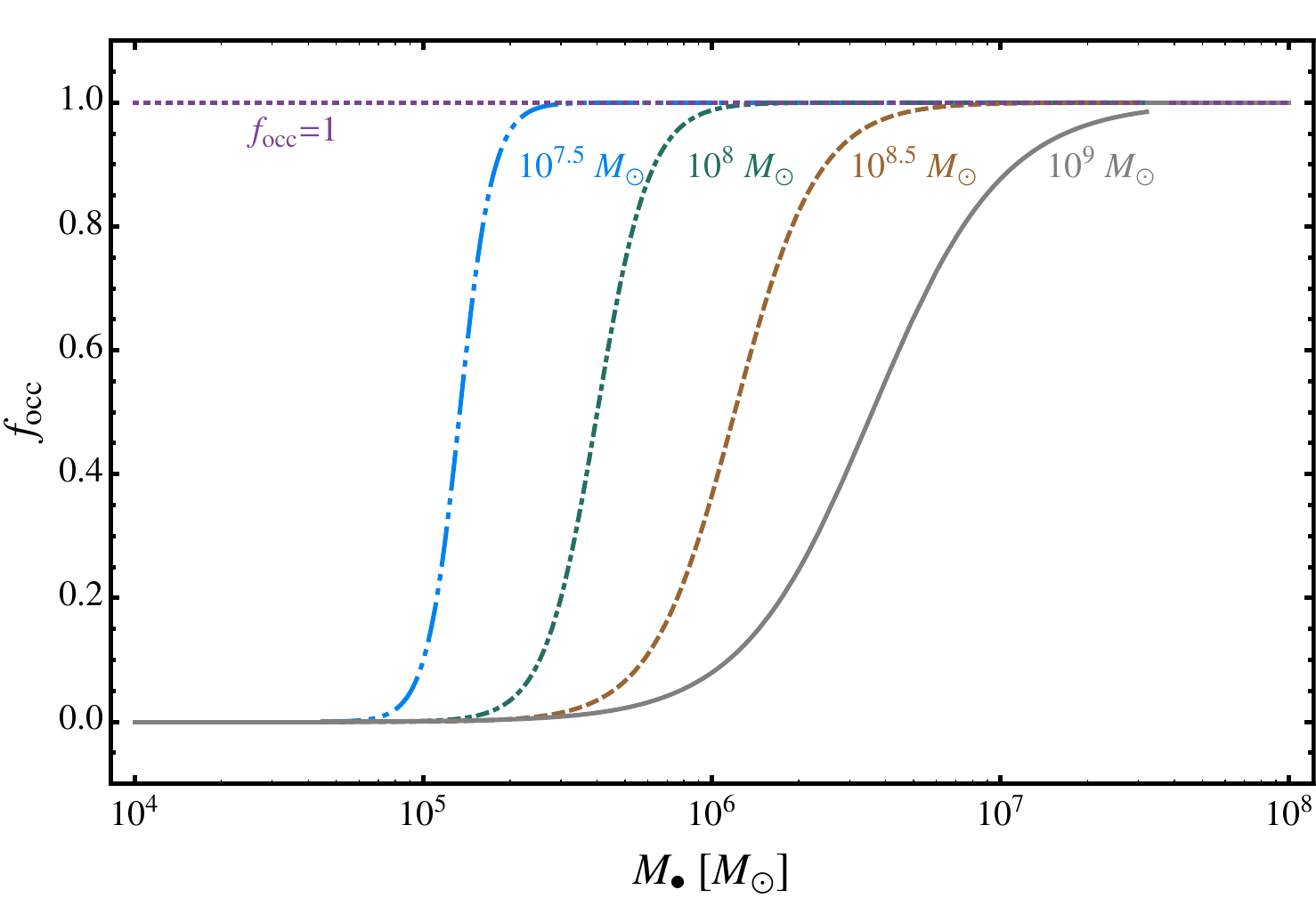}
\caption{Occupation fraction of SMBHs $f_{\rm occ}$ as a function of SMBH mass $M_{\bullet}$, based on the parameterization given in equation (\ref{eq:fOcc}) from \citet{Miller+14} for different values of the critical turnover mass $M_{\rm c}/M_{\odot} = 10^9$ ({\it gray, solid}), $10^{8.5}$ ({\it brown, dashed}), $10^{8}$ ({\it green, dot-dashed}), $10^{7.5}$ ({\it blue, dot-dot-dashed}), $\sim 0$ (uniform occupation fraction; {\it purple, dotted}).  Where appropriate, the turnover mass is labeled.}
\label{fig:fOcc}
\end{figure}

The intrinsic TDE rates are relatively robust to uncertainties in the stellar mass function and the parameterization of galaxy surface brightness profiles ($\S\ref{sec:results}$; Table \ref{tab:compare}).  However, the SMBH occupation fraction is much less certain, especially in low mass galaxies (e.g.~\citealt{Greene&Ho07}).  In order to explore the sensitivity of the TDE rate to the SMBH mass function, we follow \citet{Miller+14} in parameterizing the occupation fraction as
\begin{equation}
f_{\rm occ} = \begin{cases} \label{eq:fOcc}
0.5+0.5& \tanh\left(\ln \left(\frac{M_{\rm bulge}}{M_{\rm c}} \right) \times 2.5^{8.9-\log_{10} (M_{\rm c}/M_\odot)} \right),\\
& M_{\rm bulge} < 10^{10}M_\odot \\
1,& M_{\rm bulge} > 10^{10}M_\odot
\end{cases}
\end{equation}
where $M_{\rm bulge}$ is the bulge mass, which we relate to the SMBH mass using the $M_{\rm bulge}-M_\bullet$ relation from MM13.  The parameter $M_{\rm c}$ is the approximate mass below which the occupation fraction turns over, the value of which is not well constrained observationally but is likely to be less than $\sim 10^{8.5}M_{\odot}$ (Miller, private communication).  In what follows we consider five fiducial occupation fractions (Fig. \ref{fig:fOcc}), corresponding to $M_{\rm c}/M_\odot$ = $10^{9}$ (case A), $10^{8.5}$ (case B), $10^{8}$ (case C), and $10^{7.5}$ (case D), along with a uniform $f_{\rm occ} = 1$ (case E).

\begin{figure}
\includegraphics[width=85mm]{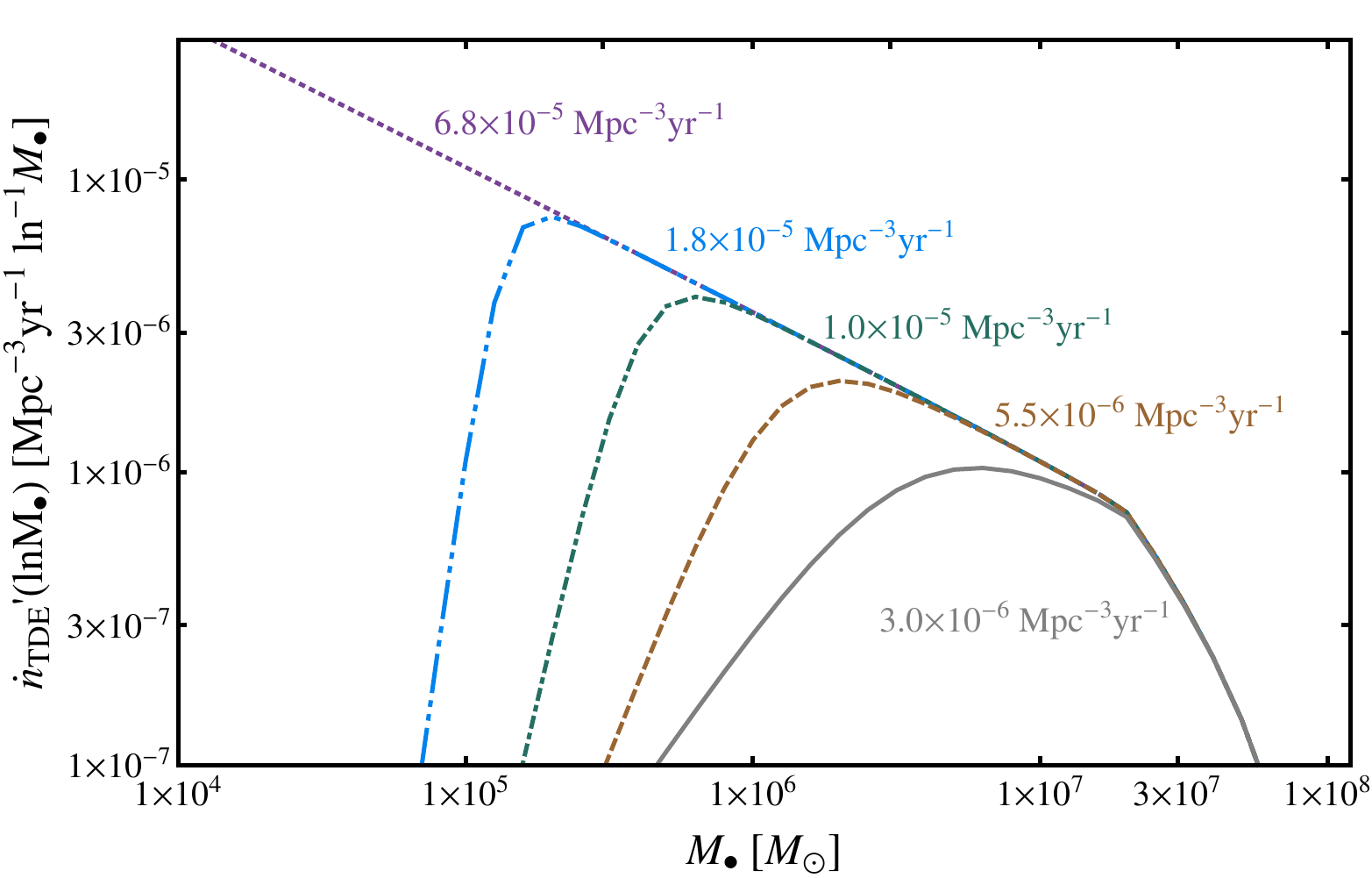}
\caption{Volumetric rate of TDEs $\dot{n}_{\rm TDE}'(\ln M_\bullet)={\rm d}\dot{n}_{\rm TDE}/{\rm d}\ln M_{\bullet}$ per unit log SMBH mass $M_{\bullet}$, as a function of $M_{\bullet}$.  Different lines correspond to different assumptions about the low mass cut-off in the SMBH occupation fraction $f_{\rm occ}$, with line styles and colors corresponding to $M_{\rm c}$ values in Fig. \ref{fig:fOcc}.  The decrease in the TDE rate at $M_{\bullet} \gtrsim 2\times 10^{7}M_{\odot}$ occurs because low mass stars (which dominate the total TDE rate for the assumed Kroupa PDMF) possess small radii and hence fall into the BH without being disrupted.  We also show the volumetric rate integrated over ${\rm d}\ln M_\bullet$ for each of our five cases, with colors corresponding to the associated curves.}
\label{fig:NDot}
\end{figure}

With the adoption of an occupation fraction, we can now calculate the volumetric rate of TDEs, $\dot{n}_{\rm TDE}$.  First we compute the differential volumetric rate of TDEs with respect to SMBH mass,
\begin{equation}
\dot{n}_{\rm TDE}'(\ln M_\bullet)= \int_{M_{\star}^{\rm min}}^{M_{\star}^{\rm max}} \dot{N}_{\rm TDE}(M_{\bullet},M_{\star})\phi_{\bullet}(M_{\bullet}) \chi_{\rm Kro}(M_{\star})dM_{\star},
\label{eq:dNdV}
\end{equation}
which is shown in Figure \ref{fig:NDot} for the different occupation fraction models.  For brevity we have written $\dot{n}_{\rm TDE}'(\ln M_\bullet)={\rm d} \dot{n}_{\rm TDE}/{\rm d}\ln M_\bullet$.  Here $\dot{N}_{\rm TDE}(M_\bullet, M_{\star})$ is given by our power law fit for $\dot{N}_{\rm TDE}$ (i.e. Eq. \ref{eq:bestfit}) if $M_\bullet < M_{\rm Hills}(M_{\star})$, but is $0$ otherwise.  Consequently, small M stars dominate the volumetric TDE rate, although their smaller Hills mass enables solar-type stars to compete for $2\times 10^7 \lesssim M_\bullet/M_\odot \lesssim 10^8$.

The full volumetric rate is then simply 
\begin{equation}
\dot{n}_{\rm TDE} = \int \frac{{\rm d}\dot{n}_{\rm TDE}}{{\rm d}{\rm ln}M_{\bullet}} {\rm d}\ln M_\bullet .
\end{equation}
One can also define a galaxy-averaged TDE rate, $\langle \dot{N}_{\rm TDE} \rangle =\dot{n}_{\rm TDE}/n_{\rm gal}$, where $n_{\rm gal} = \int \phi_{\bullet}dM_{\bullet} = 0.015~{\rm Mpc}^{-3}$ is the total spatial density of galaxies (calculated assuming a lower limit of $M_{\rm bulge} = 10^7 M_\odot$ on the bulge mass defining a ``galaxy").  The five occupation fraction distributions give different average TDE rates, with $\langle \dot{N}_{\rm TDE} \rangle$ of $2.0\times 10^{-4}~{\rm yr}^{-1}$ (case A), $3.7 \times 10^{-4}~{\rm yr}^{-1}$ (case B), $6.7 \times 10^{-4}~{\rm yr}^{-1}$ (case C), $1.2\times 10^{-3}~{\rm yr}^{-1}$ (case D), and $4.6 \times 10^{-3}~{\rm yr}^{-1}$ (case E).

\subsection{Distributions of Observables}

The TDE rate can also be translated into distributions of variables that are either directly observable, or dynamically important for TDE observables.  We calculate differential volumetric TDE rates with respect to a variable $X$, once again denoting ${\rm d}\dot{n}_{\rm TDE}/{\rm d}\ln X=\dot{n}_{\rm TDE}'(\ln X)$, by using Eq.~\eqref{eq:dNdV} and changing variables while integrating over the Kroupa PDMF $\chi_{\rm Kro}(M_{\star})$:
\begin{equation}
\dot{n}_{\rm TDE}'(\ln X) = \int_{M_\star^{\rm min}}^{M_\star^{\rm max}} \frac{{\rm d}\dot{n}_{\rm TDE}}{{\rm d}{\rm ln}M_{\bullet}} \chi_{\rm Kro} \frac{{\rm d}\ln M_\bullet}{{\rm d}\ln X} {\rm d}M_{\star}  .
\end{equation}

\begin{figure*}
\centering
\begin{tabular}{cc}
\includegraphics[width=85mm]{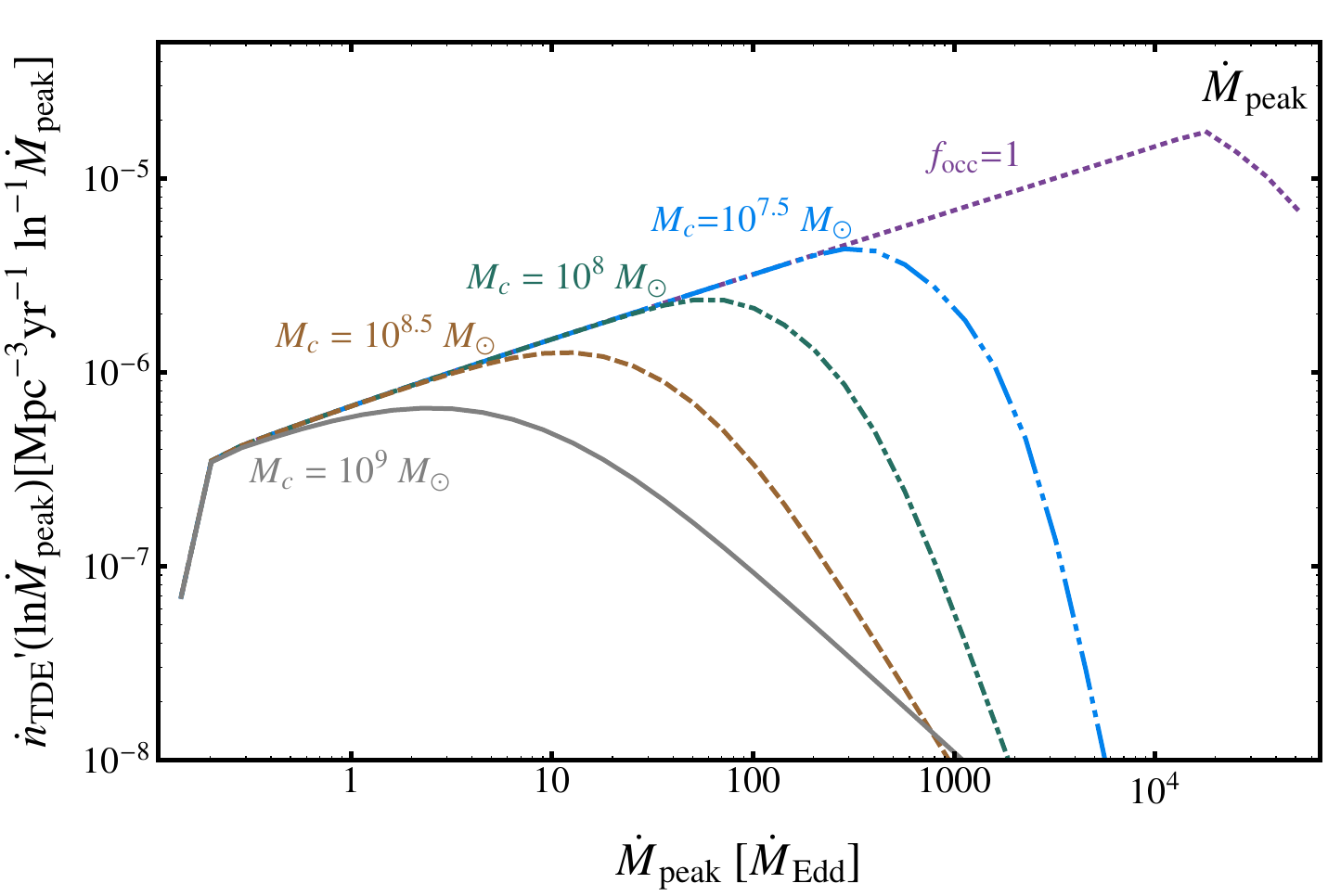} &
\includegraphics[width=85mm]{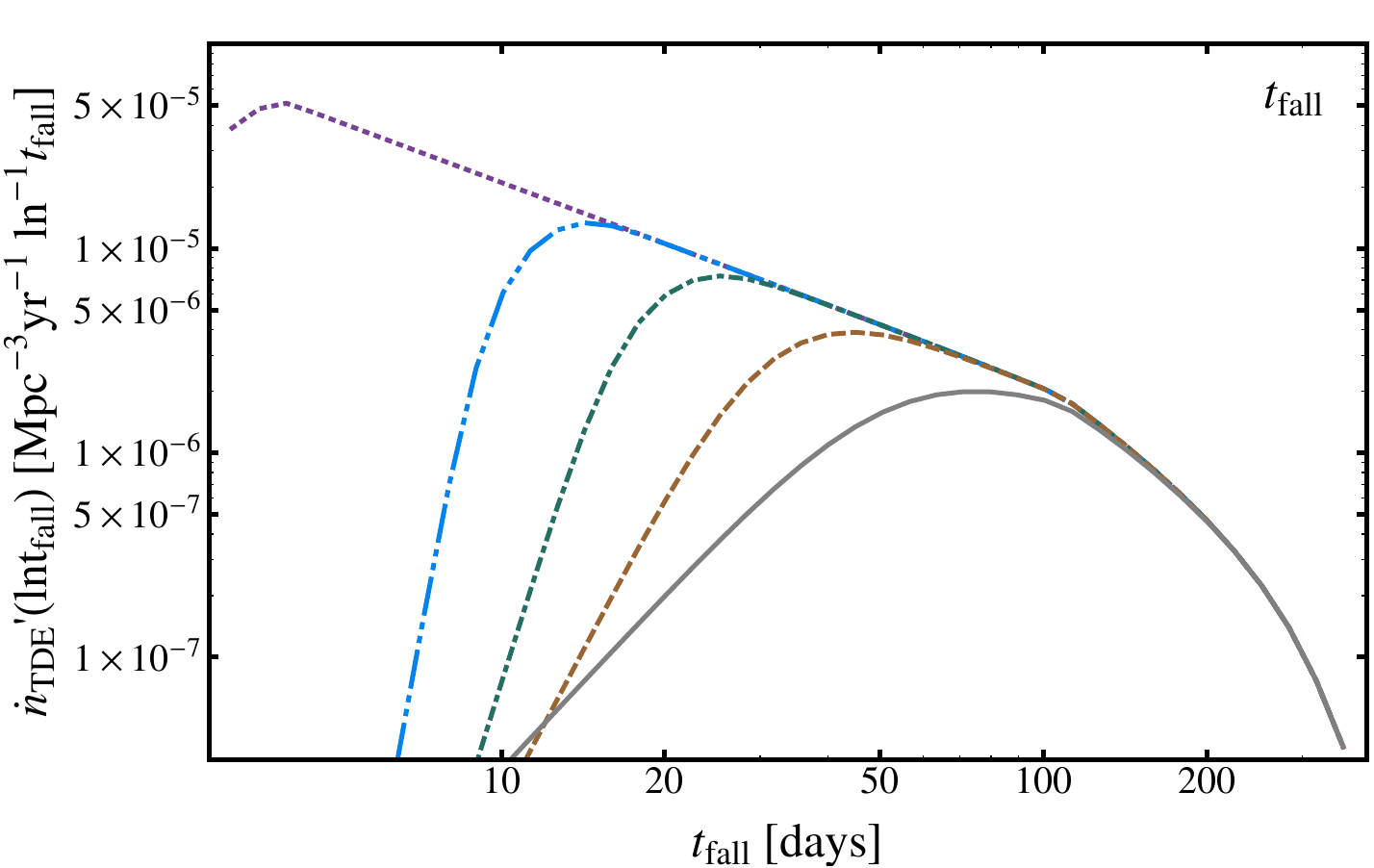} \\
\includegraphics[width=85mm]{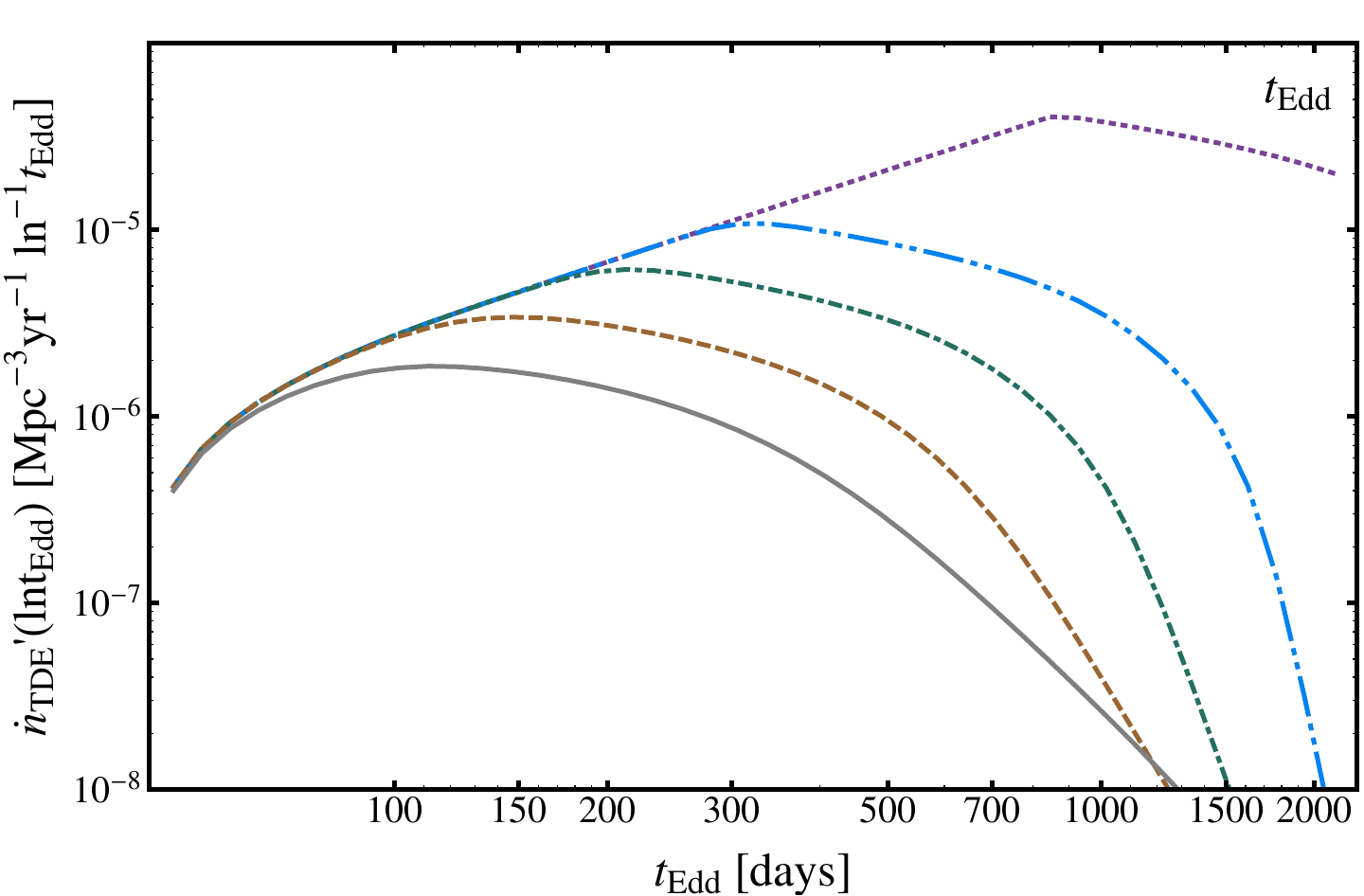} & 
\includegraphics[width=85mm]{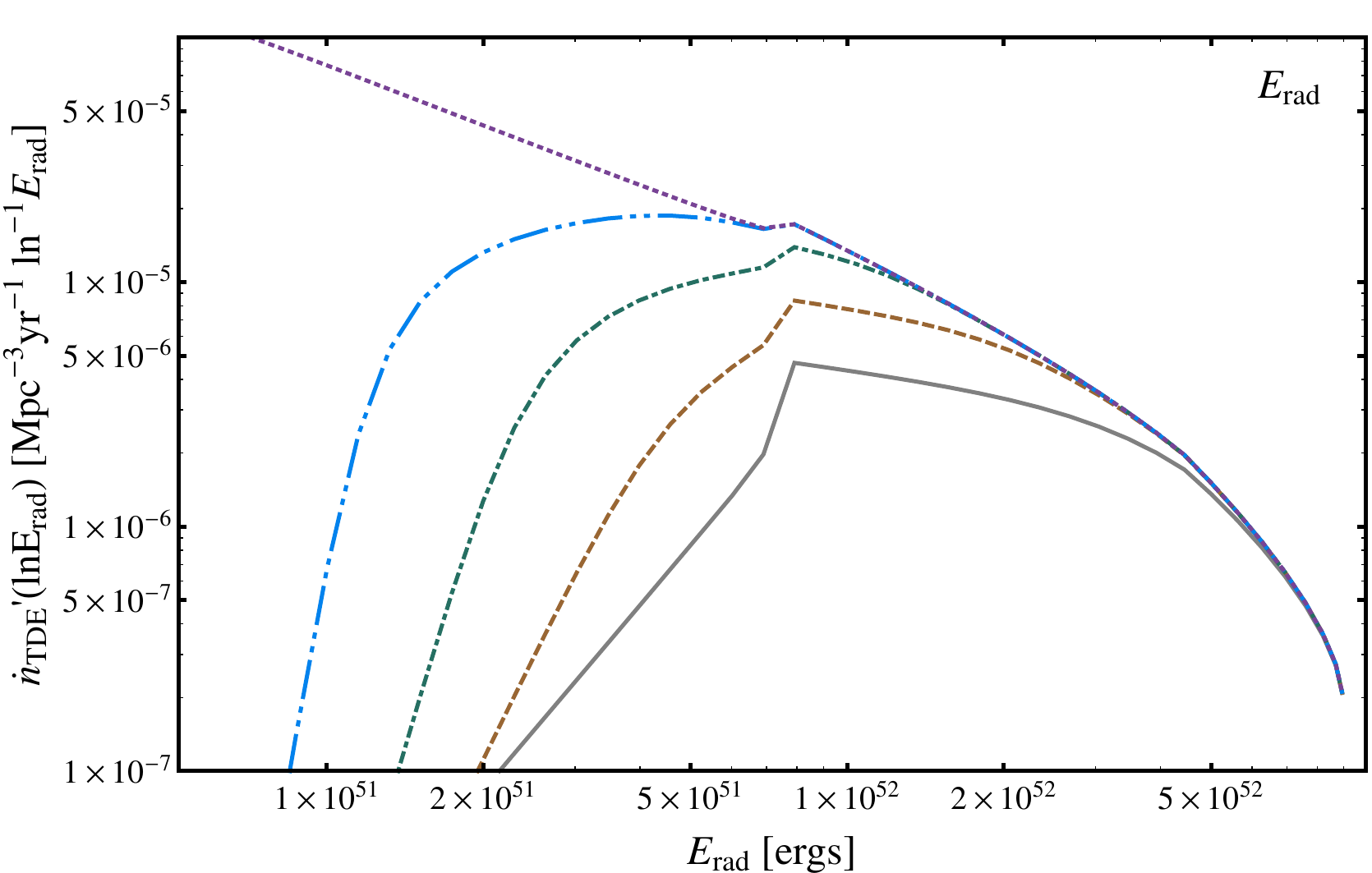}
\end{tabular}
\caption{The volumetric TDE rate, weighted by different potentially observable quantities, shown for different assumptiosn about the SMBH occupation fraction $f_{\rm occ}$, with the line styles and colors the same as in Fig.~\ref{fig:fOcc}.  Observable quantities shown include the peak fallback rate $\dot{M}_{\rm peak}/\dot{M}_{\rm Edd}$ (Eq.~\ref{eq:Mdotpeak}; {\it top left}); characteristic fallback time $t_{\rm fall}$ (Eq.~\ref{eq:tfb}; {\it top right}); Eddington timescale $t_{\rm Edd}$ (Eq.~\ref{eq:tedd}; {\it bottom left}); total Eddington-limited radiated energy $E_{\rm rad}$ (Eq.~\ref{eq:ERad}; {\it bottom right}).  }
\label{fig:NDotObs}
\end{figure*}

Figures \ref{fig:NDotObs} shows our results for TDE distribution with respect to peak fallback rates $\dot{M}_{\rm peak}/\dot{M}_{\rm edd}$ (Eq.~\ref{eq:Mdotpeak}), fallback timescales $t_{\rm fall}$ (Eq.~\ref{eq:tfb}), Eddington timescales $t_{\rm edd}$ (Eq.~\ref{eq:tedd}), and total radiated energy $E_{\rm rad}$ (Eq. \ref{eq:ERad}).  The peak Eddington ratio is very sensitive to the occupation fraction, with the most probable value of $\dot{M}_{\rm peak}/\dot{M}_{\rm edd}$ varying between $\sim 10-1000$ as the turn-over mass decreases from $M_{\rm c} = 10^{8.5}M_{\odot}$ to $10^{7.5}M_{\odot}$.  For all our choices of occupation fraction except for case A ($M_{\rm c}=10^9 M_\odot$, which is in any case disfavored by observations of nearby galactic nuclei), most TDEs are characterized by a phase of highly super-Eddington accretion.  This also implies that TDE emission mechanisms that scale with absolute accretion power, instead of those which are limited to the Eddington luminosity, provide the most sensitive probe of the SMBH occupation fraction.  

The distribution of fall-back times $t_{\rm fall}$ is less sensitive to the occupation fraction, with typical values ranging from a few weeks to a few months (except for case E, where $t_{\rm fall} \sim 1~{\rm day}$).  All models for TDE emission predict light curves with characteristic durations $\gtrsim t_{\rm fall}$.  Because the characteristic fallback time generally exceeds a couple weeks, even for a relatively low value of $M_{\rm c} = 10^{7.5}M_{\odot}$, this shows that optical surveys such as ZTF or LSST, with planned cadences of several days, should (modulo selection criteria) be limited by flux rather than cadence in TDE discovery.     

The distribution of Eddington timescales $t_{\rm edd}$ also depends only weakly on the occupation fraction; other than setting the overall normalization of the distribution, $f_{\rm occ}$ mainly determines the steepness of the cutoff at large $t_{\rm edd}$.  The value of $t_{\rm edd} \approx 500$ days inferred by the time at which the beamed X-ray emission shut off following the jetted TDE {\it Swift} J1644+57 (e.g., \citealt{Zauderer+13}; \citealt{Tchekhovskoy+13}; \citealt{Kawashima+13}) appears in line with theoretical expectations.  The distribution of (Eddington-limited) energy radiated, $E_{\rm rad}$, is more sensitive to $f_{\rm occ}$, but as discussed in \S \ref{sec:observables}, it is challenging to measure anything but lower limits on this quantity.

\subsection{Detection Rate of TDEs by Optical Surveys}
\label{sec:detection}

The population of TDEs that will be selected by optical surveys depends sensitively on which emission mechanism dominates ($\S\ref{sec:opticalmodels}$; Appendix \ref{sec:optical}).  We calculate the detection rate of TDEs, $\dot{N}_{\rm obs}$, using a simple flux threshold criterion, as is motivated by the generally long duration of TDE flares relative to the planned cadence of upcoming surveys.  Our results are normalized to those detected by an all-sky survey with a {\it g}-band limiting magnitude of $g_{\rm lim} = 19$, in order to represent the approximate sensitivity of the planned survey by ZTF (e.g.~\citealt{Rau+09}; \citealt{Kulkarni12}).  Although the $5\sigma$ limiting magnitude of PTF is formally $\sim 21$, it is hard to determine whether a detected optical transient is in fact a TDE without high signal to noise and the ability to resolve the light curve for several epochs away from peak.  All three TDE flares discovered by iPTF possess peak {\it g}-band magnitudes $g \approx 19$ (\citealt{Arcavi+14}), motivating our choice of this limit.  Absolute rates can be readily scaled to other limiting magnitudes $g$ according to $\dot{N}_{\rm obs} \propto f_{\rm sky} \times 3.95^{(g-19)}$, where $f_{\rm sky}$ is the fraction of the sky covered (e.g.,  20$\%$ for PTF).  We neglect cosmological corrections to the light curves, as well as a possible evolution in the TDE rate with redshift, because most detected events occur at $z < 1$.  These assumptions are approximately correct for PTF and ZTF, but may not be justified for the brightest emission mechanisms, when applied to LSST.  

\begin{figure*}
\centering
\begin{tabular}{cc}
\includegraphics[width=85mm]{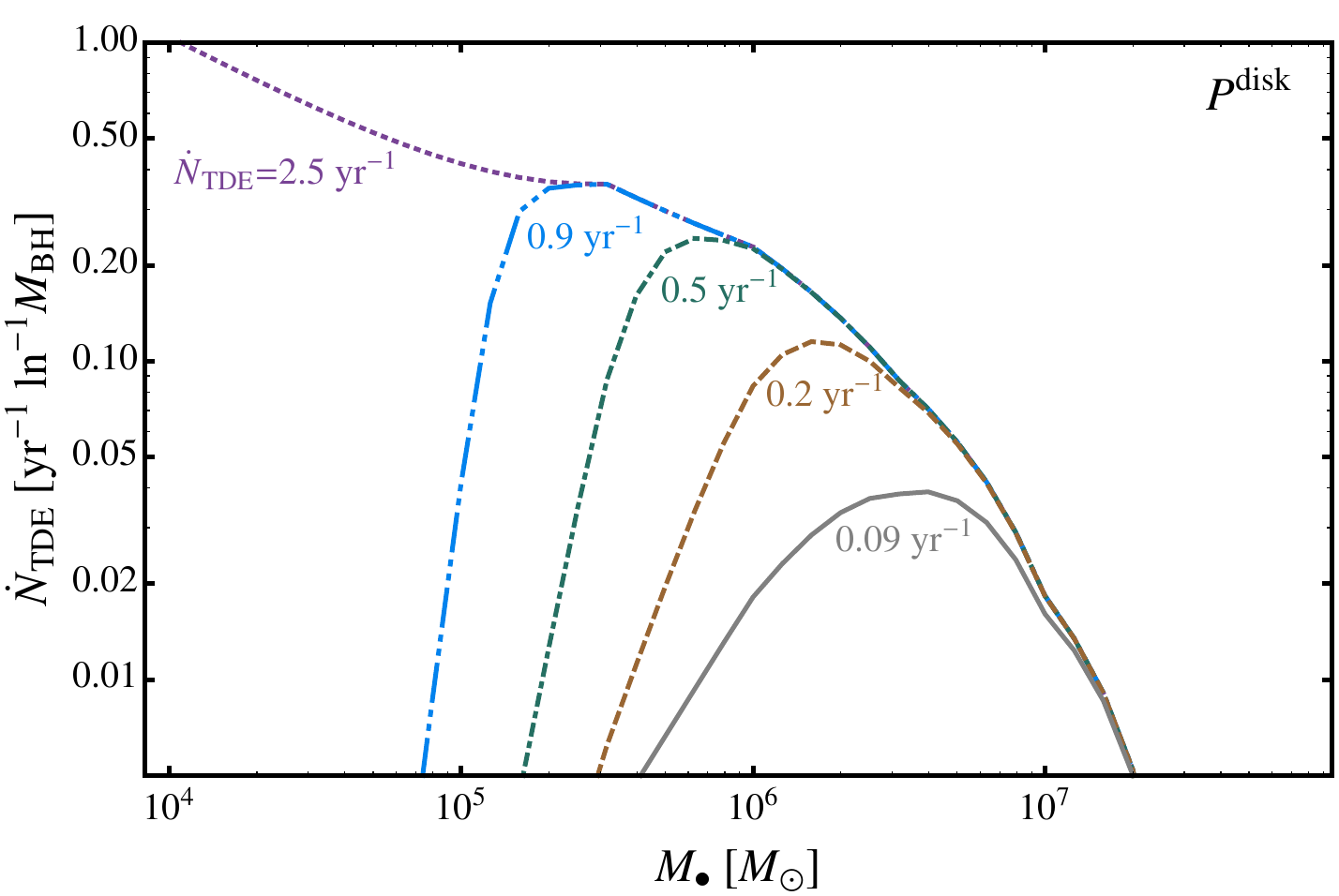} &
\includegraphics[width=85mm]{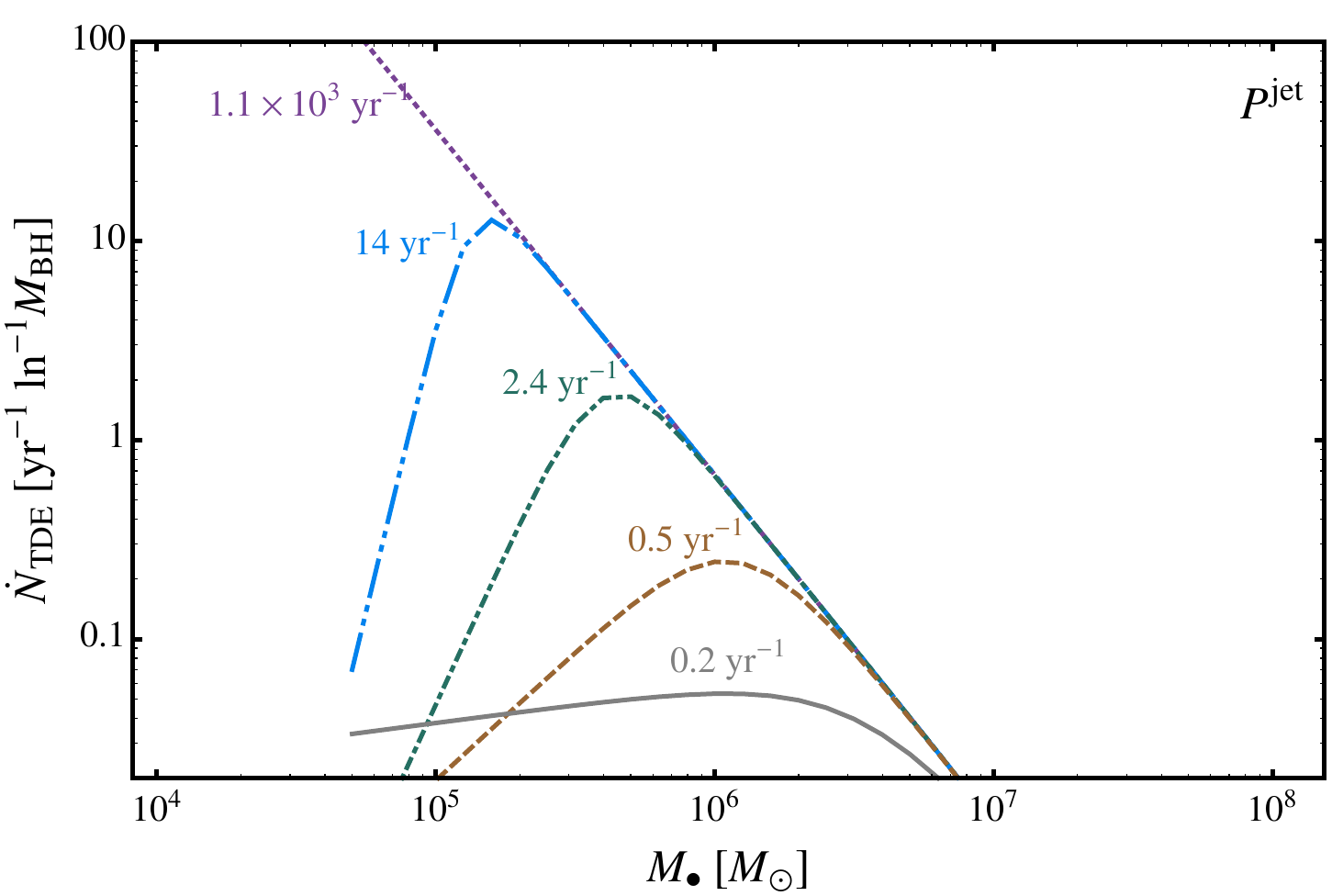} \\
\includegraphics[width=85mm]{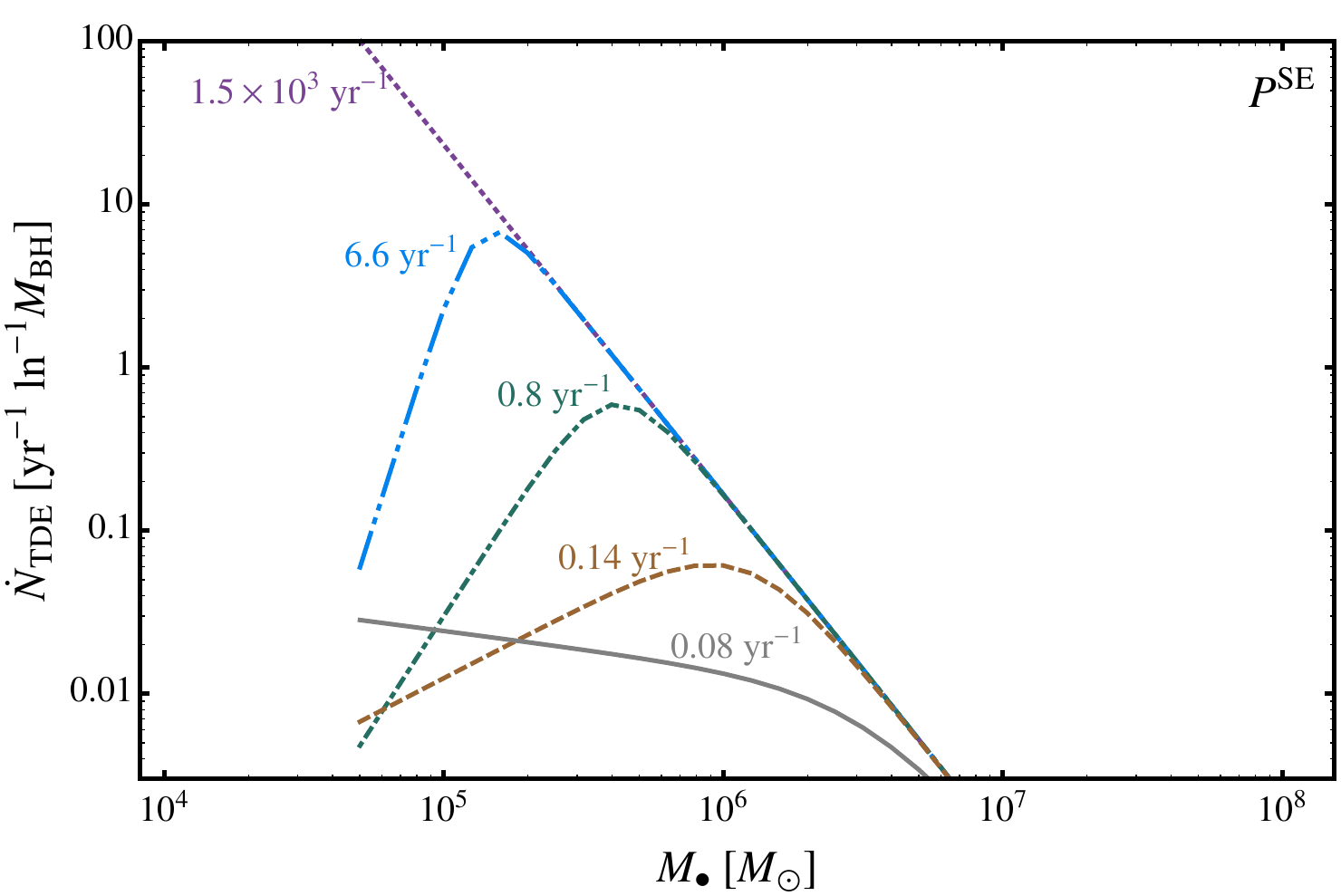} & 
\includegraphics[width=85mm]{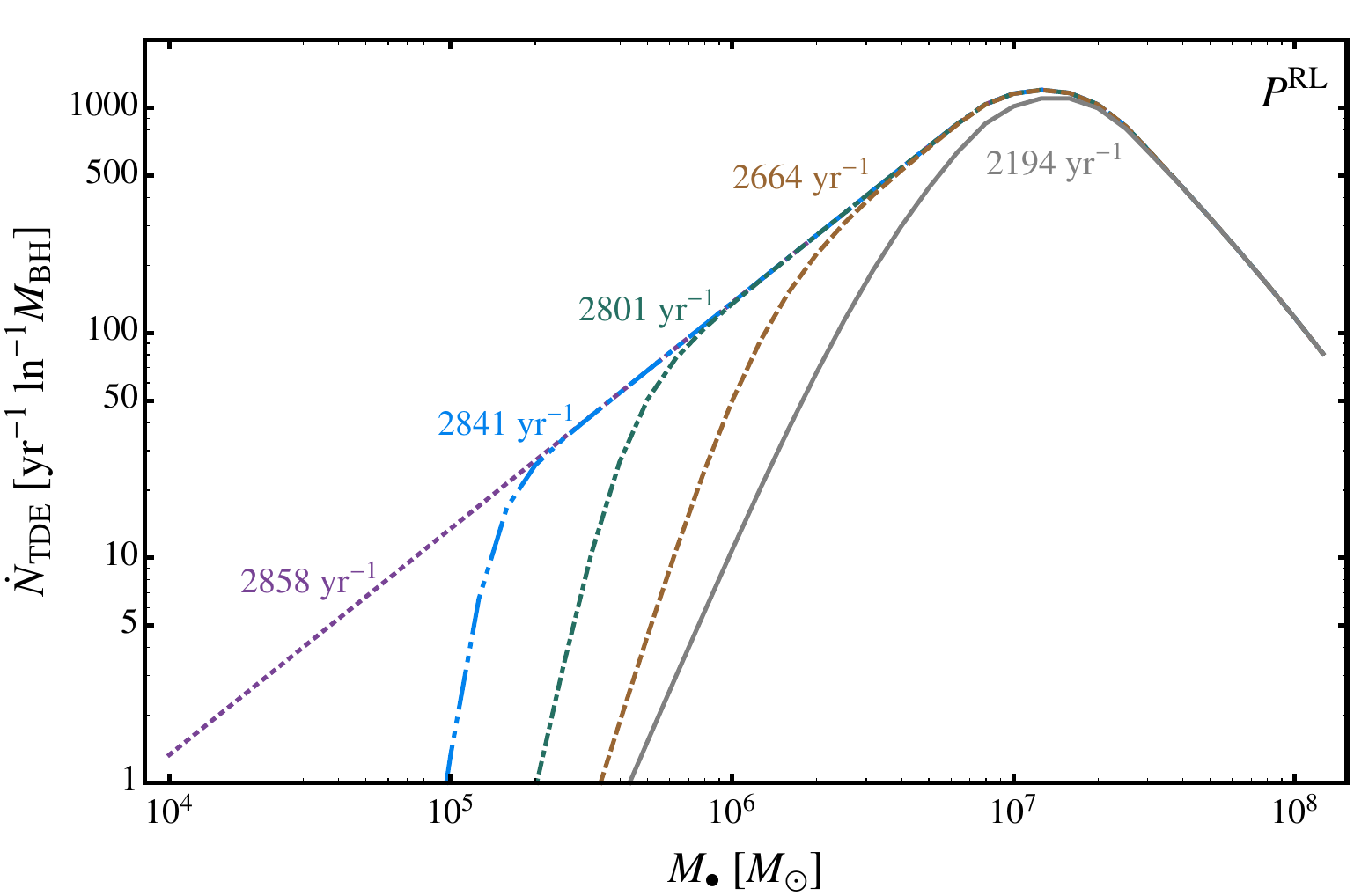}
\end{tabular}
\caption{Observed rates of TDE by an all-sky optical survey of limiting magnitude $g = 20$ per unit log SMBH mass, shown for different SMBH occupation fractions (colors and line styles are the same as in Fig.~\ref{fig:fOcc}) and for different models of the TDE optical emission mechanism (different panels).  Total rates (integrated over all SMBH masses) are marked next to each line.  Emission models shown include thermal emission from the spreading disk ({\it upper left}; $\S\ref{sec:thermal}$), synchrotron radiation from off-axis jet (assumed to accompany 1 per cent of all TDEs; {\it upper right}; $\S\ref{sec:jet}$), super-Eddington outflows ({\it lower left}; $\S\ref{sec:SE}$), and reprocessing by an optically thick layer ({\it lower right}; \S\ref{sec:reprocessing}).  Note that the off-axis jet and super-Eddington outflows strongly differentiate between SMBH occupation models, while the Eddington-limited models (spreading disks and reprocessing layers) do not.}
\label{fig:NDotObs2}
\end{figure*}

Figure \ref{fig:NDotObs2} shows the detected TDE distribution with respect to SMBH mass, assuming different models for the optical emission (and setting $f_{\rm sky}=1$).  Thermal emission from the spreading disk represents the dimmest mechanismwe  consider and the resulting detection rates ({\it upper left panel}) are correspondingly low ($\sim 0.1-3~{\rm yr}^{-1}$ for an all-sky survey).  Since the emission is Eddington limited in this scenario, the SMBH distribution measured by such a survey only depends weakly on the SMBH mass fraction, with the total number of events ranging from $0.2 - 0.9~{\rm yr}^{-1}$ as the turn off mass of the occupation fraction decreases from $M_{\rm c} = 10^{8.5}$ to $10^{7.5}M_{\odot}$.

The other scenarios result in more optimistic TDE rates.  Both super-Eddington outflows and off-axis jets predict a bottom-heavy SMBH mass distribution among detected events.  In these scenarios, the total observed TDE rate depends sensitively on the occupation fraction.  If super-Eddington outflows (synchrotron jets) are the dominant optical emission mechanism, a SMBH mass cutoff of $M_{\rm c}=10^{7.5}M_\odot$ yields 6.6 (14) detections per year, compared to a much smaller 0.14 (0.5) per year if $M_{\rm c}=10^{8.5}M_\odot$.  In both of these scenarios, case E ($f_{\rm occ}=1$) produces $\dot{N}_{\rm obs}\sim1\times10^{3}~{\rm yr}^{-1}$.  

Contrastingly, the reprocessing layer model ($\S\ref{sec:reprocessing}$) is Eddington limited, cutting off detections at low $M_\bullet$ and producing a sharp peak in $N_{\rm obs}(M_\bullet)$ near $M_{\bullet} \sim 10^{7}M_{\odot}$, with a total rate $\sim 10^3$ yr$^{-1}$ that is much less sensitive to the low-$M_{\bullet}$ occupation fraction.  Although this model provides the closest match to observed peak luminosities of TDE candidates, the predicted detection rates (after accounting for PTF's limited sky coverage) are a factor $\sim 10^{2}$ higher than what was actually found with PTF \citep{Arcavi+14}\footnote{We have conservatively set the reprocessing layer efficiency to $\epsilon_{\rm opt}=0.03$, which is already slightly low compared to the observed peak luminosities of some TDEs in Fig. \ref{fig:LPeak}.}.  As we will discuss in the subsequent section, this overprediction could be alleviated if only a small fraction, $\sim 1-10\%$, of all TDEs possess a reprocessing layer.

\section{Discussion}
\label{sec:discussion}

We have seen that volumetric tidal disruption event rates are fairly sensitive to the bottom end of the SMBH mass function.  A flux-limited TDE sample will be extremely sensitivity to choice of $f_{\rm occ}$ if optical emission is not Eddington-limited, a point raised in \citet{StrQua09} and quantified for samples of X-ray selected TDE jets in \citet{De-ColleGuillochon+:2012a}.  However, the Eddington-limited emission mechanisms we consider give detection rates $\dot{N}_{\rm obs}$ that are highly insensitive to $f_{\rm occ}$.  The current sample of TDEs is inhomogeneous and suffers from many selection effects, but is still informative because of the enormous variance in both $\int \dot{N}_{\rm TDE}(M_\bullet){\rm d}M_\bullet$ and in $\dot{N}_{\rm TDE}(M_\bullet)$ with respect to $f_{\rm occ}(M_{\bullet})$ and choice of optical emission mechanism (Fig. \ref{fig:NDotObs2}).

\subsection{Rate Tension}

We have calculated a per-galaxy TDE rate of $\langle \dot{N}_{\rm TDE}\rangle \sim {\rm few} \times 10^{-4}~{\rm yr}^{-1}$ gal$^{-1}$, which exceeds the best observationally inferred values by at least an order of magnitude.  This disagreement is with respect to the flare rate of $\sim 10^{-5}$ yr$^{-1}$ inferred by both X-ray \citep{Donley+02} and optical/UV \citep{Gezari+09} surveys\footnote{Although we do note that one analysis of X-ray TDEs inferred a rate more consistent with our calculations, of $2.3 \times 10^{-4}~{\rm yr}^{-1}~{\rm gal}^{-1}$ \citep{Esquej+08}.}.  Although many of these rate estimates are troubled by selection effects, the TDE rates inferred from the flux-limited sample of \citet{vanVelzen&Farrar14} also fall an order of magnitude below our lowest estimates.  This discrepancy is also apparent from comparing our direct estimate of the optical flare detection rate (Fig.~\ref{fig:NDotObs2}) to the optically-selected TDE sample.  For instance, for the reprocessing emission model tuned to best reproduce the observed light curves, our estimated detection rate of $\sim 100$ per year for PTF ($f_{\rm sky} = 0.2$) greatly exceeds the three accumulated TDE flare candidates reported by PTF over its three year survey (\citealt{Arcavi+14}).  

In this paper we have relaxed and updated a number of assumptions used in past theoretical rate calculations \citep{MagTre99, Wang&Merritt04} in an attempt to alleviate this rate discrepancy, but in general this has had little effect, and if anything may have only heightened the tension between theory and observation.  Our rate calculations employ a significantly larger galaxy sample than in the past\footnote{Our sample $N=144$ is significantly larger than the $N=29$ galaxy sample of \citet{MagTre99}, or the $N=41$ galaxy sample used in \citet{Wang&Merritt04}.} and we have incorporated updated galaxy scaling relations, but neither of these changes has a significant effect on the volumetric TDE rate.  Theoretically calculated TDE rates are, furthermore, relatively unchanged by the use of alternate (non-Nuker) parametrizations for galactic surface brightness profiles, or by including a realistic stellar mass function.  We emphasize that the theoretical TDE rates calculated in this paper are in most ways {\it conservative floors on the true TDE rate}, as they neglect alternative relaxational mechanisms, the non-conservation of angular momentum in aspherical potentials, and the potentially enhanced rates of angular momentum diffusion due to stellar-mass mass black holes  ($\S\ref{sec:PDMF}$).  The robustness of the tension between predicted and observed TDE rates motivates alternate ways to bring these two into alignment.  

Perhaps most conservatively, the observed flare rate could be reduced by environmental or selection effects.  Galactic nuclei can suffer from significant dust obscuration, which would reduce the optical flux and the corresponding detection rate.  Significant dust extinction was inferred for the jetted TDE {\it Swift J1644+57} (\citealt{Bloom+11}), but the SEDs of the other, thermal TDE flares show little to no evidence of reddening (Cenko, private communication).  Dust extinction cannot account for the similar rate tension present in the X-ray selected sample (\citealt{Donley+02}), although photoelectric absorption by large columns of neutral gas could in principle play a similar role.  TDE searches must also take care to distinguish actual TDE flares from impostor transients with much higher intrinsic event rates; in particular, AGN variability and nuclear supernovae must be excluded from TDE searches through careful cuts on the candidate sample.  For example, the completed PTF survey was strongly biased against TDE detection due to frequent rejection of transients in galactic nuclei (Arcavi, private communication).  Cuts such as these, and other factors related to choice of events for followup, make it clear that our ``detectable rates'' predicted in Fig. \ref{fig:NDotObs2} represent upper limits on the TDEs detectable by optical time domain surveys.  Although the large future TDE samples of optical transient surveys will resolve many of the questions raised in this section, for now it is likely more useful to compare our volumetric ($\dot{n}_{\rm TDE}$) or per-galaxy ($\langle \dot{N}_{\rm TDE} \rangle$) rates to smaller, flux-limited samples \citep{vanVelzen&Farrar14}.

If observational selection effects can be reduced in the future and this rate tension still persists, potentially more interesting explanations could exist on the theoretical side.  While almost all of our assumptions were conservative (spherical symmetry, two-body relaxation, absence of stellar mass black holes), our assumption of isotropic stellar velocities was not necessarily so.  Two-body relaxation calculations assuming isotropic velocities will overestimate the physical TDE rate in a galaxy if the true velocity distribution is significantly anisotropic in a tangentially biased way.  This is because the longer angular momentum relaxation times of tangential orbits make them less promising sources for tidal disruption.  Conversely, a radial bias in stellar orbits would increase TDE rates even further.  From both observational and theoretical perspectives, it is unclear whether galactic nuclei are sufficiently anisotropic (and overwhelmingly in the tangential direction) as to reduce TDE rates by an order of magnitude.  

Recent N-body simulations have indicated that the presence of a loss cone will bias orbits towards tangential anisotropy near the SMBH, although this bias is minor at $r_{\rm crit}$ and a radial bias appears at larger radii \citep[Figs. 9, 15]{Zhong+14}.  If steady state loss cone dynamics are indeed insufficient to provide the required tangential bias, it could arise from more exotic dynamical processes.  For example, the presence of a SMBH binary (and its ``effective loss cone'') will strongly deplete radial orbits in a galactic nucleus, and the anisotropic scar left by such a binary on stellar orbits can persist (and reduce TDE rates) for Gyr \citep{MerWan05}.  However, an extrapolation of the fitted curve in \citet{MerWan05}, Fig. 4, would indicate that the TDE rate reduction persists for $t \lesssim 10^9~{\rm yr}$ in the small galaxies that dominate $\dot{n}_{\rm TDE}$; binary-induced anisotropy would therefore be most effective at reducing $\dot{n}_{\rm TDE}$ if SMBH binaries often fail to solve the final parsec problem.

A rate discrepancy could also result from current uncertainties in the physical processes that power the observed optical emission from TDEs, in particular given that the effective temperatures $\sim 10^{4}$ K of the current sample of optical/UV flares are much lower than those predicted by simple theoretical models for the accretion disk (e.g., Appendix $\ref{sec:thermal}$).  For example, if only 10$\%$ of TDEs possess a reprocessing layer that greatly enhances their optical luminosities relative to the majority of ``unshielded" events, then current flare samples could be dominated by this minority of high luminosity events.  Observational inferences of the true event rate would then underestimate it by a factor $\sim 10$.  
Although the development of a physically-motivated model for a reprocessing layer is beyond the scope of this work, such a layer might naturally be limited to a minority of TDEs if it requires a high value of the penetration parameter $\beta$.  For instance, we estimate that $\sim 10\%$ of all TDEs should occur with $\beta \gtrsim 3$ (Fig.~\ref{fig:SamplePinhole}).  TDE debris circularization is not yet well understood, but of the two existing models for this process, both relativistic precession \citep{Hayasa+13} and hydrodynamic compression at pericenter \citep{Guillochon+14} depend sensitively on $\beta$.  This scenario is investigated in greater detail in the following subsection.

\subsection{Non-Fiducial Scenarios Limiting Flare Production}

Rates of detectable TDEs could also be reduced if TDEs from the diffusive regime of relaxation are unable to produce bright flares, instead only shedding small amounts of mass each pericenter passage as they drift towards lower angular momentum orbits.  A more detailed analysis of angular momentum diffusion suggests that the number of pericenter passages between the onset of partial disruptions and a final, full TDE is $\sim$few for $q \gtrsim 0.3$ \citep[Figs. 4.1, 4.2]{Strubb11}.  If we repeat our analysis and only count TDEs from the $q>0.3$ regime as observable, the mean TDE rate in our sample is reduced to $55\%$ of its fiducial value: not enough to explain the rate tension between theory and observation.  The differences between pinhole and diffusive TDEs may even worsen the rate discrepancy.  In many observed optically-bright TDEs, the energy release appears to be a small fraction of $0.1M_\odot c^2$, consistent with a partial disruption (Fig. \ref{fig:ERad}, see also \citealt{Campan+15}).  If partial disruptions power a fraction of the existing TDE sample, then disruptions from the diffusive regime may generate many visible flares per star, exacerbating the rate discrepancy.

\begin{figure}
\includegraphics[width=85mm]{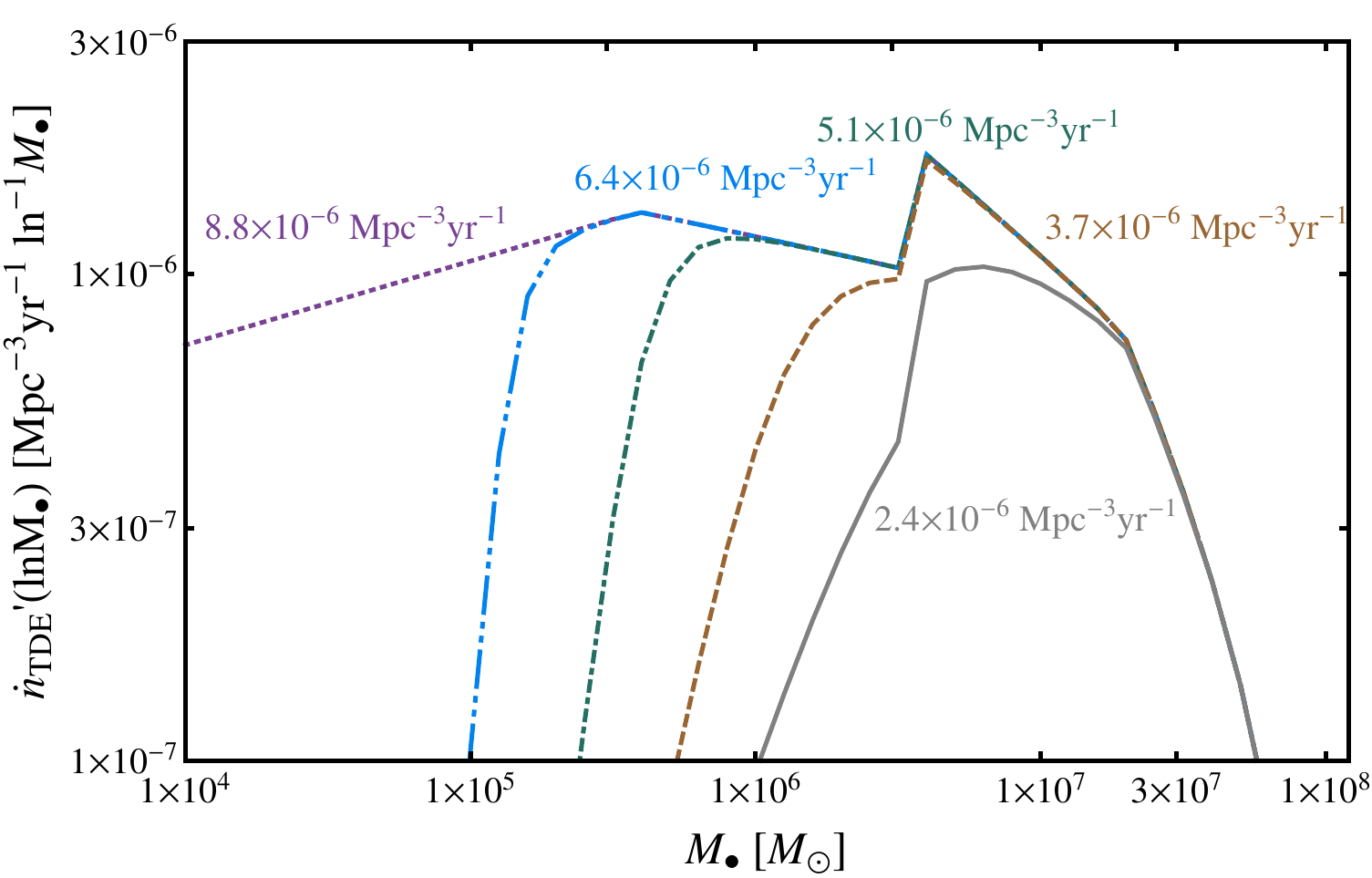}
\caption{The same as Fig. \ref{fig:NDot} (volumetric rate of TDEs, as a function of $M_{\bullet}$), but in a non-fiducial model where luminous flares are only generated by disruptions with relativistic ($R_{\rm p} < 12R_{\rm g}$) pericenters, as is motivated by \citet{Shioka+15, Hayasa+15}.  Line styles are the same as in previous figures, and the colored numbers correspond to integrated volumetric TDE rates in each scenario for $f_{\rm occ}$.  The two more conservative scenarios do not see large decreases in their integrated TDE rates, but the other three do.}
\label{fig:ratesCirc}
\end{figure}

A more extreme version of the above non-fiducial scenario is to postulate that luminous flares can only be generated assuming rapid circularization of debris streams\footnote{We are grateful to the anonymous referee for suggesting this to us.}, assuming that this circularization efficiency depends strongly on $R_{\rm p}/R_{\rm g}$.  The question of debris circularization in TDEs is very much an open one, as the dynamic range of the problem is too extensive to have been simulated from first principles in realistic TDEs.  Existing circularization simulations cheat by using one of two different non-physical limits to reduce the spatial range covered by debris streams: either (1) tidal disruption of stars on parabolic orbits by IMBHs \citep{Guillochon+14, Shioka+15}, or (2) tidal disruption of stars on eccentric orbits by SMBHs \citep{Hayasa+13, Hayasa+15, Bonner+15}.  Of the subset of these simulations that incorporate relativistic precession, efficient circularization is seen in most of the simulations of \citet{Hayasa+15, Bonner+15} but not in \citet{Shioka+15}.  The key difference is the location of the self-intersection point $R_{\rm si}$ where streams collide to dissipate kinetic energy in shocks \citep{GuiRam15, Dai+15, Stone+15}; if $R_{\rm si} \gg R_{\rm p}$, it will take many self-intersections to thermalize a large fraction of the excess specific energy, $\approx GM_\bullet / (2R_{\rm p})$.  

The simulations of \citet{Shioka+15}, which see quite inefficient circularization, have $R_{\rm si} \approx 1000R_{\rm g}$, while efficient circularization is seen in all of the \citet{Hayasa+15} simulations with an adiabatic gas equation of state and $R_{\rm si} \lesssim 250 R_{\rm g}$.   This self-intersection radius corresponds to $R_{\rm p} \approx 12.5R_{\rm g}$.  The circularization efficiency of Models 1-2 in \citet{Hayasa+15} is more ambiguous (these models have $R_{\rm si} \approx 420, 890 R_{\rm g}$), so we neglect them in this discussion.

As a conservative implementation of this idea, we postulate that luminous flares are not produced unless $R_{\rm p} < 12 R_{\rm g}$, and repeat our fiducial calculations under this assumption.  This results in all diffusive-regime TDEs being discarded when $M_\bullet \lesssim 10^7 M_\odot$, and a fraction of pinhole-regime TDEs as well.  We show the resultant volumetric TDE rates in Fig. \ref{fig:ratesCirc}.  The overall rate tension is not removed by this (rather conservative) assumption: the per-galaxy TDE rates in cases A, B, C, D, and E are reduced to $1.6\times 10^{-4}~{\rm yr}^{-1}$, $2.5\times 10^{-4}~{\rm yr}^{-1}$, $3.4\times 10^{-4}~{\rm yr}^{-1}$, $4.3\times 10^{-4}~{\rm yr}^{-1}$, and $5.8\times 10^{-4}~{\rm yr}^{-1}$, respectively\footnote{These revised rates are $81\%$, $67\%$, $50\%$, $35\%$, and $13\%$, respectively, of their fiducial values.}.  However, the distribution of $M_\bullet$ in a volume-complete TDE sample is shifted away from the smallest SMBHs, and is spread more evenly across black hole mass.  We note that the relatively low energy releases seen in optically-selected TDE flares suggest that observed flares can be accomodated by the accretion of only a small fraction of the bound mass \citep{Piran+15}, which is why we retain our earlier calculations as fiducial.  

Finally, we note two important caveats to the above discussion.  The first is the role of SMBH spin in the circularization process.  Rapid and misaligned SMBH spin induces nodal precession in tidal debris streams that winds them into different orbital planes and can retard circularization.  In the simulations of \citet{Hayasa+15}, no delay occurs in the adiabatic equation of state limit that is likely relevant for most TDEs\footnote{Delays do occur for high values of SMBH spin when the gas follows a polytropic equation of state (corresponding to efficient cooling), but order of magnitude photon diffusion timescale considerations suggest that the adiabatic limit is more physical \citep{Hayasa+15}.}; this is because heating of the debris streams increases their thickness to a size greater than the spin-induced ``gap'' at the nominal self-intersection point.  On the other hand, the semi-analytic model of \citet{GuiRam15} finds a larger role for spin-induced circularization delays, primarily due to the different analytic model employed for stream thickness.  The second caveat is the role of magnetohydrodynamic stresses in debris circularization.  These forces have not been included in any circularization simulation to date, and were initially suggested as an extra source of dissipation to aid the circularization process \citep{Guillochon+14}, but more recent work indicates that they may also be able to hinder circularization by mediating angular momentum transport \citep{Svirsk+15}.  Ultimately, the complex dynamics of TDE circularization and emission mechanisms are beyond the scope of this work, but Fig. \ref{fig:ratesCirc} provides a preliminary examination of how rates would change if highly relativistic pericenters are required to circularize debris and produce a flare.

\subsection{Black Hole Mass Distribution}

Once observational selection effects are mitigated, and our understanding of the physics of TDE emission improved, the demographics of TDE flares may prove to be a powerful probe of the occupation fraction of SMBHs in low mass galaxies.  Alternatively, given prior observational constraints on the SMBH occupation fraction, distribution of TDE flares with SMBH mass $M_{\bullet}$ could inform our knowledge of what physical processes produce TDE emission.  

\begin{figure*}
\includegraphics[width=170mm]{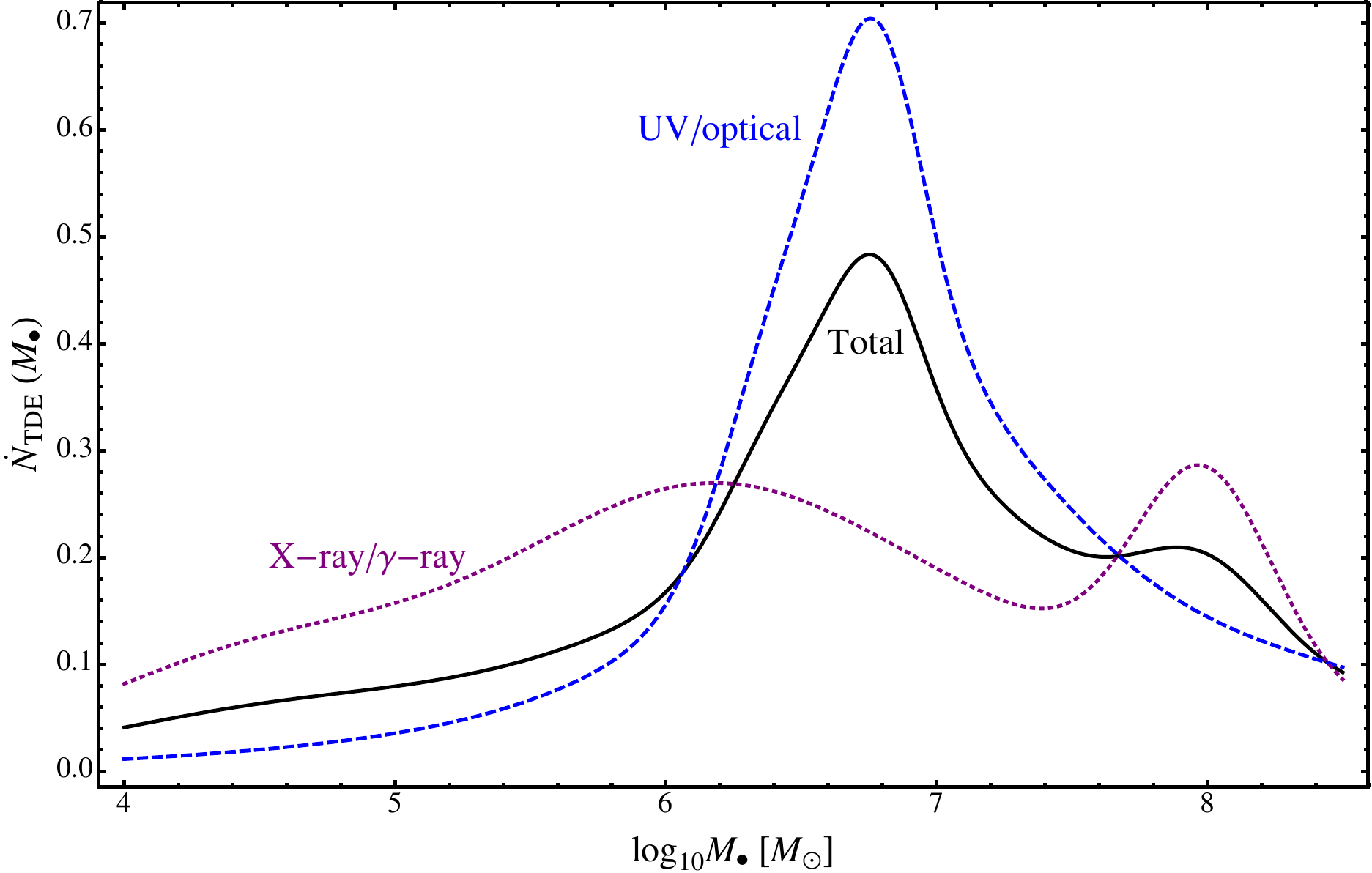}
\caption{The SMBH mass $M_\bullet$ distribution of observed TDE flares, including 11 optical/UV-selected events ({\it blue dashed}), 8 X-ray selected events ({\it purple dotted}), or the full sample ({\it black solid}).  Estimates of the SMBH mass $M_\bullet^{\rm est}$ for each flare are determined using galaxy scaling relations.  $M_\bullet - \sigma$ is used when $\sigma$ is available, but other correlations between bulge luminosity and SMBH mass are used otherwise.  Measurement errors are combined in quadrature with the intrinsic scatter of the galaxy scaling relations to produce error bars for each $M_\bullet^{\rm est}$; each individual TDE is modeled as a Gaussian probability density function $P(\log_{10}M_\bullet)=\exp(-(\log_{10}M_\bullet - \log_{10}M_\bullet^{\rm est})^2/(2s^2))/(s \sqrt{2 \pi})$, where the standard deviation is approximated as $s = \log_{10}M_\bullet^{\rm up} - \log_{10}M_\bullet^{\rm low}$.  The X-ray/$\gamma$-ray sample consists of NGC5905 \citep{Bade+96}, RXJ1420 \citep{Greine+00}, SDSS J1323 \citep{Esquej+07}, SDSS J1311 \citep{Maksym+10}, Swift J1644 \citep{Bloom+11}, SDSS J1201 \citep{Saxton+12}, WINGS J1348 \citep{Maksym+13}, and GRB060218 \citep{Shcher+13}.  The optical/UV sample consists of D1-9, D3-13 \citep{Gezari+06, Gezari+08}, D23H-1 \citep{Gezari+09}, VV-1, VV-2 \citep{vanVelzen+11}, PS1-10jh \citep{Gezari+12}, PS1-11af \citep{Chorno+14}, ASASSN-14ae \citep{Holoie+14}, and PTF09ge, PTF09axc, PTF09djl \citep{Arcavi+14}. }
\label{fig:NDotObserved}
\end{figure*}

Figure \ref{fig:NDotObserved} shows the $M_{\bullet}$ distribution of the current TDE sample, calculated using 8 X-ray and $\gamma$-ray selected tidal disruption flares, and 11 flares found in the optical or UV (see caption).  This sample is composed of all strong TDE candidates with available SMBH masses based on observed properties of the host galaxy (e.g. $M_{\rm bulge}, \sigma$) in combination with known scaling relations.  SMBH mass estimates based on theoretical fits to the observed light curves are not included, given the many uncertainties in the emission process.  The small and inhomogeneous sample used to create Figure \ref{fig:NDotObserved} is likely hampered by selection effects, thus warranting caution in its interpretation.   Nevertheless, given the huge range of predictions shown in Figure \ref{fig:NDotObs2}, even our preliminary version of this plot has significant utility for constraining uncertainties in the SMBH occupation fraction $f_{\rm occ}(M_\bullet)$, and in the nature of TDE optical emission.  

The bias towards moderately massive SMBHs visible in Figure \ref{fig:NDotObserved} is incompatible with super-Eddington outflows or blastwaves from decelerating jets, even in our case A scenario where the SMBH occupation fraction cuts off at very high values $M_\bullet \approx 10^{6.5}M_\odot$.  This suggests that Fig. \ref{fig:NDotObserved} can be interpreted as evidence that an Eddington-limited optical emission mechanism dominates the current TDE flare sample: a highly nontrivial conclusion given the enormously super-Eddington typical values of $\dot{M}_{\rm peak}$ (Fig. \ref{fig:NDotObs}) and recent numerical results on the viability of super-Eddington luminosities \citep{Sadows+14, Jiang+14}.  Alternatively, this could be seen as evidence for the non-fiducial model proposed in the prior subsection, where visible flares are only produced in events with quite relativistic pericenters ($R_{\rm p} \lesssim 12 R_{\rm g}$).

A selection effect that could influence this interpretation is the possible existence of systematic biases against detecting TDE flares in particularly low mass galaxies, for instance if such host galaxies were too dim to detect, or if the angular resolution of the telescope was insufficient to constrain the location of the TDE to the center of the galaxy.  However, the recent discovery of a potential TDE in an intermediate mass galaxy (\citealt{Donato+14}; \citealt{Maksym+13}) shows that TDEs can in fact be associated with low mass hosts in practice.  Future observational efforts will hopefully improve upon our Fig. \ref{fig:NDotObserved} by expanding the observational sample and by combining data from different surveys in a more self-consistent way. 

\subsection{E+A galaxies}

\citet{Arcavi+14} point out that all three PTF candidate TDEs in their sample occur in E+A galaxies (\citealt{Dressler&Gunn83}), which are known to be post-starburst galaxies produced by the relatively recent ($\lesssim 1$ Gyr) major merger of two galaxies \citep{Yang+08, Snyder+11}.  Because E+A galaxies represent only a fraction $\sim 10^{-3}$ of those in the low redshift Universe \citep{Goto07, Snyder+11}, if the TDE rate in E+A galaxies was the same as that in normal galaxies, then the odds of
seeing $\approx3/19$ of all TDEs in E+As would be only $\sim 10^{-2}$.  

This apparent coincidence led \citet{Arcavi+14} to suggest that the apparent rate enhancement was due to a recent SMBH merger, which can increase the TDE rate by producing a SMBH binary.  The SMBHs are brought together by dynamical friction, which can happen in less than a Hubble time so long as their mass ratio $q \lesssim 10$, as is favored for E+A progenitor mergers.  After the binary forms, it will quickly harden through three-body interactions that eject nearby stars.  Eventually, depletion of the stellar population stalls the binary hardening at $\sim {\rm pc}$ scales, giving rise to the well-known ``final parsec problem,'' but before this there is a phase where TDE rates are enhanced by many orders of magnitude, up to $\dot{N}_{\rm TDE} \sim 0.1~{\rm yr}^{-1}$ \citep{Ivanov+05, Chen+09, Chen+11}.   However, these rate enhancements are predicted to be short lived, lasting only $\sim 10^{5}-10^{6}$ years, so that the fraction of all TDEs coming from hardening SMBH binaries is $\sim 3\%$ \citep{WegBod11}.  

If we optimistically conjecture that E+A galaxies are the hosts of all hardening SMBH binaries, then the $\sim 10\%$ of TDEs associated with E+As is not too different from theoretical predictions.  However, this conjecture is difficult to reconcile with the finding of \citet{Chen+11} that the greatest enhancements to TDE rates come from $10\lesssim q \lesssim 100$.  Such mass ratios are disfavored as the origins of E+As, and furthermore will have difficulty forming binaries within 1 Gyr due to their long dynamical friction timescales \citep{Taffoni+03}. 

We propose an alternative hypothesis: that the post-starburst nature of E+As implies these galaxies have anomalously large central stellar densities and are able to produce huge TDE rates ($\dot{N}_{\rm TDE} \sim 10^{-2}~{\rm yr}^{-1}$ would be required to match the observed prevalence of E+As in the TDE sample) through enhanced two-body relaxation.  The younger stellar population may also assist in increasing the E+A TDE rates, but only by a factor $\approx 2$ for ages $\sim 100~{\rm Myr}$ (\S \ref{sec:PDMF}).  Since per-galaxy TDE rates go roughly as $\dot{N}_{\rm TDE} \propto \rho(r_{\rm crit})^2$, the stellar populations in E+A galaxies (within $r\lesssim r_{\rm crit}$) would need to be roughly an order of magnitude denser than those in normal galaxies.  Because $r_{\rm crit} \sim r_{\rm infl}$, this does not put unreasonable mass requirements on star formation during the starburst that preceded the birth of the E+A, but it does require a significant amount of star formation to be concentrated within the critical radius.  Whether this is realized in practice is unclear. 

\section{Conclusions}
\label{sec:conclusions}

We have calculated the rates of stellar tidal disruption events due to two-body relaxation in galactic nuclei, and explored the implications for current and future optical TDE flare samples.  Motivated by the substantial tension between theoretical (high) and observed (low) TDE rates, we have relaxed, updated, or improved upon several assumptions that go into theoretical rate calculations; however, the novel components of our paper have either maintained or heightened the discrepancy between theory and observation.  We stress that our neglect of alternate relaxational mechanisms, our assumption of spherical symmetry, the neglect of nuclear star clusters in \citet{Lauer+07a}, and our remnant-free stellar mass function have set a conservative floor on the true TDE rate in our sample of galaxies.  Of all our assumptions, the only one which {\it may} cause an overestimate of TDE rates in individual galaxies is that of velocity isotropy; strongly anisotropic velocities could either increase or decrease the true TDE rate.  Our major results are summarized as follows.
\begin{itemize}
\item The Nuker surface brightness profile $I_{\rm N}(R)$ is a robust choice of parametrization for use in TDE rate calculations.  Adopting the alternate core-Sersic parametrization produces little change in per galaxy TDE rates $\dot{N}_{\rm TDE}$.  Use of the Sersic parametrization will modestly decrease TDE rates, but this surface brightness profile was not designed to fit the innermost regions of galactic nuclei.  
\item Adoption of a realistic stellar PDMF will modestly increase $\dot{N}_{\rm TDE}$ relative to a calculation where all stars are taken to have mass $M_\star = M_\odot$.  Our fiducial choice of the Kroupa IMF increases the total TDE rate by a factor $\approx 1.5$.  Incorporating stellar remnants into the PDMF can produce a significantly greater increase in $\dot{N}_{\rm TDE}$, but this may be prevented by mass segregation of stellar mass black holes in the small cusp galaxies that dominate the volumetric TDE rate.  In systems where mass segregation does not occur, TDE rates will inversely correlate with nuclear metallicity.
\item A significant fraction ($\sim 30\%$) of TDEs come from the ``pinhole'' regime of relaxation, and can access large values of the penetration parameter $\beta$.  In this regime, roughly half of TDEs are partial disruptions.
\item The volumetric rate of tidal disruption events is sensitive to the uncertain occupation fraction of low-mass SMBHs, and a volume-complete sample will share that sensitivity.  However, if optical emission from TDEs is Eddington-limited, then flux-limited TDE samples will be insensitive to $f_{\rm occ}(M_\bullet)$.  Flux-limited samples of TDEs found through super-Eddington emission mechanisms will be extremely sensitive to $f_{\rm occ}(M_\bullet)$.  Sensitivity to $f_{\rm occ}$ will also be reduced if, speculatively, luminous flares require rapid circularization, and rapid circularization requires relativistic pericenters ($R_{\rm p} \lesssim 12R_{\rm g}$).
\item The current sample of observed TDEs is small and inhomogeneous, but nonetheless suggests that super-Eddington mechanisms {\it do not} dominate the optical emission of most TDEs.  Of the Eddington-limited emission mechanisms we consider here, the direct emission from viscously spreading TDE disks is too faint to produce observed tidal disruption flares.  Along with other lines of evidence, this suggests that some sort of reprocessing layer downgrades hard bolometric emission from the inner disk to softer wavelengths, but the nature of this reprocessing layer is as of yet unclear.
\item An even stronger rate tension between optically-selected TDE samples and our predictions for the size of these samples suggests that reprocessing layers may exist for only a small fraction of TDEs.  The only remaining theoretical avenues for reducing the discrepancy between theoretical and observed TDE rates appear to be this hypothesis (strong bimodality in optical emission mechanisms), or alternatively strong and predominantly tangential velocity anisotropies in galactic nuclei.
\end{itemize}

More technical results of interest can be found in our appendices; in particular, we have derived for the first time the light curves and peak luminosities of viscously spreading TDE disks, have corrected past models of super-Eddington outflows in light of new models for $\Delta \epsilon$, and have also found new closed-form analytic expressions for $\bar{\mu}(\epsilon)$ and $\mathcal{F}(\epsilon)$ in regions close to a SMBH.

In closing, we note that the reliability of our results is limited by two extrapolations we are required to make.  The first concerns the critical radius $r_{\rm crit}$ from which most TDEs are sourced: while this is marginally resolved or better for galaxies with $M_\bullet \gtrsim 10^7 M_\odot$ in the \citet{Lauer+07a} sample, it is unresolved in smaller galaxies which have an outsize effect on the TDE rate.  If small galaxies preferentially turn over to shallow density profiles at small radii $\sim r_{\rm infl}$ (in a way that large galaxies do not), our results may overestimate the true TDE rates.  We reiterate, however, that the opposite is more likely true: the Nuker fits we employ explicitly ignore excess inner light characteristic of nuclear star clusters, which are common in small galaxies.  Our second extrapolation is that our power law TDE rate fit, Eq. \eqref{eq:bestfit}, extends from the smallest galaxies in our sample ($M_\bullet \sim 10^6 M_\odot$) down to even lower masses, where the galaxy scaling relations we employ (e.g. $M_\bullet -\sigma$) are untested\footnote{We note here that in calibrating Eq. \eqref{eq:bestfit}, we explicitly excluded two very small galaxies due to concerns about our ability to estimate their SMBH masses.}.  Our two most conservative choices of occupation fraction $f_{\rm occ}$ do not significantly extrapolate in this way and should be regarded as more reliable than our three more liberal $f_{\rm occ}$ scenarios.

Overall, TDEs offer a unique and promising probe of the bottom end of the SMBH mass function.  At present we are limited by both the small size of today's TDE sample and our limited understanding of optical emission mechanisms in these events.  Amelioration of these problems offers important avenues for future observational and theoretical work, respectively, and improvements in both will allow us to fully realize the scientific potential of these dramatic events.

\section*{Acknowledgments}
We thank Chris Belczynski for providing tabulated data to map between ZAMS stellar masses and final compact remnant masses, and Tod Lauer for assistance in interpreting the Nuker data sets.  We thank Iair Arcavi and Suvi Gezari for providing useful data on the peak luminosities of observed TDEs.  We thank Brad Cenko, Jacqueline van Gorkum, Morgan MacLeod, Jeremiah Ostriker, Greg Snyder, Linda Strubbe, and Sjoert van Velzen for helpful conversations.  Finally, we also thank the anonymous referee for many useful suggestions.  BDM gratefully acknowledges support from the NSF grant AST-1410950 and the Alfred P. Sloan Foundation.  This work was supported in part by the National Science Foundation under Grant No. PHYS-1066293 and the hospitality of the Aspen Center for Physics.

\bibliographystyle{mn2e}
\bibliography{notes}

\appendix
\section{Analytic Limits}
\label{sec:analytic}

Here we derive closed-form solutions for quantities of interest in TDE rate calculations, focusing on radii sufficiently close to the SMBH that the black hole potential dominates that of the stars, and where the stellar population has a power-law spatial density $\rho_\star(r) =\rho_0 (r/r_0)^{-g}$ (e.g. Eq.~\ref{eq:nuker}).  This limit is denoted by the subscript ``near.''  Although these closed-form limits are not in general useful for calculating the energy-integrated TDE rate, $\dot{N}_{\rm TDE}$, because of a subtle feature of $q(\epsilon)$ we discuss below, they are useful for obtaining physical intution and for verifying the results of exact numerical computations.  For simplicity, we assume that all stars possess mass $M_{\star}$.

Near the SMBH, the geometric size of the loss cone is
\begin{equation}
R_{\rm LC, near}(\epsilon) = \frac{J_{\rm LC}^{2}}{J^{2}_{\rm c}(\epsilon)} = \frac{4r_{\rm t}\epsilon}{GM_\bullet},
\end{equation}
where $J_{\rm LC}^{2} = 2GM_{\bullet}r_{\rm t}$ and $J^{2}_{\rm c} = (GM_{\bullet})^{2}/(2\epsilon)$ is the angular momentum of a circular orbit with energy $\epsilon$.  The apocenter distance of a radial orbit is 
\begin{equation}
r_{\rm apo, near} = \frac{GM_\bullet}{\epsilon}.
\end{equation}

The stellar distribution function can be calculated  from Eddington's formula (Eq.~\ref{eq:Eddington}) assuming that $\psi(r) = GM_\bullet /r$, which yields
\begin{equation}
f_{\rm near}(\epsilon) = \frac{g(g-1/2)}{8^{1/2}\pi^{3/2}}\frac{\rho_0}{M_\star} \left( \frac{GM_\bullet}{r_0} \right)^{-g} \frac{\Gamma (g)}{\Gamma(g + 1/2)} \epsilon^{g - 3/2},
\end{equation}
The dimensionless change in the squared angular momentum per orbit $q(\epsilon)$, i.e. Eq.~\eqref{eq:q}, can now be estimated as:
\begin{align}
q_{\rm near}(\epsilon) =& \frac{16\pi^{1/2}G^2 M_\star \rho_0\ln{ \Lambda}}{3J_{\rm LC}^2} \left( \frac{GM_\bullet}{r_0}\right)^{-g}\\
&\times \frac{\Gamma(g)g(g-1/2)}{\Gamma(g+1/2)} (3Q_{1/2}-Q_{3/2}+2Q_0), \notag
\end{align}
where the integrals
\begin{align}
Q_0(r, \epsilon) = &\int_0^{r_{\rm apo}} \frac{r^2}{\sqrt{\psi(r)-\epsilon}} \int_0^\epsilon \tilde{\epsilon}^{g-3/2} {\rm d} \tilde{\epsilon}{\rm d}r \\
Q_{1/2}(r, \epsilon) = &\int_0^{r_{\rm apo}} \frac{r^2}{2^{1/2}(\psi(r)-\epsilon)}\\
&\times \int_\epsilon^{\psi(r)} (2\psi(r)-2\tilde{\epsilon})^{1/2}\tilde{\epsilon}^{g-3/2} {\rm d} \tilde{\epsilon}{\rm d}r \notag \\
Q_{3/2}(r, \epsilon) = &\int_0^{r_{\rm apo}} \frac{r^2}{2^{3/2}(\psi(r)-\epsilon)^2}\\
&\times \int_\epsilon^{\psi(r)} (2\psi(r)-2\tilde{\epsilon})^{3/2}\tilde{\epsilon}^{g-3/2} {\rm d} \tilde{\epsilon}{\rm d}r \notag
\end{align}
represent different contributions to the angular momentum diffusion coefficient for radial orbits.  

In the vicinity of the SMBH, evaluation of the first integral is elementary; however, evaluation of $Q_{1/2}$ and $Q_{3/2}$ is significantly more involved.  Evaluating the inner integrals (over $\int{\rm d}\tilde{\epsilon}$) yields incomplete Beta functions in the variable $\epsilon/\psi = r/r_{\rm apo}$.  Since these cannot be integrated analytically (over ${\rm d}r$), we Taylor expand the Beta functions to first order in the limit of $\epsilon/\psi=1$.  In physical terms, this is equivalent to assuming that the majority of angular momentum diffusion on a radial orbit occurs at apocenter; numerically, this is verified as accurate for radii $r<r_{\rm crit}$.  However, this approximation breaks down severely for $r \approx r_{\rm crit}$, underestimating $q(\epsilon)$ by several orders of magnitude there.  This indicates that at the critical radius denoting the transition between full and empty loss cones, angular momentum relaxation is not dominated by encounters at apocenter but instead by encounters at much smaller radii.  

Nonetheless, by assuming $r < r_{\rm crit}$, we can continue with our closed form derivation by evaluating the diffusion integrals $Q_i$ as follows:
\begin{align}
Q_0 \approx & \frac{5\pi }{8(2g - 1)}G^3M_\bullet^3\epsilon^{g - 4} \\
Q_{1/2} \approx & \pi^{1/2} \left( \frac{1811-798 g+16g^2}{120}\frac{\Gamma(4-g)}{\Gamma(15/2-g)}\right. \\ &+\left. \frac{-1+2g}{4(g-5)(g-4)} \frac{\Gamma(1/2+g)}{\Gamma(1+g)} \right)G^3M_\bullet^3\epsilon^{g -4} \notag \\
Q_{3/2} \approx & \frac{\pi}{40 \Gamma(g-3)} \left( \frac{\pi^{1/2}(-325+118g + 8g^2)\csc (\pi g)}{\Gamma(15/2-g)}\right. - \\ & \left. \times15\frac{2^{5-2g}(1-2g)^2(2g-7)(2g -5)(2g -3)\Gamma(2g - 8)}{\Gamma(2+g)} \right) \notag \\ &\times G^3M_\bullet^3 \epsilon^{g -4}. \notag
\end{align}
We can clearly see that near the SMBH, $q(\epsilon) \propto \epsilon^{g-4}$.  Finally, we can compute the flux (Eq.~\ref{eq:flux}) into the loss cone for $r<r_{\rm crit}$ and $r<r_{\rm infl}$.  Specifically,
\begin{align}
&\mathcal{F}_{\rm near}(\epsilon){\rm d}\epsilon \approx \frac{32\pi}{3\sqrt{2}} G^5M_\bullet^3\rho_0^2 \ln\Lambda \left(\frac{GM_\bullet}{r_0} \right)^{-2g} \\ &\times \left(\frac{g(g-1/2)\Gamma(g)}{\Gamma(g+1/2)}\right)^2 \frac{\epsilon^{2g -11/2}(3\tilde{Q}_{1/2}-\tilde{Q}_{3/2}+2\tilde{Q}_0) }{\ln(GM_\bullet/(4r_{\rm t}\epsilon))}{\rm d}\epsilon. \notag
\end{align}
Here $\tilde{Q}_{\rm x}=Q_{\rm x}/(G^3M_\bullet^3\epsilon^{g-4})$.  We have neglected the logarithmic contribution of $\exp(q(\epsilon))$ because our Taylor expansion is only valid in the limit $q(\epsilon) < 1$.

\section{Optical Emission Models}
\label{sec:optical}

\subsection{Thermal Disk Emission}
\label{sec:thermal}

The most physically secure source of optical luminosity in TDEs is the accretion disk formed around the black hole following circularization of bound stellar debris.  The disk emission is often modeled as a multicolor blackbody, in which case the distinction between theoretical models boils down to differences in the disk structure and evolution.  

At early times when the accretion rate is super-Eddington, the disk is radiatively inefficient and is often modeled as a slim disk (\citealt{Abramowicz+88}).  Many initial works (e.g., \citealt{StrQua09}; \citealt{LodRos11}) assume that the disk terminates at the outer radius corresponding to the circularization radius $R_{\rm c}= 2r_{t}/\beta$ set by angular momentum conservation.  For $\dot{M} \gg \dot{M}_{\rm edd}$ the disk radiates locally at the Eddington luminosity $L_{\rm edd}$, in which case the outer temperature is given by,
\begin{eqnarray}
T_{\rm d}   \simeq  \left(\frac{L_{\rm edd}}{2\pi\sigma_{\rm SB} R_{\rm c}^{2}}\right)^{1/4} \approx 2.2\times 10^{5}M_{6}^{1/12}m_{\star}^{-7/15}\beta^{1/2}\,\,{\rm K}
\label{eq:TSQ}
\end{eqnarray}
corresponds to a spectral peak in the far-UV, such that the peak {\it g}-band optical luminosity ($\nu = 5\times 10^{14}$ Hz),
\begin{equation}
 \nu L_{\nu}^{\rm th} = \frac{8\pi^{2} R_{\rm c}^{2}kT_{\rm d}\nu^{3}}{c^{2}} \approx 6\times 10^{40}m_{\star}M_{6}^{3/4}\beta^{-5/6}\,{\rm erg\,s^{-1}}.
\label{eq:LSQ}
\end{equation}
resides on the Rayleigh-Jeans tail.  \citet{LodRos11} emphasized that at late times ($t \gg t_{\rm fall}$, once $\dot{M} \lesssim \dot{M}_{\rm edd}$) the optical light curve $L_{\rm opt} \propto \dot{M}^{1/4} \propto t^{-5/12}$ declines more gradually than the mass fall-back rate $\propto t^{-5/3}$.

Models assuming a fixed outer disk radius are incompatible with recent well-sampled optical TDE light curves (e.g., \citealt{Gezari+12, Chorno+14, Arcavi+14}), which instead possess much higher peak luminosities $L_{\rm peak} \approx 10^{43-44}$ erg s$^{-1}$ and are characterized by much lower blackbody temperatures $\sim 2-3\times 10^{4}$ K than predicted above.  One potential reason for the discrepency is the outwards viscous spreading of the disk (e.g.~\citealt{Cannizzo+90}; \citealt{SheMat14}).  At early times the disk is advective and domianted by radiation pressure.  During this phase the viscous time at the circularization radius is much shorter than the characteristic fallback time $t_{\rm fall}$ (Eq.~\ref{eq:tfb}), resulting in the disk spreading rapidly outwards from the initial circularization radius.  The thick disk is truncated at the radius $R_{\rm edd}$ where the accretion rate $\dot{M}$ decreases below the local Eddington limit $\dot{M}_{\rm edd}^{\rm loc} \simeq \dot{M}_{\rm edd}(R_{\rm edd}/R_{\rm iso}$), where $R_{\rm iso}$ is the innermost stable orbit.   At $R = R_{\rm edd}(t)$ the disk collapses\footnote{The intermediate regime of a radiation pressure dominated, radiatively efficient disk is thought to be thermally and viscously unstable (\citealt{Lightman&Eardley74}).} to a radiatively efficient (geometrically thin), gas pressure-dominated state (\citealt{SheMat14}); this causes a sudden and large increase in the viscous time, reducing the disk temperature and luminosity to a much lower value than in the thick disk state and effectively shutting off the emission.\footnote{The disk will continue to viscously spread on the longer viscous timescale associated with this gas pressure-dominated state, producing much dimmer emission lasting for years or longer (e.g.~\citealt{Cannizzo+90}).} 

A simplified model for optical emission from the spreading disk is developed as follows.  The surface density $\Sigma$ evolves according to the diffusion equation for angular momentum transport by viscosity with a delta function mass source at the circularization radius $R_{\rm c}$.  This results in the following solution (\citealt{SheMat14}; see Appendix B of \citet{Metzger+12b} for derivation)
\begin{align}
\Sigma(R,t)= &\frac{2}{2-n}\frac{R_{\rm c}^{5/4}R^{-n-1/4}}{3\nu_{\rm c}}\label{eq:Sig_sing} \\
&\times \int\limits_0^T S_t(T^\prime)G(w(R),w(R_{\rm c}),T-T^\prime)dT^\prime, \notag
\end{align}
where 
\begin{eqnarray}
S_g(R,t)=\frac{\dot M(R=R_{\rm c},t)}{2\pi R_{\rm c}}\delta(R-R_{\rm c}),
\label{eq:mdot_sing}
\end{eqnarray}
is the mass source function from fallback accretion and 
\begin{eqnarray}
G(w,w^\prime,z)\equiv\frac{1}{2z}\exp\left(-\frac{w^2+w^{\prime 2}}{4z}
\right)I_\ell \left(\frac{w w^\prime}{2z}\right)
\label{eq:Green}
\end{eqnarray}
is the Green's function, with $I_\ell$ a Modified Bessel Function of order $\ell = 1/2(2 - n)$.  Here $T = 3\nu_{\rm c} t$ and $w(R) = \frac{2}{2-n}R^{1-n/2}$ are new time and spatial coordinates, respectively, where $\nu = \nu_{\rm c} R^{n}$ is the viscosity, $n = 1/2$ and $\nu_{\rm c} = \alpha (H/R)^{2}(GM_{\bullet}/R_{\rm c})^{1/2}$ for a slim disk, where $\alpha = 0.1$ is the effective kinematic viscosity and $H \sim 0.5R$ for a thick disk.

\begin{figure}
\includegraphics[width=85mm]{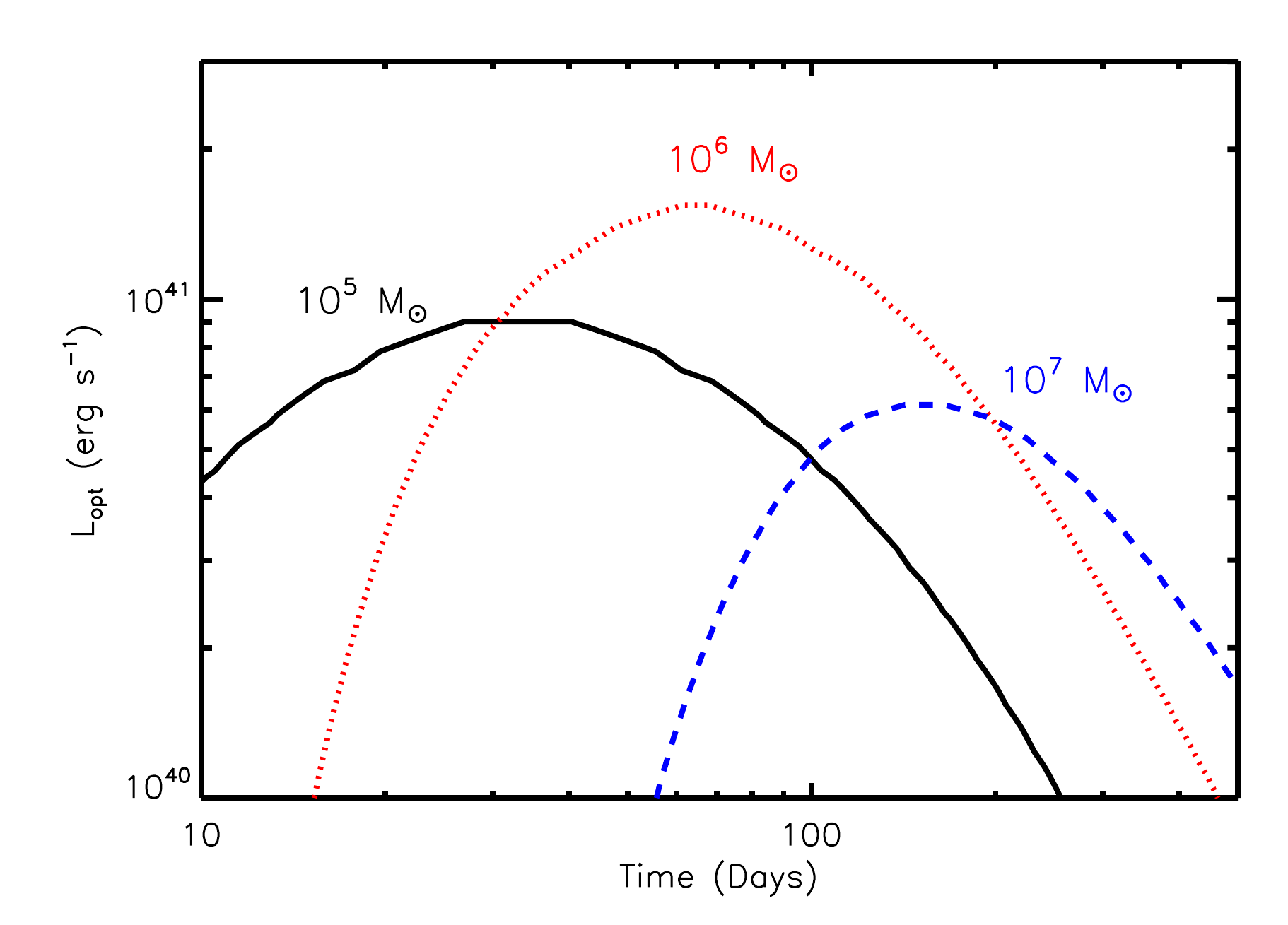}
\caption{Optical light curves of spreading disk, calculated for a $\beta = 1$ disruption of a solar mass star and shown for different SMBH masses, $M_{\bullet} = 10^{5}M_{\odot}$ ({\it black solid}), $10^{6}M_{\odot}$ ({\it red dotted}), and $10^{7}M_{\odot}$ ({\it blue dashed}).}
\label{fig:spreading}
\end{figure}

At each timestep the outer edge of the slim disk $R_{\rm edd}(t)$ is estimated as the location where $\dot{M} = 3\pi \nu \Sigma = \dot{M}_{\rm edd}^{\rm loc}$.  Optical light curves are then calculated under the assumption that most emission originates from the outer radius of the thick disk (which we verify), where the outer disk temperature is calculated from Eq. (\ref{eq:TSQ}) but replacing $R_{\rm c}$ with the larger value of $R_{\rm edd}(t)$.  Figure \ref{fig:spreading} shows light curve calculations for different black hole masses.  The peak optical luminosity of the spreading disk is higher than for a disk with a fixed outer radius, but is still insufficient to explain those observed from recent TDEs.  

An additional model for the super-Eddington evolution of TDE disks, which accounts for the (sometimes highly) sub-Keplerian nature of low angular momentum gas flows was recently introduced by \citet{Coughlin&Begelman14} (see also \citealt{Loeb&Ulmer97}).  Although its physical motivations differ, this model predicts an extend quasi-spherical atmosphere that reprocesses the accretion power, similar to the predictions of the scenario described next.  

\subsection{Reprocessing Layer}
\label{sec:reprocessing}
The low temperatures, high optical luminosities, and slow time evolution of many recently observed TDEs have led to the suggestion that the high-energy disk emission is generally reprocessed by a loosely bound layer of debris that orbits at large radii and obscures the SMBH (\citealt{Guillochon+14}).  Although more detailed models for this emission are considered in \citet{Guillochon+14}, this model contains many free parameters that could change from event to event and may be unnecessarily complicated if our goal is to capture the fundamental properties that characterize many of the current optical TDE sample, namely $\sim$ constant temperature emission with a light curve that appears to track the expected fall-back accretion rate (in particular declining $\propto t^{-\alpha}$ after the peak, with $\alpha \gtrsim 5/3$).  

For sake of simplicity, we thus introduce an ad hoc, phenomenological model in which the optical luminosity is a fixed fraction of the expected disk luminosity, $L_{\rm opt} = \epsilon_{\rm opt}L_{\rm bol}$, where
\begin{equation}
L_{\rm bol} = \begin{cases} \label{eq:reprocessing}
L_{\rm edd}
& \dot{M} > \dot{M}_{\rm edd} \\
0.1\dot{M}c^{2},& \dot{M} \le \dot{M}_{\rm edd}
\end{cases}
\end{equation}
and $\epsilon_{\rm opt} < 1$ is the reprocessing efficiency.  We take $\epsilon_{\rm opt} = 0.03$ as a fiducial value, motivated by the desire to reproduce the observed peak optical luminosities of the PTF TDE sample of \citet{Arcavi+14}, for which many of the events possess black body luminosities of $L_{\rm opt} \sim $ few 10$^{43}$ erg s$^{-1}$ in galaxies with inferred SMBH masses $M_{\bullet} \sim $ few 10$^{6} M_{\odot}$ (for which $L_{\rm edd} \sim $ few $10^{44}$ erg s$^{-1}$).  Though extremely crude, this model captures the essence of the reprocessing layer until its physical nature is better understood.

This model succeeds much better than the others at reproducing observed peak luminosities in TDE candidates (Fig. \ref{fig:LPeak}), but these luminosities are so high as to predict enormous detection rates, strongly at odds with the existing TDE sample.  

\subsection{Super-Eddington Outflows} 
\label{sec:SE}

During the early phase of highly super-Eddington fall-back, the accretion disk is likely to be advective and dominated by radiation pressure.  Such disks are believed to drive mildly relativistic outflows, a process that has been modeled analytically \citep{BlaBeg99, StrQua09} and numerically \citep{Ohsuga+05, Sadows+14}.  The adiabatically expanding outflow grows in size, but its luminosity is limited by the eventually receding radius of its photosphere.  We adopt here the simple model of \citet{LodRos11} to describe optical emission from these outflows, which we have however modified to incorporate the correct parametrization for the energy spread of the debris $\Delta \epsilon$ determined by recent work (\citealt{GuiRam13}, \citealt{Stone+13}).

The bulk of the super-Eddington outflow (assumed to be spherical) is launched from the circularization radius $R_{\rm c} = 2r_{t}/\beta$ with a velocity $v_{\rm w}= \sqrt{2GM_\bullet/R_{\rm L}}$ taken to equal the local escape velocity.  The temperature $T_{\rm L}$ at the base of the wind can be found by energy conservation, 
\begin{equation}
4\pi R_{\rm L}^2v_{\rm W} \left( \frac{4}{3}a_{\rm R}T_{\rm L}^4 \right ) = \frac{1}{2} \dot{M}f_{\rm out}v_{\rm w}^2,
\label{eq:windenergy}
\end{equation}
where $f_{\rm out} \sim 0.1$ is the fraction of the inflowing mass $\dot{M}$ redirected into the outflowing wind.

The temperature at the radius of the photosphere $R_{\rm ph}$ is related to $T_{\rm L}$ by the assumption of adiabatic expansion,
\begin{equation}
T_{\rm ph}=T_{\rm L} \left( \frac{R_{\rm ph}}{R_{\rm L}} \right)^{-2/3} \left( \frac{f_{\rm out}}{f_{\rm v}} \right)^{1/3}.
\end{equation}
Initially, the high optical depth of the outflow places the photosphere approximately at the outer edge, $R_{\rm edge}=tv_{\rm w}$.  However, after a time
\begin{equation}
t_{\rm edge}=5.3~{\rm days}~M_6^{-1/8}\beta^{-3/8}r_\star^{3/4}f_{\rm v}^{-3/4}f_{\rm out, -1}^{3/8},
\end{equation}
the outer edge becomes optically thin, and the photosphere moves inward to a new radius
\begin{equation}
R_{\rm ph}=1.1\times 10^{14}~{\rm cm}~ m_\star^{1/6}r_\star^{3/2}\beta^{-1/2}f_{\rm out, -1}f_{\rm v}^{-1} t_{\rm mon}^{-5/3},
\end{equation}
where $t_{\rm mon}=t/(30~{\rm days})$ and $f_{\rm out, -1}=f_{\rm out}/0.1$.  The temperature at the wind launching radius from equation (\ref{eq:windenergy}) is 
\begin{equation}
T_{\rm L}=2.3\times10^5~{\rm K}~f_{\rm out, -1}^{1/4}f_{\rm v}^{1/4} t_{\rm mon}^{-5/12}\beta^{5/8}m_\star^{7/24}r_\star^{-3/8},
\end{equation}
such that
\begin{align}
T_{\rm ph} = \begin{cases}
1.2 \times 10^3~{\rm K}~&f_{\rm out, -1}^{7/12}f_{\rm v}^{-3/4}\beta^{-3/8}m_\star^{-1/24}\\
&\times r_\star^{5/8}t_{\rm mon}^{-13/12}, t<t_{\rm edge} \\ 
2.7\times 10^4~{\rm K}~&f_{\rm out, -1}^{-1/12}f_{\rm v}^{7/12}\beta^{7/24}M_6^{2/9}m_\star^{-1/24}\\
&\times r_\star^{-17/24}t_{\rm mon}^{25/36}, t>t_{\rm edge}.
\end{cases}
\end{align}
Although past work \citep{StrQua09} considered both of these regimes, more recent theoretical revisions to $\Delta \epsilon$ \citep{Stone+13, GuiRam13} imply that almost all TDEs will have $t_{\rm edge} \ll t_{\rm fall}$, making only the latter regime relevant.  We therefore focus on it for the remainder of this section.

The bolometric luminosity of the outflow,
\begin{align}
L^{\rm SE} = 4\pi R_{\rm ph}^{2}\sigma T_{\rm ph}^{4} \approx &4.5 \times 10^{42}~{\rm erg~s}^{-1}~f_{\rm out, -1}^{5/3}f_{\rm v}^{1/3} \\
&\times \beta^{1/6}M_6^{8/9}m_\star^{1/6}r_\star^{1/6}t_{\rm mon}^{-5/9}, \notag
\end{align}
declines as $\propto t^{-5/9}$, but the optical luminosity on the Rayleigh-Jeans tail, 
\begin{align}
\nu L_{\nu}^{\rm SE} = &5\times 10^{41}~{\rm erg~s}^{-1}~f_{\rm out, -1}^{23/12}f_{\rm v}^{17/12}\beta^{17/24} \\
&\times M_6^{2/9}m_\star^{7/24}r_\star^{55/24}t_{\rm mon}^{-95/36} \nu_5^3, \notag
\end{align}
declines significantly faster ($\propto t^{-95/36}$), where $\nu_{5}=\nu/(5\times10^{14}~{\rm Hz})$.  Assuming that the luminosity peaks on a timescale $\sim t_{\rm fall}$ (Eq.~\ref{eq:tfb}), the peak luminosity is given by
\begin{align}
\nu P_{\nu}^{\rm SE} = &2.2 \times 10^{41}~{\rm erg~s^{-1}}f_{\rm out, -1}^{23/12}f_{\rm v}^{-17/12}\beta^{-17/24} \\
&\times M_6^{-79/72} m_\star^{211/72}r_\star^{-5/3}\nu_{500}^3. \notag
\end{align}
Emission decreases rapidly once $R_{\rm ph}=R_{\rm L}$, i.e. once the entire outflow becomes optically thin.  This occurs at a time
\begin{equation}
t_{\rm thin} = 9.0\times 10^6~{\rm s}~ f_{\rm out, -1}^{3/5}f_{\rm v}^{-3/5} \beta^{3/10} M_6^{-1/5}m_\star^{3/10}r_\star^{3/10}.
\end{equation}
In our TDE rate calculations ($\S\ref{sec:detection}$) we assume $f_{\rm v} = f_{\rm out,-1} = 1$ and $\beta = 1$. 

\subsection{Synchrotron Radiation from Decelerating Off-Axis Jet}
\label{sec:jet}

A final source of optical luminosity in TDEs is synchrotron radiation from a relativistic jet powered by accretion onto the black hole.  The non-thermal X-ray transient {\it Swift J1644+57} was interpreted as a jetted TDE viewed along the jet axis (\citealt{Bloom+11}; \citealt{Levan+11}; \citealt{Burrows+11}; \citealt{Zauderer+11}), with bright radio emission produced by the shock interaction between the jet and the dense circumnuclear medium surrounding the black hole (e.g.~\citealt{Metzger+12}).  The majority of TDEs are not viewed on-axis (probably less than one percent, given the inferred beaming angle for {\it Swift J1644+57}) and hence do not produce bright non-thermal X-rays due to relativistic beaming.  Nevertheless, the afterglow still becomes visible as a radio source once the shocked gas decelerates to mildly relativistic speeds.   

At late times, the blastwave will become mildly relativistic and spherical, asymptoting to a Sedov-Taylor blastwave with total energy $E_{\rm j} = 10^{52}E_{52}$ erg similar to that inferred for {\it Swift J1644+57} (e.g.~\citealt{Berger+12}).  The radius and velocity evolve with time $t = t_{\rm yr}$ yr according to
\begin{align}
r_{s} &= 1.0E_{52}^{1/5}n_{1}^{-1/5}t_{\rm yr}^{2/5} {\rm pc}; \\
\beta &\equiv v_{\rm s}/c = 0.4r_{\rm s}/ct = 0.71 E_{\rm 52}^{1/5}n_{1}^{-1/5}t_{\rm yr}^{-3/5},
\end{align}
where $n = 1 n_{1}$ cm$^{-3}$ is the density of the circumnuclear medium on subparsec scales of relevance to jet deceleration.  We assume that the shock accelerates electrons to an energy spectrum $f(E) \propto E^{-p}$ where $p = 2.5-3$ (as inferred in e.g. radio supernovae or gamma-ray burst afterglows), and that a fraction $\epsilon_{e} = 0.1\epsilon_{\rm e,-1}$ and $\epsilon_{\rm B} = 0.01\epsilon_{\rm B,-2}$ of the shock energy goes into accelerating relativistic electrons and amplifying the magnetic field, respectively.  The former implies a minimum Lorentz factor for the electrons $\gamma_{\rm m} = \frac{p-2}{p-1}\frac{m_p}{m_e}\epsilon_e\beta^{2}$, while the latter implies a post-shock magnetic field strength $B = \sqrt{6\pi \epsilon_B m_p c^{2}\beta^{2}}$.  The characteristic (or peak) synchrotron frequency (e.g.~\citealt{Nakar&Piran11}),
\begin{equation}
\nu_{\rm m} = \frac{eB\gamma_{\rm m}^{2}}{2\pi m_e c} \approx 3\times 10^{7} n_{1}^{-1/2} E_{\rm 52} \epsilon_{\rm B,-2}^{1/2}\epsilon_{\rm e,-1}^{2}t_{\rm yr}^{-3}{\rm Hz},
\end{equation}
is typically at radio frequencies, while the characteristic cooling frequency
\begin{equation}
\nu_{\rm c} = \frac{eB\gamma_{\rm c}^{2}}{2\pi m_e c} \approx 5\times 10^{14}\epsilon_{B,-2}^{-3/2}E_{52}^{-3/5}n_{1}^{3/5}t_{\rm yr}^{-1/5}  {\rm Hz}
\end{equation}
is typically in the IR to UV, depending most sensitively on $\epsilon_{B}$ (where $\gamma_{\rm c} = 2m_e/3\sigma_T \epsilon_B t m_p c \beta^{2}$).  If one assumes that the optical waveband resides between the characteristic and cooling frequencies $\nu_{\rm m} <\nu_{\rm opt} < \nu_{\rm c}$, then the flux evolution is given by (\citealt{Nakar&Piran11})
\begin{eqnarray}
\nu L_{\nu}^{\rm jet} =& \nu_{\rm opt} L_{\nu,\rm m}\left(\frac{\nu_{\rm opt}}{\nu_{\rm m}}\right)^{-(p-1)/2} \nonumber \\
\underset{p = 2.5}\approx & 2.5\times 10^{42}{\rm erg\,s^{-1}}n_{1}^{0.33}\epsilon_{B,-2}^{0.88}\epsilon_{e,-1}^{1.5} \label{eq:Ljet}\\ 
&\times E_{52}^{1.55}t_{\rm yr}^{-1.65}, \nonumber
\end{eqnarray}
where $L_{\nu, \rm m}$ is the flux at $\nu_{\rm m}$ (eq.~6 of \citealt{Nakar&Piran11}).  Equation (\ref{eq:Ljet}) is valid only at times $t > t_{\rm nr}$ after the initially ultra-relativistic jet slows to mildly relativistic velocities.  If the reverse shock is very strong, then the jet is immediately decelerated to mildly relativistic speeds and this emission can occur as early as the reverse shock crossing time $\simeq 2 t_{\rm fall}$ (\citealt{Sari&Piran95}), resulting in a peak luminosity
\begin{align}
\nu P_{\nu}^{\rm jet}(t_{\rm nr}) = &2.5\times 10^{43}{\rm erg\,s^{-1}}n_{1}^{0.33}\epsilon_{B,-2}^{0.88}\epsilon_{e,-1}^{1.5}M_{6}^{-0.8}\notag  \\
&\times E_{52}^{1.55}m_{\star}^{-0.32}(t_{\rm nr}/2t_{\rm fall})^{-1.65}.
\label{eq:Ljetpeak}
\end{align}
However, if the jet is still relativistic after the reverse shock crossing, then the onset of the off-axis emission could be delayed for a time $\sim$ years or longer, resulting in dimmer emission $\propto t_{\rm nr}^{-1.65}$.

In our TDE rate calculations ($\S\ref{sec:detection}$) we assume $\epsilon_{e} = 0.1$, $\epsilon_{B} = 0.01$, $n = 1$ cm$^{-3}$, and $E_{52} = 1(m_{\star}/0.5)$ to match radio modeling of {\it Swift J1644+57} (\citealt{Metzger+12}; \citealt{Berger+12}), assuming it to be the result of the disruption of a solar mass star.  We also take $t_{\rm nr} = 10t_{\rm fall}$, even though we expect that $t_{\rm nr}$ could be significantly shorter if the circumnuclear density is significantly higher.  Finally, we assume that only a fraction $0.3\%$ of TDEs produce relativistic jets, a number which is justified as follows.  That {\it Swift} detected 1 jetted TDE in approximately ten years of observations, out to a redshift $z \approx 0.35$ (comoving volume of $V = 11$ Gpc$^{-3}$), implies a volumetric detection rate of $\sim 10^{-11}$ yr$^{-1}$ Mpc$^{-3}$.  Assuming a beaming fraction of $\approx 100$ (\citealt{Metzger+12}), the event rate is $\sim 10^{-9}$ yr$^{-1}$ Mpc$^{-3}$, which for a galaxy density $n_{\rm gal} \sim 10^{-2}$ Mpc$^{-3}$ corresponds to a per-Galaxy rate $\sim 10^{-7}$ Mpc$^{-3}$, or $\sim 0.1-1\%$ of the average per galaxy TDE rate $\langle \dot{N}_{\rm TDE} \rangle \sim 10^{-5}-10^{-4}$ yr$^{-1}$ (depending on whether observationally inferred or theoretically predicted rates are taken as more trustworthy).     

\subsection{Other Models}

Other possible sources of optical emission that we do not consider include recombination transients from the unbound stellar debris \citep{StrQua09, KasRam10}.  Though relatively dim to begin with, this emission process appears to be even less promising now due to the lower energy spread $\Delta \epsilon$ of the debris found by recent studies (e.g.~\citealt{Stone+13}), and because self-gravity may confine unbound ejecta to a tiny solid angle for $\beta \lesssim 3$ events \citep{Guillochon+14}.  High-$\beta$ disruptions of white dwarfs by intermediate mass black holes (IMBHs) may also lead to a supernova-like transient powered by the radioactive decay of $^{56}$Ni in the ejecta \citep{Rosswog+09}.  Given the uncertain distribution of IMBHs, the rates of such events are difficult to gauge.  

\section{Full Results}
\label{sec:fullResults}

\begin{figure}
\includegraphics[width=85mm]{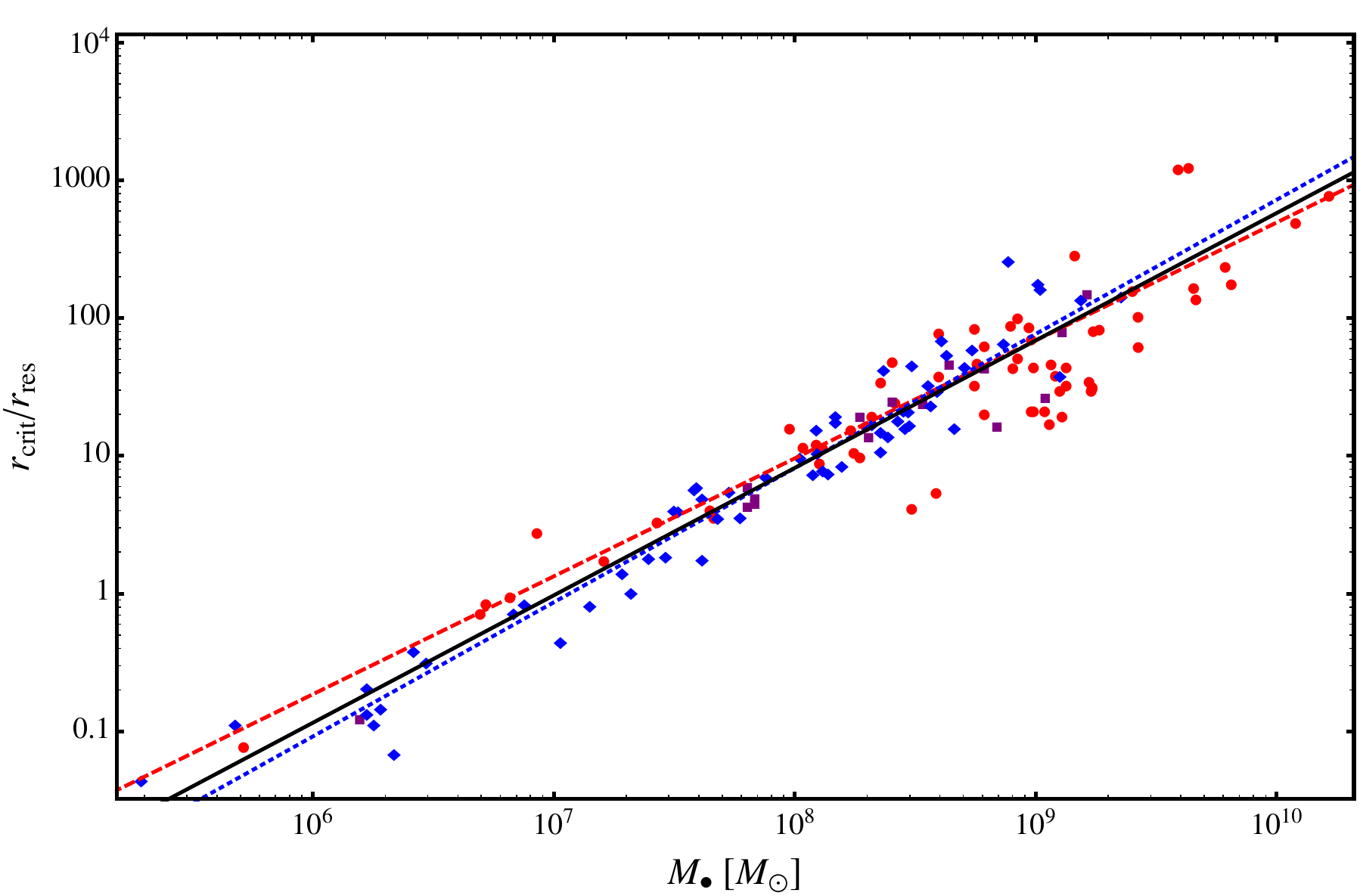}
\includegraphics[width=85mm]{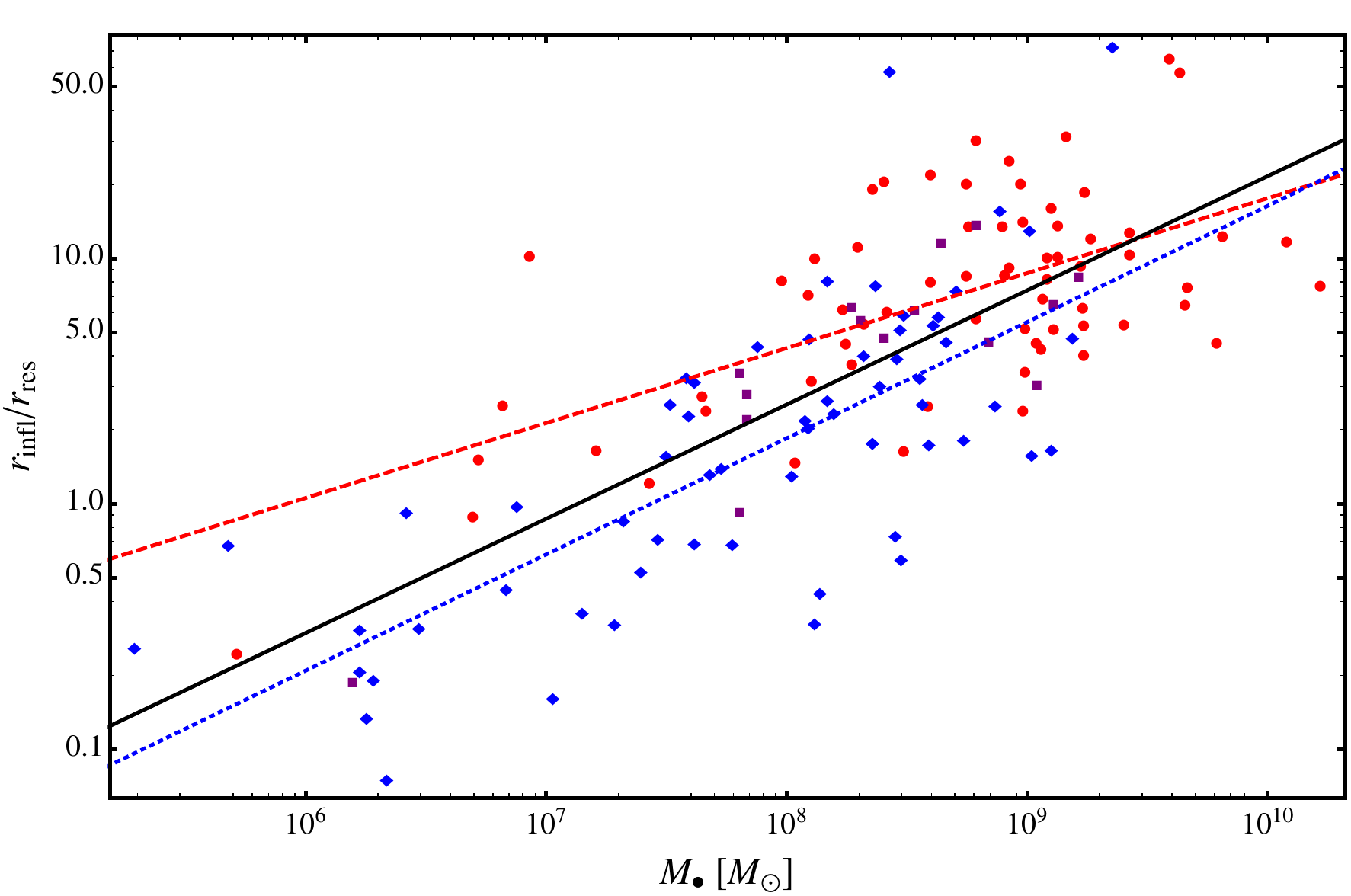}
\includegraphics[width=85mm]{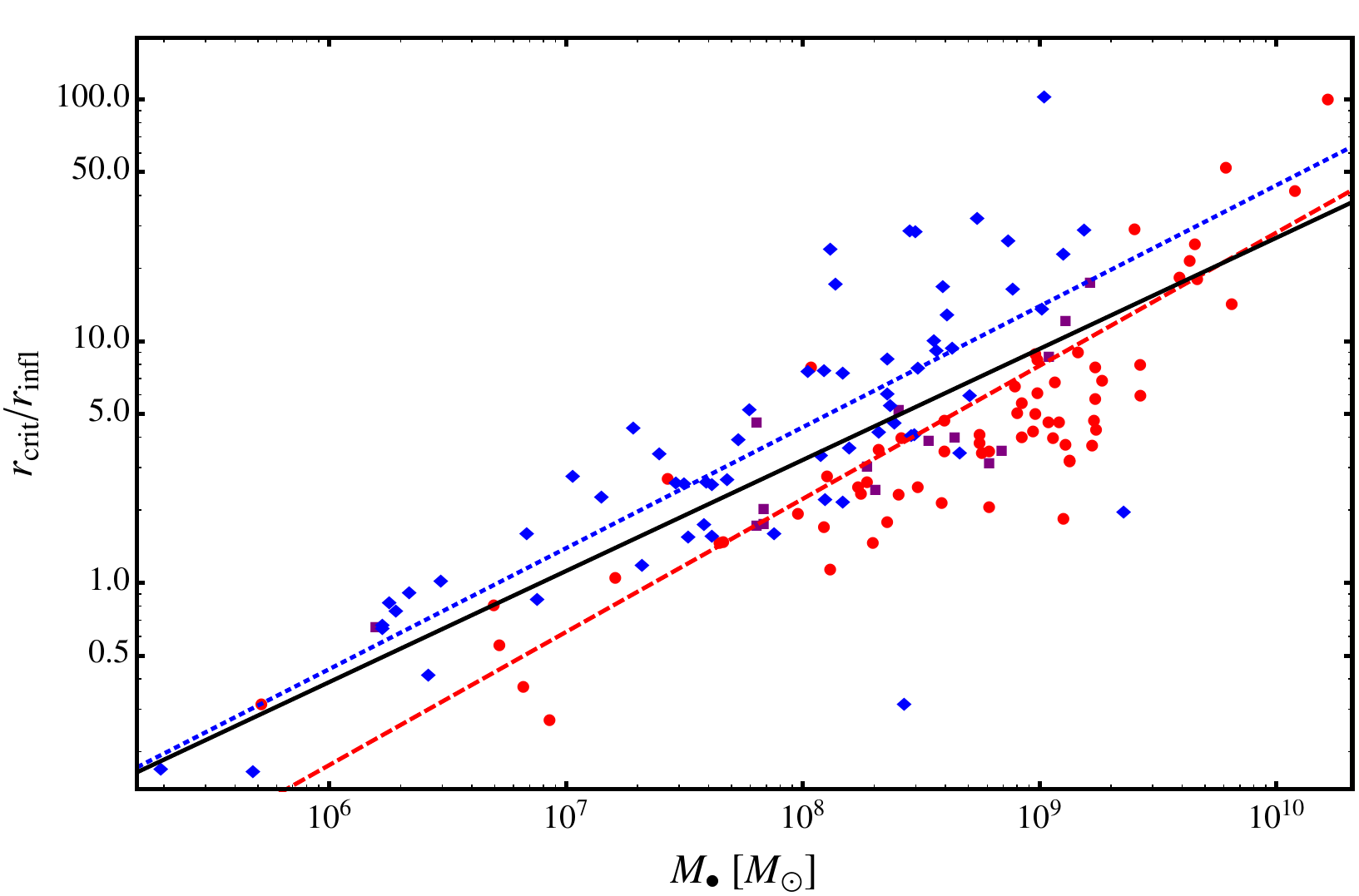}
\caption{The critical radius $r_{\rm crit}$ and the SMBH influence radius $r_{\rm infl}$.  The top and middle panels show $r_{\rm crit}$ and $r_{\rm infl}$ normalized by $r_{\rm res}$, the resolution limit of HST.  We conservatively take the angular resolution limit to be $0.04''$, although for a subset of our galaxy sample it is actually $0.02''$ \citep{Lauer+07a}.  Most galaxies with $M_\bullet > 10^7 M_\odot$, and a few with smaller SMBH masses, possess resolved critical radii.  The influence radius is moderately harder to resolve.  In the bottom panel, we plot the ratio $r_{\rm crit}/r_{\rm infl}$ against $M_\bullet$: this ratio is of order unity for $M_\bullet \sim 10^6 M_\odot$ and increases with increasing SMBH mass.  As in Fig. \ref{fig:SampleNDotMBH}, we represent cusp, core, and intermediate galaxies with blue diamonds, red circles, and purple squares, respectively.  All three panels show best fit power laws to the cusp, core, and full samples as dotted blue, dashed red, and solid black lines, respectively.}
\label{fig:rCritInfl}
\end{figure}

In this appendix, we present the full tabulated results of our rate calculations, as well as other results of more technical interest.  From our initial sample of 217 galaxies, we discard all those with $\gamma < 0$ (i.e. those that cannot be deprojected assuming spherical symmetry), leaving 168 galaxies.  We further discard those whose distribution functions $f(\epsilon)$ are not positive-definite (indicating an incompatibility with the assumption of velocity isotropy), leaving 146 galaxies in the final sample.  Both of these steps preferentially remove very massive galaxies with SMBHs above the Hills limit for main sequence stars.  Finally, there are two galaxies with such low values of $\sigma$ that application of the $M_\bullet - \sigma$ relation implies $M_\bullet \ll 10^5 M_\odot$.  This is so far outside the SMBH mass range over which galaxy scaling relations are calibrated that we eliminate these two outliers, leaving a final count of 144 galaxies in our sample.

In Fig. \ref{fig:rCritInfl}, we plot different normalizations of the critical radius $r_{\rm crit}$ from which most stellar flux into the loss cone originates.  We find that the HST photometry of \citet{Lauer+07a} is able to resolve $r_{\rm crit}$ for a large majority of galaxies in our sample, including almost every galaxy with $M_\bullet > 10^7 M_\odot$.  However, most galaxies with $M_\bullet < 10^7 M_\odot$ possess unresolved critical radii.  Resolved critical radii greatly increase our confidence in estimates of per-galaxy TDE rates (since otherwise we are extrapolating surface brightness profiles into the region that determines the TDE rate), which is why we include even high mass galaxies in the power law fit for $\dot{N}_{\rm TDE}(M_\bullet)$ in Eq. \ref{eq:bestfit}.

Although we find very limited scatter in the $r_{\rm crit}(M_\bullet)$ relation, $r_{\rm infl}(M_\bullet)$ is notably less precise.  This can be seen in the middle and bottom panels of Fig. \ref{fig:rCritInfl}, which also demonstrate that the influence radius is unresolved for most cusp galaxies in our sample, and that $r_{\rm crit} \sim r_{\rm infl}$ for galaxies beneath the Hills limit.  For significantly larger galaxies, however, $r_{\rm crit}$ can often be an order of magnitude or more greater than the influence radius.

Because these results may be of more general interest, we present here our best-fit power laws for $r_{\rm crit}/r_{\rm infl}$ and $r_{\rm infl}$, as functions of $M_\bullet$.  In particular, we write
\begin{align}
\frac{r_{\rm crit}}{r_{\rm infl}} = & A_1 \left( \frac{M_\bullet}{10^8M_\odot} \right)^{B_1} \\
\frac{r_{\rm infl}}{\rm pc}= & A_2 \left( \frac{M_\bullet}{10^8M_\odot} \right)^{B_2}.
\end{align}
For core, cusp, and all galaxies, $\{A_1, B_1, A_2, B_2\} = \{2.2, 0.55, 27, 0.60\}$, $\{A_1, B_1, A_2, B_2\} = \{4.4, 0.50, 11, 0.58\}$, and $\{A_1, B_1, A_2, B_2\} = \{3.22, 0.46, 16, 0.69 \}$, respectively.

\begin{center}
\begin{table*}
 \begin{minipage}{180mm}
   \caption{Our full, tabulated results for the 144 galaxies taken from \citet{Lauer+07a}.  We present critical radii $r_{\rm crit}$, SMBH influence radii $r_{\rm infl}$, TDE rates $\dot{N}_{\rm TDE}$, and fractions $f_{\rm pinhole}$ of TDEs fed to the black hole from the pinhole regime.  We also tabulate the two galactic structural parameters of greatest relevance for our derived results: SMBH mass $M_\bullet$ and Nuker law inner slope $\gamma$.}
   \begin{tabular}{r|r|r|r|r|r|r|r|r|r|r|r|r|r|r|r|r|r|r} \\
     \hline
     Galaxy$^a$ & $\gamma$$^b$ & $\log M_{\bullet}$$^c$ & $\log_{10} (r_{\rm infl}/{\rm pc})$$^d$ & $\log_{10} (r_{\rm crit}/{\rm pc})$$^e$ & $\dot{N}_{\rm TDE}$$^{f}$ & $f_{\rm pinhole}$$^g$  \\
     \hline
NGC1331 & 0.57 & 5.29 & 0.0722 & -0.708 & -4.96 & 0.805 \\
NGC4467 & 0.94 & 5.68 & 0.359 & -0.433 & -4.72 & 0.504 \\
NGC2636 & 0.06 & 5.71 & 0.257 & -0.247 & -5.14 & 0.974 \\
NGC7743 & 0.5 & 6.20 & -0.0995 & -0.283 & -3.71 & 0.704 \\
NGC3599 & 0.75 & 6.22 & -0.0466 & -0.244 & -3.70 & 0.526 \\
NGC4150 & 0.58 & 6.22 & -0.0758 & -0.260 & -3.71 & 0.658 \\
NGC4121 & 0.85 & 6.25 & -0.195 & -0.285 & -3.39 & 0.396 \\
NGC4474 & 0.72 & 6.28 & -0.108 & -0.235 & -3.57 & 0.524 \\
MCG08-27-18 & 0.79 & 6.34 & -0.162 & -0.212 & -3.40 & 0.425 \\
NGC3605 & 0.59 & 6.42 & 0.602 & 0.210 & -4.76 & 0.729 \\
NGC2685 & 0.73 & 6.47 & -0.0751 & -0.0786 & -3.47 & 0.456 \\
NGC4458 & 0.16 & 6.69 & 0.485 & 0.392 & -4.39 & 0.901 \\
NGC4387 & 0.1 & 6.72 & 0.718 & 0.460 & -4.73 & 0.766 \\
VCC1440 & 0.14 & 6.82 & 0.940 & 0.511 & -4.82 & 0.655 \\
NGC4742 & 1.04 & 6.83 & 0.143 & 0.336 & -3.39 & 0.143 \\
NGC4503 & 0.64 & 6.88 & 0.520 & 0.440 & -4.17 & 0.561 \\
VCC1545 & 0.05 & 6.93 & 1.55 & 0.980 & -6.02 & 0.895 \\
NGC3900 & 0.51 & 7.03 & -0.0410 & 0.391 & -3.15 & 0.254 \\
NGC7332 & 0.62 & 7.15 & 0.216 & 0.561 & -3.48 & 0.342 \\
NGC4464 & 0.14 & 7.21 & 0.757 & 0.777 & -4.37 & 0.732 \\
NGC4417 & 0.71 & 7.28 & 0.0366 & 0.667 & -3.09 & 0.182 \\
NGC5370 & 0.62 & 7.32 & 0.862 & 0.926 & -4.44 & 0.520 \\
NGC1351 & 0.78 & 7.39 & 0.321 & 0.845 & -3.46 & 0.193 \\
NGC3377 & 0.03 & 7.43 & 0.437 & 0.867 & -3.67 & 0.327 \\
NGC2699 & 0.84 & 7.46 & 0.585 & 0.988 & -3.83 & 0.206 \\
NGC2549 & 0.67 & 7.50 & 0.597 & 0.997 & -3.86 & 0.304 \\
NGC6340 & 0.59 & 7.52 & 0.905 & 1.09 & -4.36 & 0.517 \\
NGC3384 & 0.71 & 7.58 & 0.858 & 1.09 & -4.14 & 0.326 \\
NGC3056 & 0.9 & 7.59 & 0.745 & 1.15 & -3.95 & 0.165 \\
NGC4494 & 0.52 & 7.62 & 1.02 & 1.21 & -4.50 & 0.505 \\
NGC6278 & 0.55 & 7.62 & 0.716 & 1.11 & -3.99 & 0.367 \\
NGC0596 & 0.16 & 7.65 & 1.07 & 1.23 & -4.56 & 0.612 \\
NGC1426 & 0.26 & 7.66 & 1.09 & 1.26 & -4.58 & 0.615 \\
NGC1439 & 0.74 & 7.68 & 0.819 & 1.24 & -4.09 & 0.261 \\
NGC4564 & 0.8 & 7.73 & 0.674 & 1.26 & -3.82 & 0.167 \\
NGC3065 & 0.79 & 7.77 & 0.581 & 1.29 & -3.64 & 0.133 \\
NGC1427 & 0.3 & 7.80 & 1.14 & 1.38 & -4.56 & 0.589 \\
NGC2778 & 0.33 & 7.80 & 0.635 & 1.30 & -3.75 & 0.193 \\
NGC0474 & 0.37 & 7.83 & 1.10 & 1.40 & -4.45 & 0.553 \\
NGC5831 & 0.33 & 7.83 & 1.19 & 1.44 & -4.63 & 0.510 \\
UGC4551 & 0.51 & 7.88 & 1.35 & 1.55 & -4.85 & 0.534 \\
NGC4697 & 0.22 & 7.98 & 1.32 & 1.61 & -4.74 & 0.497 \\
NGC1553 & 0.74 & 8.02 & 0.711 & 1.58 & -3.71 & 0.109 \\
NGC4026 & 0.15 & 8.03 & 0.648 & 1.54 & -3.61 & 0.144 \\
NGC2634 & 0.81 & 8.08 & 1.16 & 1.68 & -4.33 & 0.172 \\
NGC2950 & 0.82 & 8.09 & 0.785 & 1.65 & -3.73 & 0.0854 \\
NGC7213 & 0.06 & 8.09 & 1.49 & 1.72 & -4.90 & 0.812 \\
NGC3266 & 0.64 & 8.09 & 1.42 & 1.75 & -4.80 & 0.372 \\
NGC5576 & 0.01 & 8.10 & 1.22 & 1.66 & -4.47 & 0.432 \\
NGC4168 & 0.17 & 8.12 & 1.86 & 1.91 & -5.52 & 0.856 \\
NGC5017 & 1.12 & 8.12 & 0.393 & 1.76 & -2.89 & 0.00675 \\
IC0875 & 1.07 & 8.14 & 0.535 & 1.76 & -3.08 & 0.0100 
     \label{tab:rates}
   \end{tabular}
\\ $^{a}$ Galaxy name; $^b$ Power law slope $\gamma$ for inner regions of Nuker $I(R)$ profile; $^c$ SMBH mass in units of $M_\odot$; $^{d}$SMBH influence radius $r_{\rm infl}$ as calculated from spherically deprojected Nuker profiles; $^{e}$Critical radius $r_{\rm crit}$ as calculated from $q(\epsilon_{\rm crit})=1$ and $\psi(r_{\rm crit})=\epsilon_{\rm crit}$; $^f$ TDE rate, calculated using Nuker parameterization; $^{g}$ fraction $f_{\rm pinhole}$ of all TDEs in the pinhole regime of disruption.
 \end{minipage}
\end{table*}
\end{center}

\begin{table*}
\renewcommand\thetable{C1}
 \begin{minipage}{180mm}
   \caption{Full results (continued).}
   \begin{tabular}{r|r|r|r|r|r|r|r|r|r|r|r|r|r|r|r|r|r|r} \\
     \hline
     Galaxy & $\gamma$ & $\log M_{\bullet}$ & $\log_{10} (r_{\rm infl}/{\rm pc})$ & $\log_{10} (r_{\rm crit}/{\rm pc})$ & $\dot{N}_{\rm TDE}^{N}$$^{b}$ & $f_{\rm pinhole}$  \\
     \hline
NGC2434 & 0.75 & 8.17 & 1.54 & 1.87 & -4.92 & 0.297 \\
NGC4660 & 0.91 & 8.17 & 0.950 & 1.81 & -3.93 & 0.0685 \\
UGC6062 & 0.8 & 8.20 & 1.27 & 1.82 & -4.42 & 0.158 \\
NGC3608 & 0.09 & 8.23 & 1.44 & 1.83 & -4.74 & 0.481 \\
NGC3193 & 0.01 & 8.25 & 1.49 & 1.86 & -4.82 & 0.516 \\
NGC5198 & 0.23 & 8.27 & 1.44 & 1.86 & -4.72 & 0.437 \\
NGC7727 & 0.43 & 8.27 & 1.42 & 1.90 & -4.71 & 0.342 \\
NGC5903 & 0.13 & 8.30 & 1.86 & 2.02 & -5.35 & 0.862 \\
NGC2902 & 0.49 & 8.31 & 1.55 & 1.94 & -4.87 & 0.449 \\
NGC0821 & 0.1 & 8.32 & 1.43 & 1.98 & -4.71 & 0.248 \\
NGC5812 & 0.59 & 8.32 & 1.33 & 1.95 & -4.52 & 0.251 \\
ESO507-27 & 0.7 & 8.36 & 1.22 & 2.00 & -4.32 & 0.131 \\
NGC4128 & 0.71 & 8.36 & 1.06 & 1.98 & -4.06 & 0.104 \\
NGC4636 & 0.13 & 8.36 & 1.82 & 2.07 & -5.25 & 0.820 \\
NGC1023 & 0.74 & 8.37 & 1.25 & 1.97 & -4.32 & 0.161 \\
UGC4587 & 0.72 & 8.39 & 1.42 & 2.07 & -4.60 & 0.123 \\
NGC3379 & 0.18 & 8.40 & 1.67 & 2.03 & -4.98 & 0.603 \\
NGC3585 & 0.31 & 8.40 & 1.29 & 2.00 & -4.39 & 0.246 \\
NGC1052 & 0.18 & 8.42 & 1.38 & 1.98 & -4.49 & 0.358 \\
A2247-M1 & 0.85 & 8.43 & 3.25 & 2.74 & -6.93 & 0.101 \\
NGC5308 & 0.82 & 8.45 & 0.663 & 2.11 & -3.31 & 0.0155 \\
ESO443-39 & 0.75 & 8.46 & 1.54 & 2.14 & -4.73 & 0.177 \\
NGC3595 & 0.75 & 8.47 & 1.53 & 2.13 & -4.68 & 0.175 \\
ESO378-20 & 0.86 & 8.48 & 0.726 & 2.17 & -3.34 & 0.00900 \\
NGC4143 & 0.59 & 8.49 & 1.24 & 2.12 & -4.27 & 0.141 \\
NGC7578B & 0.1 & 8.49 & 1.74 & 2.13 & -5.01 & 0.708 \\
NGC5898 & 0.41 & 8.53 & 1.60 & 2.19 & -4.81 & 0.320 \\
NGC4648 & 0.92 & 8.55 & 1.23 & 2.22 & -4.09 & 0.0443 \\
ESO447-30 & 0.84 & 8.56 & 1.30 & 2.25 & -4.27 & 0.0719 \\
A2040-M1 & 0.16 & 8.59 & 1.93 & 2.26 & -5.25 & 0.723 \\
NGC2907 & 0.58 & 8.59 & 1.05 & 2.27 & -3.93 & 0.0448 \\
NGC3607 & 0.26 & 8.60 & 1.67 & 2.21 & -4.84 & 0.441 \\
NGC4589 & 0.21 & 8.60 & 1.59 & 2.26 & -4.75 & 0.247 \\
NGC4621 & 0.75 & 8.61 & 1.26 & 2.35 & -4.21 & 0.0374 \\
NGC2974 & 0.62 & 8.63 & 1.39 & 2.36 & -4.43 & 0.0748 \\
NGC1316 & 0.35 & 8.64 & 1.70 & 2.31 & -4.91 & 0.289 \\
A0189-M1 & 0.85 & 8.66 & 2.06 & 2.59 & -5.46 & 0.0848 \\
NGC5845 & 0.51 & 8.70 & 1.60 & 2.37 & -4.71 & 0.193 \\
NGC3414 & 0.83 & 8.74 & 0.963 & 2.46 & -3.66 & 0.0179 \\
NGC3348 & 0.09 & 8.75 & 1.83 & 2.41 & -5.01 & 0.418 \\
NGC4278 & 0.06 & 8.75 & 1.81 & 2.43 & -5.00 & 0.294 \\
NGC5813 & 0.05 & 8.76 & 1.88 & 2.41 & -5.06 & 0.447 \\
NGC0545 & 0.08 & 8.79 & 1.91 & 2.45 & -5.07 & 0.531 \\
NGC0720 & 0.06 & 8.79 & 2.23 & 2.55 & -5.59 & 0.893 \\
NGC4709 & 0.32 & 8.79 & 1.99 & 2.49 & -5.26 & 0.452 \\
A0419-M1 & 0.33 & 8.84 & 2.14 & 2.69 & -5.49 & 0.122 \\
NGC3078 & 0.95 & 8.87 & 1.25 & 2.65 & -3.96 & 0.0166 \\
NGC3115 & 0.52 & 8.89 & 1.48 & 2.69 & -4.42 & 0.0410 \\
NGC0524 & 0.03 & 8.90 & 1.82 & 2.63 & -4.94 & 0.173 \\
NGC5557 & 0.02 & 8.91 & 1.94 & 2.64 & -5.12 & 0.237 \\
NGC4365 & 0.07 & 8.92 & 2.02 & 2.62 & -5.17 & 0.398 \\
NGC5077 & 0.23 & 8.92 & 1.90 & 2.64 & -5.03 & 0.258 \\
NGC2300 & 0.07 & 8.97 & 2.07 & 2.70 & -5.26 & 0.335 \\
IC0613 & 0.24 & 8.98 & 1.78 & 2.72 & -4.76 & 0.151 
   \end{tabular}
 \end{minipage}
\end{table*}

\begin{table*}
\renewcommand\thetable{C1}
 \begin{minipage}{180mm}
   \caption{Full results (continued)}
   \begin{tabular}{r|r|r|r|r|r|r|r|r|r|r|r|r|r|r|r|r|r|r} \\
     \hline
     Galaxy & $\gamma$ & $\log M_{\bullet}$ & $\log_{10} (r_{\rm infl}/{\rm pc})$ & $\log_{10} (r_{\rm crit}/{\rm pc})$ & $\dot{N}_{\rm TDE}^{N}$$^{b}$ & $f_{\rm pinhole}$  \\
     \hline
NGC2986 & 0.18 & 8.98 & 2.01 & 2.71 & -5.15 & 0.267 \\
NGC1500 & 0.08 & 8.99 & 1.95 & 2.73 & -5.04 & 0.208 \\
NGC7014 & 0.08 & 8.99 & 1.85 & 2.78 & -4.91 & 0.120 \\
NGC2592 & 0.92 & 9.01 & 1.82 & 2.95 & -4.82 & 0.0160 \\
NGC5838 & 0.93 & 9.02 & 0.821 & 2.82 & -3.21 & 0.00353 \\
NGC3551 & 0.13 & 9.04 & 2.06 & 2.72 & -5.15 & 0.426 \\
A0912-M1 & 0.48 & 9.04 & 2.02 & 2.95 & -5.25 & 0.0500 \\
A3144-M1 & 0.07 & 9.06 & 2.16 & 2.75 & -5.30 & 0.397 \\
NGC7052 & 0.16 & 9.06 & 1.95 & 2.78 & -4.97 & 0.286 \\
A0496-M1 & 0.1 & 9.08 & 2.39 & 10.6 & -5.67 & 0.00 \\
IC4931 & 0.09 & 9.08 & 2.12 & 2.79 & -5.23 & 0.474 \\
A3558-M1 & 0.05 & 9.10 & 2.75 & 3.01 & -6.24 & 0.917 \\
IC2738 & 0.53 & 9.10 & 1.68 & 3.03 & -4.65 & 0.0158 \\
A0376-M1 & 0.19 & 9.11 & 2.29 & 2.86 & -5.49 & 0.561 \\
NGC7626 & 0.36 & 9.11 & 1.85 & 2.93 & -4.85 & 0.0708 \\
A2147-M1 & 0.18 & 9.13 & 2.44 & 2.94 & -5.74 & 0.590 \\
NGC4874 & 0.12 & 9.13 & 2.45 & 2.95 & -5.75 & 0.696 \\
NGC4374 & 0.13 & 9.16 & 2.04 & 2.99 & -5.07 & 0.0955 \\
NGC2872 & 1.01 & 9.19 & 1.63 & 3.08 & -4.31 & 0.00994 \\
NGC5796 & 0.41 & 9.21 & 1.87 & 3.11 & -4.87 & 0.0320 \\
A0397-M1 & 0.07 & 9.22 & 2.38 & 2.95 & -5.54 & 0.671 \\
IC0115 & 0.09 & 9.23 & 2.32 & 2.99 & -5.45 & 0.308 \\
A0533-M1 & 0.06 & 9.23 & 2.27 & 3.03 & -5.39 & 0.193 \\
A3564-M1 & 0.05 & 9.23 & 2.16 & 3.05 & -5.25 & 0.126 \\
NGC0741 & 0.1 & 9.24 & 2.41 & 3.04 & -5.60 & 0.303 \\
NGC1016 & 0.09 & 9.26 & 2.25 & 3.08 & -5.36 & 0.176 \\
A0261-M1 & 0.76 & 9.35 & 3.40 & 3.68 & -6.57 & 0.0156 \\
ESO507-45 & 0.16 & 9.40 & 1.88 & 3.35 & -4.73 & 0.0228 \\
NGC3842 & 0.11 & 9.42 & 2.38 & 3.28 & -5.43 & 0.143 \\
A0295-M1 & 0.13 & 9.43 & 2.53 & 3.30 & -5.65 & 0.180 \\
NGC4649 & 0.17 & 9.59 & 2.35 & 3.62 & -5.33 & 0.0288 \\
NGC1399 & 0.12 & 9.63 & 2.37 & 3.70 & -5.36 & 0.0214 \\
IC0712 & 0.17 & 9.66 & 2.25 & 3.65 & -5.10 & 0.0377 \\
A3556-M1 & 0.09 & 9.67 & 2.43 & 3.69 & -5.38 & 0.0342 \\
IC1695 & 0.23 & 9.79 & 2.22 & 3.94 & -4.99 & 0.00606 \\
A3736-M1 & 0.11 & 9.81 & 2.67 & 3.82 & -5.69 & 0.0641 \\
A3532-M1 & 0.18 & 10.1 & 2.67 & 4.29 & -5.61 & 0.00831 \\
A3528-M1 & 0.18 & 10.2 & 2.49 & 4.48 & -5.15 & 0.00350 
   \end{tabular}
 \end{minipage}
\end{table*}

\end{document}